\documentclass[11pt]{article}
\usepackage[T1]{fontenc}
\usepackage[sc]{mathpazo}
\usepackage{times}
\usepackage{amsmath,amssymb,fullpage,graphicx}
\usepackage[colorlinks=true,linkcolor=blue,citecolor=red,bookmarksnumbered=true]{hyperref}
\usepackage{bookmark}
\usepackage{gbmacros,tocloft}
\usepackage[varg]{txfonts}

\makeatletter
\newcommand*\titlesize{\@setfontsize\titlesize{16}{1.2}}
\def\title#1{{\noindent\titlesize\bf\sffamily\color{blue}#1\par}\vglue2ex}
\def\author#1{\par{\noindent\sffamily\large#1}\par\vglue1.4ex}
\def\address#1{{\it\noindent#1}\par\medskip}

\makeatother

\advance\textwidth -4mm
\advance\hoffset 2mm
\advance\footskip 4pt
\advance\parskip 2pt
\def\eref#1{(\ref{#1})}
\def\tr{{\mathrm{tr}}}

\def\Res{\mathop{\rm Res}\nolimits}
\def\im{\mathrm{im}}
\def\ee{\mathrm{e}}
\def\phiad{\~{\phi}}
\def\half{{\textstyle\frac12}}
\let\@=\mathbf
\let\eref=\eqref
\allowdisplaybreaks

\bookmarksetup{startatroot}

\def\be{\begin{equation}}
\def\ee{\end{equation}}
\def\bse{\begin{subequations}}
\def\ese{\end{subequations}}
\def\q{\mathbf{q}}
\def\r{\mathbf{r}}
\def\btheta{\boldsymbol{\theta}}
\def\02{\boldsymbol{0}_{2 \times 2}}
\def\re{\mathrm{re}}
\def\im{\mathrm{im}}

\def\bnu{{\boldsymbol\nu}}
\def\bpsi{{\boldsymbol\psi}}
\def\brho{{\boldsymbol\rho}}
\def\vsigm{\varsigma}
\def\phiad{\tilde\phi}
\def\G{{\mathbf{G}}}
\def\z{{\hat{z}}}

\let\~=\tilde

\def\phiad{{\tilde{\phi}}}
\allowdisplaybreaks

\let\.=\dot
\let\^=\hat
\let\~=\tilde
\let\==\bar

\def\_#1{{\mathsf{#1}}}
\def\z{\hat{z}}
\def\w{\hat{w}}

\def\I{\mathrm{I}}
\def\II{\mathrm{II}}

\def\punct{^{\,[\raise0.08ex\hbox{\scriptsize$\slash$}\kern-0.34em0]}}

\newtheorem{remark}[theorem]{Remark}
\def\proof#1{\par\noindent\textbf{#1}}

\cftaftertoctitleskip \medskipamount
\begin{document}

\title{The three-component defocusing nonlinear Schr\"odinger equation with nonzero boundary conditions}
\author{Gino Biondini$^1$, Daniel K. Kraus$^1$ and Barbara Prinari$^{2,3}$}
\address{$^1$ State University of New York at Buffalo, Department of Mathematics, Buffalo, NY\\
$^2$ University of Colorado at Colorado Springs, Department of Mathematics, Colorado Springs, CO\\
$^3$ Universit\`a del Salento, Department of Physics and Mathematics, and Sezione INFN, Lecce, Italy
}

\begingroup
\small\noindent
\textbf{Abstract}
\\[1ex]
We present a rigorous theory of the inverse scattering transform (IST) for the three-component defocusing
nonlinear Schr\"odinger (NLS) equation with initial conditions approaching constant values with the same amplitude as
$x\to\pm\infty$.
The theory combines and extends to a problem with non-zero boundary conditions three fundamental ideas:
(i) the tensor approach used by Beals, Deift and Tomei for the $n$-th order scattering problem,
(ii) the triangular decompositions of the scattering matrix used by Novikov, Manakov, Pitaevski and Zakharov
for the $N$-wave interaction equations,
and
(iii) a generalization of the cross product via the Hodge star duality, which, to the best of our knowledge,
is used in the context of the IST for the first time in this work.
The combination of the first two ideas allows us to rigorously obtain a fundamental set of analytic eigenfunctions.
The third idea allows us to establish the symmetries of the eigenfunctions and scattering data.
The results are used to characterize the discrete spectrum and to obtain
exact soliton solutions, which describe generalizations of the
so-called dark-bright solitons of the two-component NLS equation.
\endgroup

\medskip
\noindent{\small\today}


\bigskip

\section{Introduction}


The vector nonlinear Schr\"{o}dinger (VNLS) equation,
$i \~{\@q}_t + \~{\@q}_{xx} - 2 \nu \|\~{\@q}\|^2 \~{\@q}= \@0$
(where $\~{\@q}:\Real\times\Real^+\to\Complex^N$, $\|\cdot\|$ is the Euclidean norm,
subscripts $x$, $t$, and $z$ denote partial differentiation
and the values $\nu = -1$ and $\nu = 1$ denote, respectively, the focusing and defocusing cases),
belongs to the class of the infinite-dimensional completely integrable integrable systems,
which possess a remarkably rich mathematical structure (e.g., see \cite{AC1991,AS1981,BBEIM,FT1987,Gesztesy,NMPZ1984} and references therein).
In particular, the scalar ($N=1$) version is referred to as the nonlinear Schr\"odinger (NLS) equation,
and the two-component version as the Manakov system.
Scalar and vector NLS systems also appear in many physical contexts, such as deep water waves, nonlinear optics, acoustics,
and Bose-Einstein condensation (e.g., see \cite{AS1981,InfeldRowlands,Sulem,Whitham} and references therein).
As a result, these systems have been the object of considerable study over the last fifty years.
The majority of works in the literature deal with ``localized'' initial conditions, i.e., situations such that
$\~{\@q}(x,t)\to\@0$ as $x\to\pm\infty$.
On the other hand,
problems in which the limit of $\~{\@q}(x,t)$ as $x\to\pm\infty$ is non-zero are also of great applicative interest in most fields
in which the NLS and VNLS equations arise.
For example, recent experiments in Bose-Einstein condensates have shown a proliferation of dark-dark solitons
in the dynamics of two miscible superfluids experiencing fast counterflow in a narrow (cigar-shaped) condensate \cite{hoefer,jpb45p115301}.
Moreover, cases in which the solutions do not vanish at infinity are relevant for the study of modulational instability \cite{lombardo2,naturephys,ZO2009,zhaoliu}
in the focusing dispersion regime.
Such problems are less well characterized from a mathematical point of view, and their study is the subject of this work.
Specifically, we consider the above VNLS equation with the following boundary conditions (BC):
$\lim\nolimits_{x \to \pm \infty}\~{\@q}(x,t) = \@q_o \,\e^{i \theta_\pm+2i\nu q_o^2t}$,
with $\theta_\pm \in \Real$ arbitrary
and with $q_o = \| \@q_o \|\ge0$.
(Note that the time dependence of the BC is dictated by the partial differential equation.)
We refer to $q_o=0$ as the case of zero boundary conditions (ZBC)
and to $q_o\ne0$ as that of nonzero boundary conditions (NZBC).
When $q_o\ne0$, it is convenient to consider a slightly modified VNLS equation, namely
\begin{equation}
\label{e:vnls}
i \@q_t + \@q_{xx} - 2 \nu ( \|\@q\|^2 - q_o^2 ) \@q = \@0,
\end{equation}
obtained via the simple change of dependent variable $\~{\@q}(x,t) = \@q(x,t) \e^{2 i \nu q_o^2 t}$.
The corresponding BC are
\begin{equation}
\label{ENZBC}
\lim_{x \to \pm \infty} \@q(x,t) = \@q_\pm = \@q_o \,\e^{i \theta_\pm},
\end{equation}
and are now independent of time, which simplifies the analysis that follows.

Recall that the initial-value problem (IVP) for integrable nonlinear partial differential equations (PDEs)
can be solved via the inverse scattering transform (IST).
In turn, the IST is based on formulating the PDE as the compatibility condition of a Lax pair \cite{lax}.
In particular, \eref{e:vnls} is associated with the following Lax pair:
\begin{gather}
\phi_x = \@X\,\phi,\qquad
\phi_t = \@T\,\phi,
\label{e:Laxpair}
\end{gather}
where
\bse
\begin{gather}
\@X(x,t,k) = -ik \@J + \@Q\,,\qquad
\@T(x,t,k) = 2ik^{2} \@J - i \@J (\@Q_x - \@Q^2 +q_o^2 ) - 2k \@Q\,,
\\
\@J= \begin{pmatrix} 1&\@0 \\ \@0&-\@I \end{pmatrix}\!,\qquad
\@Q(x,t) = \begin{pmatrix} 0& \r^T \\  \q & \@0 \end{pmatrix}\!,
\end{gather}
\ese
and $\r = \nu\q^*$.
[In other words, \eref{e:vnls} is equivalent to the zero-curvature condition
$\@X_t - \@T_x + \@X\@T - \@T\@X = \@0$, which in turn is equivalent to
the compatibility condition $\phi_{xt}=\phi_{tx}$ of~\eref{e:Laxpair}.]
The first of~\eref{e:Laxpair}
[which amounts to the eigenvalue problem for the first-order matrix ordinary differential operator $i\@J(\partial_x - \@Q)$]
is called the scattering problem,
$\phi(x,t,k)$ is referred to as the scattering eigenfunction,
$k\in \Complex$ as the scattering parameter,
and the solution $\@q(x,t)$ of the VNLS equation \eref{e:vnls} appearing in both $\@X$ and $\@T$ above
as the scattering potential.
The second of~\eref{e:Laxpair} is simply called the time evolution equation.

The IST for the focusing NLS equation with ZBC was developed in~\cite{ZS1972}
(see also Refs.~\cite{AC1991,AS1981,NMPZ1984}).
The theory was then made rigorous in \cite{BealsCoifman} (see also \cite{BDT,DeiftTrubowitz}).
The IST for the defocusing scalar case with NZBC was developed in~\cite{ZS1973}, and was subsequently rigorously revisited in
\cite{FT1987,prinari2013},
while the focusing case with NZBC was recently done in \cite{BK2014,demontis}.
The IST for the Manakov system for the focusing case with ZBC was developed in 1974 \cite{Manakov},
and the theory can be extended to an arbitrary number of components in a straightforward way \cite{APT2003}.
The vector case of NZBC is very different, however.
Note that, already in the scalar case, the IST with NZBC is considerably more involved than in the case of ZBC
\cite{prinari2013,FT1987},
and even though the IST with ZBC was settled a long time ago, some questions are still outstanding in the case of NZBC (e.g., see \cite{BF2014,BP2014}).
Moreover, the IST for vector systems with NZBC is significantly more challenging than that for scalar case.
As a result, the IST for VNLS systems with NZBC remained by and large open for a long time.

The IST for the defocusing Manakov system with NZBC was outlined in~\cite{PAB2006},
making use of an idea introduced in \cite{Kaup} to solve the IVP for the three-wave interaction equations.
The problem was then rigorously revisited in~\cite{gbdkk2015}, and
the focusing Manakov system with NZBC was studied in \cite{KBK2015} using similar methods.
Unfortunately, the approach used in \cite{gbdkk2015,KBK2015,PAB2006} 
does not extend to the VNLS equation in more than two components.

A significant step towards
the IST for the defocusing $N$-component VNLS equation with NZBC with $N >2$
was recently presented in \cite{PBT2011}.
There, the methods of \cite{BDT} were generalized to systems with NZBC and degenerate eigenvalues.
More precisely, by formulating extended eigenvalue problems for suitably defined tensors,
existence and uniqueness of solutions of such extended problems were rigorously proved.
In turn, solutions of the original scattering problem were recovered in terms of these tensors.
In this way,
a complete set of meromorphic eigenfunctions was rigorously constructed.
On the other hand, several issues were left open in \cite{PBT2011}, most notably
the symmetries of the eigenfunctions and scattering data, a complete characterization of the spectral data and of the discrete spectrum,
and a discussion of the resulting soliton solutions.
The purpose of this work is to overcome those difficulties and
present a complete and rigorous theory of IST for the defocusing VNLS equation with NZBC.


It turns out that a combination of three fundamental ideas is needed in order to develop a full IST
for the multi-component VNLS equation with NZBC.
The first idea is the use of tensors to construct a fundamental set of
meromorphic eigenfunctions, as discussed above.
The second idea is the use of triangular decompositions of the scattering matrix (generalizing the ideas of \cite{NMPZ1984}
to systems with NZBC and degenerate eigenvalues)
in order to construct fundamental analytic eigenfunctions (FAE) in terms of the meromorphic ones.
Indeed, a first contribution of this work is to combine these two ideas, which allows for a rigorous
characterization of the analyticity properties of the minors of the scattering matrix.
As it will be shown, these properties are significantly more involved than for $2\times2$ or $3\times3$ scattering problems.

For the problem at hand, however, even a combination of the above two ideas is not enough to formulate a complete theory of IST.
Indeed, it is well known that a full characterization of the symmetries of the eigenfunctions and scattering data
is needed in the inverse problem in order to correctly recover the actual solution of the PDE from the IST.
Until now, such a characterization was still missing for the multicomponent VNLS equation with NZBC.
The third and final idea, which is introduced here for the first time in the context of the IST
to the best of our knowledge,
is the use of a generalized cross product (based on the Hodge star duality)
to determine the symmetry properties of the eigenfunctions
and, as a result, of the scattering data.
As it will be shown, these properties are also significantly more involved than for $2\times2$ or $3\times3$ scattering problems.

In this work we then combine all of these three ideas to develop a complete, rigorous theory of IST
for the defocusing three-component VNLS equation with NZBC.
Several results are obtained:
(i)~The precise analyticity properties of the Jost eigenfunctions and scattering matrix are given;
(ii)~A complete set of fundamental analytic eigenfunctions is obtained for each half plane;
(iii)~The behavior of the Jost eigenfunctions and scattering coefficients at the branch points is discussed;
(iv)~The symmetries of the eigenfunctions are derived rigorously using a generalized version of the familiar cross product;
(v)~These symmetries are used to rigorously characterize the discrete spectrum and to obtain the associated
norming constants in certain cases;
(vi)~The asymptotic behavior of the Jost eigenfunctions is derived systematically;
(vii)~The inverse problem is formulated in terms of a Riemann-Hilbert problem (RHP);
(viii)~Explicit relations among all reflection coefficients are given, and all entries of the scattering matrix are determined
in the case of reflectionless solutions;
(ix)~Explicit dark-bright soliton solutions are obtained;
(x)~A comparison between the small-deviation limit of the IST and the linearization of the VNLS equation is presented;
(xi)~The reduction of the present approach to the defocusing Manakov system with NZBC is presented
and compared to the approach presented in \cite{gbdkk2015,PAB2006}.

The following notation will be used throughout this work:
asterisk denotes complex conjugation,
boldface denotes vectors and matrices of appropriate size,
superscripts $T$ and $\dag$ denote respectively matrix transpose and matrix adjoint;
$\@A_d$, $\@A_o$, $\@A_{bd}$, and $\@A_{bo}$ denote respectively the diagonal, off-diagonal,
block-diagonal, and block-off-diagonal parts of a square matrix $\@A$.
In addition,
we use the notation $(A_1,\dots,A_N)$ to denote the columns of a matrix $\@A$,
and $\@I$ and $\@0$ denote appropriately-sized identity and zero matrices, respectively.
Importantly, \textit{subscripts} $\pm$ will be used to denote quantities normalized as $x \to \pm \infty$,
whereas \textit{superscripts} $\pm$ will be used to denote regions of analyticity --- or, more generally, meromorphicity.
More precisely (with a lone exception that will be pointed out in Section~\ref{s:FME})
the superscripts $\pm$ will denote quantities that are analytic (or, more generally, meromorphic)
in the upper-half plane (UHP) or the lower-half plane (LHP) of the complex plane, denoted respectively as $\Complex^\pm$.
The proofs of all theorems, lemmas, corollaries, etc. are found in the appendix.

\def\q{\mathbf{q}}
\def\r{\mathbf{r}}
\def\btheta{\boldsymbol{\theta}}
\def\02{\boldsymbol{0}_{2 \times 2}}
\def\re{\mathrm{re}}
\def\im{\mathrm{im}}

\def\bnu{{\boldsymbol\nu}}
\def\bpsi{{\boldsymbol\psi}}
\def\brho{{\boldsymbol\rho}}
\def\vsigm{\varsigma}
\def\phiad{\tilde\phi}
\def\G{{\mathbf{G}}}
\def\z{{\hat{z}}}

\section{Direct problem}
\label{s:Laxpair}

For simplicity, in this work we will consider the case in which the asymptotic polarization vectors $\@q_\pm$ at $x\to\pm\infty$ are parallel.
(Note that the case in which $\@q_\pm$ are not parallel is still an open problem even in the 2-component case, cf.~\cite{gbdkk2015}.)\,
In this case,
thanks to the $U(N)$ invariance of~\eref{e:vnls},
without loss of generality they can be chosen to be of the form
\begin{equation}
\q_\pm = (0,0,q_\pm)^T\,,
\label{e:NZBC}
\end{equation}
with $q_\pm = q_o\,\e^{i \theta_\pm}$.
We will formulate the IST in a way that the reduction $q_o\to0$ can be taken explicitly throughout.

\subsection{Riemann surface, uniformization, Jost solutions and scattering matrix}

In order to introduce the Jost eigenfunctions, one must first study the asymptotic scattering problems as $x\to\pm\infty$,
which are given by
\begin{equation}
\label{e:phixasympscattering}
\phi_x = \@X_\pm\, \phi,
\end{equation}
where $\@X_\pm = \lim_{x\to\pm\infty}\@X = -ik\@J+\@Q_\pm$.
The eigenvalues of $\@X_\pm$ are $ik$ (with multiplicity 2) and $\pm i \lambda$, where
\begin{equation}
\lambda(k) = (k^2 - q_o^2)^{1/2}.
\label{e:Riemann}
\end{equation}
As in the scalar case \cite{ZS1973}, these eigenvalues have branching.
The branch points are the values of $k$ for which $\lambda(k)=0$, i.e., $k= \pm q_o$.
We deal with this issue as in \cite{gbdkk2015,BP2014,FT1987,PAB2006}
by introducing the two-sheeted Riemann surface defined by~\eref{e:Riemann},
taking the branch cut on $(-\infty,-q_o] \cup [q_o,\infty)$,
and we define $\lambda(k)$ so that $\Im\lambda\ge0$ on sheet~$\Complex_I$ and $\Im\lambda(k)\le0$ on sheet~$\Complex_II$
(see~\cite{PAB2006} for further details).
Next, we introduce a uniformization variable via the conformal mapping
\begin{equation}
z=k+\lambda\,.
\end{equation}
Importantly, the inverse map yields both $k$ and $\lambda$ as rational functions of $z$:
\begin{equation}
\label{e:uniform}
k = \half\, (z+q_o^2 / z), \qquad
\lambda = \half\, (z-q_o^2 / z).
\end{equation}
One can then express all $k$-dependence of eigenfunctions and scattering data (including the one resulting from~$\lambda$)
in terms of $z$,
thereby eliminating all branching.
The branch cuts on the two sheets of the Riemann surface are mapped onto the real $z$-axis;
$\Complex_\I$ is mapped onto the upper-half plane of the complex $z$-plane;
$\Complex_\II$ is mapped onto the lower-half plane of the complex $z$-plane;
$z(\infty_\I) = \infty$ if $\Im(k)>0$;
$z(\infty_\I) = 0$ if $\Im(k)<0$;
$z(\infty_\II) = 0$ if $\Im(k)>0$;
$z(\infty_\II) = \infty$ if $\Im(k)<0$;
$z(k,\lambda_\I)z(k,\lambda_\II)=q_o^2$;
$|k| \to \infty$ in the upper-half plane of $\Complex_\I$ corresponds to $z \to \infty$ in the upper-half $z$-plane;
$|k| \to \infty$ in the lower-half plane of $\Complex_\II$ corresponds to $z \to \infty$ in the lower-half $z$-plane;
$|k| \to \infty$ in the lower-half plane of $\Complex_\I$ corresponds to $z \to 0$ in the upper-half $z$-plane; and
$|k| \to \infty$ in the upper-half plane of $\Complex_\II$ corresponds to $z \to 0$ in the lower-half $z$-plane.
Finally, the segments $k\in[-q_o,q_o]$ in each sheet correspond, respectively, to the upper-half and lower-half of the
circle $C_o$ of radius $q_o$ centered at the origin in the complex $z$-plane.


We are now ready to introduce the Jost solutions over the
continuous spectrum $\Sigma$, which consists of all values of $k$ (in either sheet) such that $\lambda(k) \in \Real$;
that is, $k \in \Real \setminus (-q_o,q_o)$.
In the complex $z$-plane, the corresponding set is the whole real axis.
We write the eigenvalues and the corresponding eigenvector matrices
of the asymptotic scattering problems~\eref{e:phixasympscattering}
as
\begin{equation}
\label{e:Epm}
i \@{\Lambda}(z) = \mathrm{diag}(-i \lambda,ik,ik,i \lambda),
\qquad \@E_\pm(z) = \begin{pmatrix} 1 & -(i/z) \q_\pm^\dag \\ (i/z) \q_\pm & \@I \end{pmatrix}
= \@I - (i/z) \@J \@Q_\pm,
\end{equation}
respectively, so that
\begin{equation}
\@X_\pm \@E_\pm = \@E_\pm i \@{\Lambda}.
\end{equation}
This normalization is the generalization of the one used in \cite{gbdkk2015,BP2014} for the scalar case.
For future reference, note that
\begin{equation}
\det\@E_\pm(z) = 1-q_o^2/z^2 := \gamma(z),\qquad
\@E_\pm^{-1}(z) = \frac{1}{\gamma(z)} \left[\diag (1,\gamma(z),\gamma(z),1) + (i/z) \@J \@Q_\pm\right].
\label{e:einverse}
\end{equation}
We now discuss the asymptotic time dependence.
As $x \to \pm \infty$, the time evolution of the solutions of the Lax pair is asymptotic to
\begin{equation}
\phi_t = \@T_\pm \phi,
\end{equation}
where $\@T_\pm = \lim_{x\to\pm\infty} \@T = 2ik^2 \@J + i\@J \@Q_\pm^2 - iq_o^2 \@J - 2k\@Q_\pm$.
The eigenvalues of $\@T_\pm$ are $-i(k^2 + \lambda^2)$ (with multiplicity two) and $\pm 2ik \lambda$.
Since the BC are constant, the consistency of the Lax pair \eref{e:Laxpair} implies
$\@X_\pm\@T_\pm=\@T_\pm\@X_\pm$, so $\@X_\pm$ and $\@T_\pm$ admit common eigenvectors.
Indeed,
\begin{equation}
\label{e:simultaneous}
\@T_\pm \@E_\pm = -i \@E_\pm\@{\Omega}, \qquad \@{\Omega}(z) = \mathrm{diag}(-2k \lambda,k^2 + \lambda^2, k^2+\lambda^2, 2k \lambda)\,.
\end{equation}
Thus, $\forall z \in \Real$, we can define the Jost solutions $\phi_\pm(x,t,z)$
as the \textit{simultaneous} solutions of both parts of the Lax pair satisfying the BC
\begin{equation}
\label{e:phixasymp}
\phi_\pm (x,t,z) = \@E_\pm (z) \e^{i \@{\Theta} (x,t,z)} + o(1), \qquad x \to \pm \infty,
\end{equation}
where 
\begin{equation}
\label{e:thethetamatrix}
\@{\Theta}(x,t,z)= \@{\Lambda}(z) x - \@{\Omega}(z) t = \mathrm{diag}(\theta_1(x,t,z),\theta_2(x,t,z),\theta_2(x,t,z),-\theta_1(x,t,z)).
\end{equation}
As usual, the advantage of introducing simultaneous solutions of both parts of the Lax pair is that
all of the scattering data will be independent of time.

Of course one must still rigorously prove that such Jost eigenfunctions are well-defined.
The above discussion can be made rigorous by defining the Jost eigenfunctions
as the solutions of appropriate Volterra linear integral equations.
One can remove the asymptotic exponential oscillations and introduce modified eigenfunctions:
\begin{equation}
\mu_\pm (x,t,z) = \phi_\pm(x,t,z) \e^{-i \@{\Theta}(x,t,z)},
\label{e:modified}
\end{equation}
so that
\begin{equation}
\displaystyle\lim_{x \to \pm \infty} \mu_\pm (x,t,z) = \@E_\pm(z).
\end{equation}
Using the modified eigenfunctions, in Appendix~\ref{a:direct} we show that,
for all $z\in\Real\setminus\{0,\pm q_o\}$,
$\mu_-(x,t,z)$ exists, is unique, and is uniformly continuous over the interval $x\in(-\infty,a]$
for all $a\in\Real$
if  $q(x,t)-q_-\in L^1(-\infty,a)$.
Similarly for $\mu_+(x,t,z)$ over $[a,\infty)$
if  $q(x,t)-q_+\in L^1(a,\infty)$.
Moreover, the columns $\phi_{-,1}(x,t,z)$ and $\phi_{+,4}(x,t,z)$ can be analytically continued onto the upper-half $z$-plane, with continuous projection to the real $z$-axis (except possibly at the branch points),
while $\phi_{+,1}(x,t,z)$ and $\phi_{-,4}(x,t,z)$ can be analytically continued onto the lower-half $z$-plane, again with continuous projection to the real axis.
Finally, if  $(1+|x|)(q(x,t)-q_\pm)\in L^1(\Real^\pm)$, the Jost solutions are also well-defined at
both of the branch points $\pm q_o$.

As explained before, the point $z=0$ is the image of the points at infinity in the
$k,\lambda$ variables, and therefore the behavior of the Jost solutions as $z\rightarrow 0$ will be analyzed separately,
together with the behavior as $z\rightarrow \infty$. Anticipating the results in Sec.~\ref{s:asymptoticsoriginal},
$\mu_{\pm,1} (x,t,z)$ and $\mu_{\pm,4} (x,t,z)$ behave like $1/z$ as $z\rightarrow 0$, while the remaining columns of
$\mu_\pm(x,t,z)$ and the scattering coefficients are all finite as $z\rightarrow 0$ [cf. \eref{e:mu+1inf}, \eref{e:zasympreal} and \eref{e:Ainf0}].

\begin{remark}
As in the focusing and defocusing Manakov system with NZBC \cite{PAB2006,gbdkk2015,KBK2015},
in general, the remaining columns of the Jost eigenfunctions do not admit analytic continuation
off the real $z$-axis.
The difference between the 2-component case and the 3-component case is that,
for the Manakov system, this defect of analyticity only affected one of the columns of
each of $\phi_\pm$,
and it was obviated by using the adjoint scattering problem
to obtain two auxiliary analytic eigenfunctions \cite{gbdkk2015,PAB2006}.
That approach, though, cannot be extended to the three-component case.
The resolution of this problem will be discussed in Section~\ref{s:FME}.
\end{remark}

The scattering matrix is introduced in a standard way.
If $\phi(x,t,z)$ solves~\eref{e:Laxpair},
Abel's formula implies $\partial_x (\det\phi) = \tr\@X \,\det\phi$
and $\partial_t (\det\phi) = \tr\@T\,\det\phi$.
Since $\tr\@X=2ik$ and $\tr\@T= -2i(k^2+\lambda^2)$, we have
\begin{equation*}
\partialderiv{ }x \det(\phi_\pm(x,t,z)\e^{-i \@{\Theta}(x,t,z)}) =
\partialderiv{ }t \det(\phi_\pm(x,t,z)\e^{-i \@{\Theta}(x,t,z)}) = 0.
\end{equation*}
Then~\eref{e:phixasymp} implies
\begin{equation}
\label{e:det}
\det \phi_\pm(x,t,z) = \gamma(z) \,\e^{2i\theta_2(x,t,z)}, \qquad
(x,t) \in \Real^2\,,\quad
z\in\Real\,.
\end{equation}
Thus, for all $z\in\Real\setminus\{\pm q_o\}\,$,
$\phi_-(x,t,z)$ and $\phi_+(x,t,z)$ are both fundamental matrix solutions of the Lax pair,
so there exists a $4 \times 4$ matrix $\@A(z)$ such that
\begin{equation}
\phi_-(x,t,z) = \phi_+(x,t,z) \@A(z), \quad z \in \Real\setminus\{\pm q_o\}\,.
\label{e:scattering}
\end{equation}
As usual, $\@A(z) = (a_{ij}(z))$ is referred to as the \textit{scattering matrix}, and its entries
as \textit{scattering coefficients}.
Note that with our normalizations for the Jost eigenfunctions, $\@A(z)$ is independent of time.
Moreover, \eref{e:det} and \eref{e:scattering} imply
\begin{equation}
\label{e:unitdet}
\det\@A(z) = 1\,,\qquad z \in \Real\setminus\{\pm q_o\}\,.
\end{equation}
It is also convenient to introduce $\@B(z) := \@A^{-1}(z) = (b_{ij}(z))$.
In the scalar case, the analyticity of the diagonal scattering coefficients follows from their
integral representations in terms of analytic eigenfunctions,
while for the Manakov system, alternative integral representations of the Jost eigenfunctions can be used to
prove analogous results (see \cite{gbdkk2015} for more details).
In the $N$-component case, however, it will be necessary to use a different approach,
which was introduced in Ref.~\cite{PBT2011},
and which will also yield more general results
(that are also applicable to the Manakov system, as we will see in Section~\ref{s:manakovsec}).

So far, the setup is essentially the same as for the Manakov system,
with the deficiency in the number of analytic Jost eigenfunctions being the most glaring difference.
As mentioned above, however,
the method of circumventing this defect of analyticity here is completely different from the one used in the Manakov case.
We turn to this issue next.

\subsection{Tensors and fundamental meromorphic eigenfunctions}
\label{s:FME}

In this section and the next we show how a combination of the methods of~\cite{BDT}
and the approach of~\cite{NMPZ1984} can be used
to construct a complete set of analytic eigenfunctions in each half plane.
We begin by briefly recalling some results from \cite{PBT2011}, where
generalizing the approach introduced in \cite{BDT} for the $N$-th order scattering operator,
a rigorous formalism was derived to obtain a fundamental set of meromorphic eigenfunctions.
First, following \cite{PBT2011} we introduce the following quantities:
\begin{definition}
\label{d:fg}
Given the columns of the matrices $\mu_\pm(x,t,z)$,
for all $n = 1,\dots,4$ and all $z \in \Real$
we define the totally antisymmetric fundamental tensors
\begin{gather*}
f_n^+(x,t,z) =
\mu_{-,1}(x,t,z) \wedge \cdots \wedge \mu_{-,n}(x,t,z),
\qquad
f_n^-(x,t,z) =
\mu_{-,n}(x,t,z) \wedge \cdots \wedge \mu_{-,4}(x,t,z),
\\
g_n^+(x,t,z) =
\mu_{+,n}(x,t,z) \wedge \cdots \wedge \mu_{+,4}(x,t,z),
\qquad
g_n^-(x,t,z) =
\mu_{+,1}(x,t,z) \wedge \cdots \wedge \mu_{+,n}(x,t,z),
\end{gather*}
where the symbol ``$\wedge$" denotes the wedge product as in~\cite{PBT2011}.
\end{definition}
The following result was proved in~\cite{PBT2011}:
\begin{theorem}
For all $n=1,\dots,4$, the fundamental tensors $f_n^\pm(x,t,z)$ and $g_n^\pm(x,t,z)$ can be extended analytically to the following regions:
\begin{gather*}
f_n^+(x,t,z), g_n^+(x,t,z): \quad \Im z > 0,
\qquad
f_n^-(x,t,z), g_n^-(x,t,z): \quad \Im z < 0,
\label{t:tensoranalyticity}
\end{gather*}
with continuous projection to the real axis.
\end{theorem}

\begin{remark}
The definitions of the tensors $f_n^-(x,t,z)$ and $g_n^-(x,t,z)$
differ slightly from those in~\cite{PBT2011}.
More precisely, in \cite{PBT2011} the ordering of the eigenvalues of $\@X_\pm$ was reversed in the LHP, and
as a result the same definition was used for $f_n^\pm$ and $g_n^\pm$ in the UHP and LHP.
It will be apparent later how these changes affect the analysis to our benefit.
\end{remark}
Theorem~\ref{t:tensoranalyticity} also leads to the following result~\cite{PBT2011}:
\begin{theorem}
There exist scalar functions $\Delta_1^\pm(z)$, $\Delta_2^\pm(z)$, and $\Delta_3^\pm(z)$, analytic on $\Complex^\pm$,
respectively, with smooth extensions to $\Real \setminus \{ \pm q_o \}$ from their respective regions of analyticity,
such that the following hold for all $n =1,2,3$:%
\bse
\begin{gather}
f_n^+(x,t,z) \wedge g_{n+1}^+(x,t,z) =
\Delta_n^+(z) \gamma_n(z) \,\@e_1 \wedge \cdots \wedge \@e_4, \qquad \Im z > 0,
\\
g_n^-(x,t,z) \wedge f_{n+1}^-(x,t,z) =
\Delta_n^-(z) \gamma_n^*(z^*) \,\@e_1 \wedge \cdots \wedge \@e_4, \qquad \Im z < 0,
\end{gather}
\ese
where
\begin{equation}
\gamma_n(z) = \det (\@E_{-,1}(z),\dots,\@E_{-,n}(z),\@E_{+,n+1}(z),\dots,\@E_{+,4}(z))
\label{e:gammandef}
\end{equation}
and $\{\@e_j\}_{j=1}^4$ is the standard basis for $\Real^4$.
\end{theorem}
(Again, note the slight difference between what is presented here and what was presented in~\cite{PBT2011}.)
In Section~\ref{s:FAE} we will establish a precise connection between these functions and the scattering matrix,
which will allow us to determine the analyticity properties of the scattering data.
We will also see that the functions $\Delta^\pm(z)$ will provide part of the spectral data of the problem.
\begin{definition}
(Discrete spectrum)
Define
\begin{equation}
Z^\pm = \{z \in \Complex^\pm : \prod_{j=1}^3 \Delta^\pm_j(z) = 0\}.
\end{equation}
\end{definition}
We will show in Section~\ref{s:discretespectrum} that $Z^\pm$ is in fact the discrete spectrum.
First, however, we need to construct a fundamental set of analytic eigenfunctions.
To this end, we need the following results \cite{PBT2011}:
\begin{theorem}
\label{t:mwedgedef}
(Fundamental meromorphic eigenfunctions)
For all $z \in \Complex^\pm \setminus Z^\pm$, there exist unique analytic fundamental matrix solutions
$\Xi^\pm(x,t,z)$ of the scattering problem, with columns $\Xi_n^\pm(x,t,z)$ for $n=1,\dots 4$,
defined by
\bse
\begin{gather}
f_n^+(x,t,z) = f_{n-1}^+(x,t,z) \wedge \Xi_n^+(x,t,z), \qquad \Xi_n^+(x,t,z) \wedge g_n^+(x,t,z) = \@0,
\\
f_{n-1}^-(x,t,z) = \Xi_{n-1}^-(x,t,z) \wedge f_n^-(x,t,z), \qquad g_n^-(x,t,z) \wedge \Xi_n^-(x,t,z) = \@0,
\end{gather}
\ese
together with the weak boundary conditions
\bse
\begin{gather}
\lim_{x \to -\infty} \@E_{-,1}(z) \wedge \cdots \wedge \@E_{-,n-1}(z) \wedge \Xi_n^+(x,t,z)
= \@E_{-,1}(z) \wedge \cdots \wedge \@E_{-,n}(z),
\\
\lim_{x \to -\infty} \Xi_n^-(x,t,z) \wedge \@E_{-,n+1}(z) \wedge \cdots \wedge \@E_{-,4}(z)
= \@E_{-,n}(z) \wedge \cdots \wedge \@E_{-,4}(z),
\end{gather}
\ese
and with the following asymptotic behavior at the opposite infinity:
\bse
\begin{gather}
\lim_{x \to \infty} \Xi_n^+(x,t,z) \wedge \@E_{+,n+1}(z) \wedge \cdots \wedge \@E_{+,4}(z)
= \frac{\gamma_n(z)}{\gamma_{n-1}(z)} \frac{\Delta_n^+(z)}{\Delta_{n-1}^+(z)}
\@E_{+,n}(z) \wedge \cdots \wedge \@E_{+,4}(z),
\\
\lim_{x \to \infty} \@E_{+,1}(z) \wedge \cdots \wedge \@E_{+,n-1}(z) \wedge \Xi_n^-(x,t,z)
= \frac{\gamma_{n-1}^*(z^*)}{\gamma_n^*(z^*)} \frac{\Delta_{n-1}^-(z)}{\Delta_n^-(z)}
\@E_{+,1}(z) \wedge \cdots \wedge \@E_{+,n}(z),
\end{gather}
\ese
with $\gamma_n(z)$ given by \eref{e:gammandef} as before.
\end{theorem}
\begin{theorem}
\label{t:meromorphic}
For all $z \in \Complex^\pm \setminus Z^\pm$,
the matrices $\@m^\pm(x,t,z) = \Xi^\pm(x,t,z)\,\e^{-i\@\Theta(x,t,z)}$ have
the following asymptotic behavior:
\bse
\label{e:auxxasymp}
\begin{gather}
\lim_{x \to -\infty} \@m^+(x,t,z) = \@E_-(z) \alpha_o^+(z), \quad \lim_{x \to \infty} \@m^+(x,t,z) = \@E_+(z) \beta_o^+(z),
\\
\lim_{x \to -\infty} \@m^-(x,t,z) = \@E_-(z) \beta_o^-(z), \quad \lim_{x \to \infty} \@m^-(x,t,z) = \@E_+(z) \alpha_o^-(z),
\end{gather}
\ese
where $\alpha_o^\pm(z)$ and $\beta_o^\pm(z)$
have the following structure:
\begin{gather*}
\alpha_o^+(z) = \begin{pmatrix} 1&0&0&0 \\ 0&1&\alpha_{o,23}^+&0 \\ 0&0&1&0 \\ 0&0&0&1 \end{pmatrix},
\quad
\beta_o^-(z) = \begin{pmatrix} 1&0&0&0 \\ 0&1&0&0 \\ 0&\beta_{o,32}^-&1&0 \\ 0&0&0&1 \end{pmatrix},
\\
\alpha_o^-(z) = \begin{pmatrix} {\=\gamma^*(z^*)}/{\Delta_1^-} &0&0&0 \\
  0&{\Delta_1^-}/{\Delta_2^-}&\alpha_{o,23}^-&0 \\
  0&0&{\Delta_2^-}/{\Delta_3^-}&0 \\ 0&0&0&{\Delta_3^-}/{\=\gamma^*(z^*)} \end{pmatrix},
\quad
\beta_o^+(z) = \begin{pmatrix} {\Delta_1^+}/{\=\gamma} &0&0&0 \\
  0&{\Delta_2^+}/{\Delta_1^+}&0&0 \\
  0 & \beta_{o,32}^+ & {\Delta_3^+}/{\Delta_2^+}&0 \\ 0&0&0& {\=\gamma}/{\Delta_3^+}
\end{pmatrix},
\end{gather*}
where the $z$-dependence in the right-hand side was omitted for brevity,
and
\begin{equation}
\=\gamma(z) = \gamma(z)/(1 - q_- r_+/z^2)\,.
\label{e:gammabardef}
\end{equation}
Moreover, the matrices $\alpha_o^+(z)$, $\beta_o^-(z)$, $\alpha_o^-(z)\@\Delta^-(z)$ and $\beta_o^+(z)\@\Delta^+(z)$
are analytic for all $z\in\Complex^\pm \setminus \Real$,
with
\begin{equation}
\@\Delta^-(z) = \diag(\Delta_1^-,\Delta_2^-,\Delta_3^-,1)\,,
\qquad
\@\Delta^+(z) = \diag(1,\Delta_1^+,\Delta_2^+,\Delta_3^+)\,.
\end{equation}
Finally, both $\@m^\pm(x,t,z)$ and $\alpha_o^\pm(z)$ and $\beta_o^\pm(z)$
have smooth projections to $\Real \setminus \{\pm q_o\}$ \cite{PBT2011}.
\end{theorem}
Comparing this result to~\cite{PBT2011}, the slight difference is now clear:
Essentially, all that is done is the switching of the behaviors as $x \to \pm \infty$ in the lower-half $z$-plane.
%
Taking the limit $\Im z\to0^\pm$ one has \cite{PBT2011}:

\begin{theorem}
\label{t:2011_2}
(Projection to the real axis)
For $z \in \Real$,
\bse
\begin{gather}
\@m^+(x,t,z) = \@E_-(z) \e^{i \@\Theta(x,t,z)} \alpha^+(z) \,\e^{-\i \@\Theta(x,t,z)} + o(1), \quad x \to -\infty,
\\
\@m^+(x,t,z) = \@E_+(z) \e^{i \@\Theta(x,t,z)} \beta^+(z) \,\e^{- i \@\Theta(x,t,z)} + o(1), \quad x \to \infty,
\\
\@m^-(x,t,z) = \@E_-(z) \e^{i \@\Theta(x,t,z)} \beta^-(z) \,\e^{- i \@\Theta(x,t,z)} + o(1), \quad x \to -\infty,
\\
\@m^-(x,t,z) = \@E_+(z) \e^{i \@\Theta(x,t,z)} \alpha^-(z) \,\e^{- i \@\Theta(x,t,z)} + o(1), \quad x \to \infty,
\end{gather}
\ese
where
\be
\alpha^\pm(z) = \alpha_o^\pm + \begin{pmatrix} 0&\alpha_{12}^\pm&\alpha_{13}^\pm& \alpha_{14}^\pm \\ 0&0&0&\alpha_{24}^\pm \\ 0&0&0&\alpha_{34}^\pm \\ 0&0&0&0 \end{pmatrix},
\quad
\beta^\pm(z) = \beta_o^\pm + \begin{pmatrix} 0&0&0&0 \\ \beta_{21}^\pm &0&0&0 \\ \beta_{31}^\pm&0&0&0 \\
  \beta_{41}^\pm &\beta_{42}^\pm &\beta_{43}^\pm&0 \end{pmatrix},
\label{e:alphapmdef}
\ee
and where the $z$-dependence in the right-hand side was omitted for brevity.
\end{theorem}
\begin{remark}
We emphasize that that not all entries of $\alpha^\pm(z)$ and $\beta^\pm(z)$ are obtained from
the entries of $\alpha_o^\pm(z)$ and $\beta_o^\pm(z)$.
This is because 
the limits $x \to \pm \infty$ and $z \to \Real$ do not commute in general.
Also, the additional entries of $\alpha^\pm(z)$ and $\beta^\pm(z)$
cannot in general be extended off the real $z$-axis.
In other words,
these additional entries are not projections to the real axis of analytic functions.
\end{remark}

Thus, for the matrix entries of $\alpha^\pm(z)$ and $\beta^\pm(z)$
that appear explicitly in the right-hand side of~\eref{e:alphapmdef}, the superscripts $\pm$ do not denote analyticity.
This is the only instance in which we will deviate from the convention.

Next, we note that $\@m^\pm(x,z,t) \e^{i \@\Theta(x,t,z)}$ solve the Lax pair in \eref{e:Laxpair}.
We then compare the asymptotics in Theorem \ref{t:2011_2} with those in \eref{e:phixasymp} to explicitly obtain
the decomposition of the fundamental meromorphic eigenfunctions in terms of the Jost eigenfunctions for $z \in \Real$:
\bse
\label{e:Mdef}
\begin{gather}
\@m^+(x,t,z) \e^{i \@\Theta(x,t,z)} = \phi_-(x,t,z) \alpha^+(z) = \phi_+(x,t,z) \beta^+(z),
\\
\@m^-(x,t,z) \e^{i \@\Theta(x,t,z)} = \phi_-(x,t,z) \beta^-(z) = \phi_+(x,t,z) \alpha^-(z).
\end{gather}
\ese
We will show how the matrices $\@m^\pm(x,t,z)$ and relations~\eref{e:Mdef}
provide all the information needed to formulate the inverse problem.

\subsection{Triangular decompositions and fundamental analytic eigenfunctions}
\label{s:FAE}
\def\minorset#1#2{{\big(\substack{#1\\#2}\big)}}

The results of the previous section provide the framework to obtain a set of fundamental analytic eigenfunctions.
Namely, generalizing ideas presented in~\cite{NMPZ1984} for a first-order matrix system with ZBC,
we use~\eref{e:Mdef} to write triangular decompositions of the scattering matrix
and use these decompositions to obtain the analyticity properties
of the scattering matrix and a complete set of analytic eigenfunctions for each half plane.
Indeed, in Appendix~\ref{a:direct} we show the following:

\begin{lemma}
\label{L:triang}
(Triangular decomposition of the scattering matrix)
The scattering matrix $\@A(z)$, defined for all $z \in \Real\setminus\{\pm q_o\}$ by \eref{e:scattering},
admits the following triangular decompositions:
\begin{equation}
\@A(z) = \beta^+(z) [\alpha^+(z)]^{-1} = \alpha^-(z) [\beta^-(z)]^{-1},
\label{e:triangular}
\end{equation}
with $\alpha^\pm(z)$ and $\beta^\pm(z)$ as in~\eref{e:alphapmdef}.
\end{lemma}
Hereafter, we will use the notation of~\cite{gantmacher}:
\begin{definition}(Minors)
Let $\@C=(c_{ij})$ be an $N \times N$ matrix.
A minor of $\@C$ is a determinant of the form
\begin{equation}
C_{\minorset{i_1,i_2,\dots,i_p}{k_1,k_2,\dots,k_p}} =
\det \begin{pmatrix} c_{i_1k_1} & c_{i_1k_2} & \dots & c_{i_1k_p}
\\
c_{i_2k_1} & c_{i_2k_2} & \cdots & c_{i_2k_p}
\\
\vdots & \vdots & \ddots & \vdots
\\
c_{i_pk_1} & c_{i_pk_2} & \cdots & c_{i_pk_p} \end{pmatrix},
\end{equation}
where $1 \leq i_1 < i_2 < \cdots < i_p \leq N$ and $1 \leq k_1 < k_2 < \cdots < k_p \leq N$.
The upper and lower principal minors of $\@C$ are, respectively, determinants of the form
\begin{equation}
C_{[1,\dots,p]} = C_{\minorset{1,\dots,p}{1,\dots,p}}, \qquad
C_{[p,\dots,N]} = C_{\minorset{p,p+1,\dots,N}{p,p+1,\dots,N}},
\qquad 1 \leq p \leq N.
\end{equation}
\end{definition}
The minors of an $N \times N$ matrix $\@A$ and those of its inverse $\@B= \@A^{-1}$ are related as follows~\cite{gantmacher}:
\begin{lemma}
\label{L:relation}
For arbitrary $1 \leq i_1 < i_2 < \cdots < i_p \leq N$ and $1 \leq k_1 < k_2 < \cdots < k_p \leq N$,
\begin{equation}
B_{\minorset{i_1,i_2,\dots,i_p}{k_1,k_2,\dots,k_p}}
  = (-1)^{\sum_{j=1}^{p} [i_j + k_j]} A_{\minorset{k_1',k_2',\dots,k_{N-p}'}{i_1',i_2',\dots,i_{N-p}'}},
\end{equation}
where $i_1 < i_2 < \cdots < i_p$ and $i_1' < i_2' < \cdots < i_{N-p}'$
(and similarly for the $k_j$ and $k_j'$)
form a complete set of indices $1,\dots,N$.
\end{lemma}
In particular, for the scattering matrix $\@A(z)$ and its inverse, Lemma~\ref{L:relation} implies the following identities:
\bse
\label{e:AB}
\begin{gather}
A_{[1]}(z) = B_{[2,3,4]}(z)\,,\qquad
A_{[1,2]}(z) = B_{[3,4]}(z)\,,\qquad
A_{[1,2,3]}(z) = B_{[4]}(z)\,,\\
A_{[4]}(z) = B_{[1,2,3]}(z)\,,\qquad
A_{[3,4]}(z) = B_{[1,2]}(z)\,,\qquad
A_{[2,3,4]}(z) = B_{[1]}(z)\,.
\end{gather}
\ese
In other words, both $\@A(z)$ and $\@B(z)$ have three nontrivial upper principal minors and three nontrivial lower principal minors
[since $A_{[1,2,3,4]}(z) = \det\@A(z) = 1 = \det\@B(z) = B_{[1,2,3,4]}(z)$];
however, only three among these six minors are independent.
In Appendix~\ref{a:direct} we use these results to characterize the
decomposition~\eref{e:triangular} of the scattering matrix $\@A(z)$
while taking into consideration the constraints given in Theorem~\ref{t:meromorphic}:
\begin{lemma}
\label{L:decompexplicit}
The matrices $\alpha^\pm(z)$ and $\beta^\pm(z)$ in Theorem~\ref{t:2011_2}
are related to the minors of the scattering matrices $\@A(z)$ and $\@B(z)$ as follows:
\bse
\label{e:ludecomp}
\begin{gather}
\alpha^+(z)\@D^+(z) = \begin{pmatrix}
1 & - A_{\minorset{1}{2}} & A_{\minorset{1,2}{2,3}} & - A_{\minorset{1,2,3}{2,3,4}}
\\
0 & A_{[1]} & - A_{\minorset{1,2}{1,3}} & A_{\minorset{1,2,3}{1,3,4}}
\\
0 & 0 & A_{[1,2]} & - A_{\minorset{1,2,3}{1,2,4}}
\\
0 & 0 & 0 & A_{[1,2,3]}
\end{pmatrix}
= \begin{pmatrix}
1 & B_{\minorset{1,3,4}{2,3,4}} & B_{\minorset{1,4}{3,4}} & B_{\minorset{1}{4}}
\\
0 & B_{[2,3,4]} & B_{\minorset{2,4}{3,4}} & B_{\minorset{2}{4}}
\\
0 & 0 & B_{[3,4]} & B_{\minorset{3}{4}}
\\
0 & 0 & 0 & B_{[4]}
\end{pmatrix},
\\
\beta^+(z)\@D^+(z) = \begin{pmatrix}
A_{[1]} & 0 & 0 & 0
\\
A_{\minorset{2}{1}} & A_{[1,2]} & 0 & 0
\\
A_{\minorset{3}{1}} & A_{\minorset{1,3}{1,2}}
& A_{[1,2,3]} & 0
\\
A_{\minorset{4}{1}} & A_{\minorset{1,4}{1,2}}
& A_{\minorset{1,2,4}{1,2,3}} & 1
\end{pmatrix}
= \begin{pmatrix}
B_{[2,3,4]} & 0 & 0 & 0
\\
- B_{\minorset{2,3,4}{1,3,4}} & B_{[3,4]} & 0 & 0
\\
B_{\minorset{2,3,4}{1,2,4}} & - B_{\minorset{3,4}{2,4}} & B_{[4]} & 0
\\
- B_{\minorset{2,3,4}{1,2,3}} & B_{\minorset{3,4}{2,3}} & - B_{\minorset{4}{3}} & 1
\end{pmatrix},
\\
\alpha^-(z)\@D^-(z) = \begin{pmatrix} 1 & A_{\minorset{1,3,4}{2,3,4}} & A_{\minorset{1,4}{3,4}} & A_{\minorset{1}{4}}
\\
0 & A_{[2,3,4]} & A_{\minorset{2,4}{3,4}} & A_{\minorset{2}{4}}
\\
0 & 0 & A_{[3,4]} & A_{\minorset{3}{4}}
\\
0 & 0 & 0 & A_{[4]} \end{pmatrix}
= \begin{pmatrix} 1 & - B_{\minorset{1}{2}} & B_{\minorset{1,2}{2,3}} & - B_{\minorset{1,2,3}{2,3,4}}
\\
0 & B_{[1]} & - B_{\minorset{1,2}{1,3}} & B_{\minorset{1,2,3}{1,3,4}}
\\
0 & 0 & B_{[1,2]} & - B_{\minorset{1,2,3}{1,2,4}}
\\
0 & 0 & 0 & B_{[1,2,3]} \end{pmatrix},
\\
\beta^-(z)\@D^-(z) = \begin{pmatrix}
A_{[2,3,4]} & 0 & 0 & 0
\\
- A_{\minorset{2,3,4}{1,3,4}} & A_{[3,4]} & 0 & 0
\\
A_{\minorset{2,3,4}{1,2,4}} & - A_{\minorset{3,4}{2,4}} & A_{[4]} & 0
\\
- A_{\minorset{2,3,4}{1,2,3}} & A_{\minorset{3,4}{2,3}} & - A_{\minorset{4}{3}} & 1
\end{pmatrix}
= \begin{pmatrix}
B_{[1]} & 0 & 0 & 0
\\
B_{\minorset{2}{1}} & B_{[1,2]} & 0 & 0
\\
B_{\minorset{3}{1}} & B_{\minorset{1,3}{1,2}} & B_{[1,2,3]} & 0
\\
B_{\minorset{4}{1}} & B_{\minorset{1,4}{1,2}} & B_{\minorset{1,2,4}{1,2,3}} & 1
\end{pmatrix},
\end{gather}
\ese
where
\bse
\begin{gather}
\@D^+(z) = \diag\left(1,A_{[1]},A_{[1,2]},A_{[1,2,3]}\right)
= \diag\left(1,B_{[2,3,4]},B_{[3,4]},B_{[4]}\right),
\\
\@D^-(z) = \diag\left(A_{[2,3,4]},A_{[3,4]},A_{[4]},1\right)
= \diag\left(B_{[1]},B_{[1,2]},B_{[1,2,3]},1\right),
\end{gather}
\ese
and where the $z$-dependence in the right-hand side was omitted for brevity.
\end{lemma}
Comparing the matrices~\eref{e:ludecomp} with their definitions in Theorem~\ref{t:meromorphic}
we then obtain:
\begin{corollary}
\label{c:deltas}
The scalar analytic functions $\Delta_j^\pm(z)$ ($j=1,2,3$) are given by
\bse
\begin{gather}
\Delta_1^+(z)/\=\gamma(z) =  A_{[1]}(z) = B_{[2,3,4]}(z)\, ,
\quad
\Delta_1^-(z)/\=\gamma^*(z^*)  = A_{[2,3,4]}(z) = B_{[1]}(z)\,,
\\
\Delta_2^+(z)/\=\gamma(z) = A_{[1,2]}(z) = B_{[3,4]}(z)\, ,
\quad
\Delta_2^-(z)/\=\gamma^*(z^*) = A_{[3,4]}(z) = B_{[1,2]}(z)\, ,
\\
\Delta_3^+(z)/\=\gamma(z) = A_{[1,2,3]}(z) = B_{[4]}(z)\, ,
\quad
\Delta_3^-(z)/\=\gamma^*(z^*) = A_{[4]}(z) = B_{[1,2,3]}(z)\, ,
\end{gather}
\ese
with $\=\gamma(z)= \gamma(z)/(1-q_-r_+/z^2)$, as in~\eref{e:gammabardef}.
\end{corollary}
Again in Appendix~\ref{a:direct},
comparing these results with the rest of Theorem~\ref{t:meromorphic}, we obtain the following important result regarding the
analyticity of certain minors of $\@A(z)$ and $\@B(z)$:
\begin{theorem}
\label{t:gantmacher}
All the upper principal minors of $\@A(z)$ and all the lower principal minors of $\@B(z)$ are analytic for $\Im z > 0$,
and all the lower principal minors of $\@A(z)$ and all the upper principal minors of $\@B(z)$ are analytic for $\Im z < 0$:
\bse
\begin{gather}
A_{[1]}(z),~A_{[1,2]}(z),~A_{[1,2,3]}(z):\quad \Im z > 0\,,
\qquad
B_{[1]}(z),~B_{[1,2]}(z),~B_{[1,2,3]}(z):\quad \Im z < 0.
\end{gather}
In addition, the following non-principal minors of $\@A(z)$ and $\@B(z)$ are also analytic:
\begin{gather}
A_{\minorset{1,2}{1,3}}(z),~A_{\minorset{1,3}{1,2}}(z):\quad \Im z > 0,
\qquad
B_{\minorset{1,2}{1,3}}(z),~B_{\minorset{1,3}{1,2}}(z):\quad \Im z < 0.
\end{gather}
\ese
\end{theorem}
Lemma~\ref{L:relation} also yields the following identities for the analytic off-diagonal entries of $\alpha^\pm(z)$ and $\beta^\pm(z)$:
\bse
\begin{gather}
A_{\minorset{1,2}{1,3}}(z) = - B_{\minorset{2,4}{3,4}}(z),
\quad
A_{\minorset{1,3}{1,2}}(z) = - B_{\minorset{3,4}{2,4}}(z),
\qquad
\Im z \geq 0,
\\
B_{\minorset{1,2}{1,3}}(z) = - A_{\minorset{2,4}{3,4}}(z),
\quad
B_{\minorset{1,3}{1,2}}(z) = - A_{\minorset{3,4}{2,4}}(z),
\qquad
\Im z \leq 0.
\end{gather}
\ese

\begin{remark}
The connection between this work and~\cite{NMPZ1984} should now be clear.
Namely, above we established the analyticity of the principal minors of the scattering matrix,
results which were claimed (but not proved)
for a general $N\times N$ scattering problem with ZBC~\cite{NMPZ1984}.
We also proved the analyticity of certain non-principal minors.
In doing so, we have constructed the proper tools to generate a fundamental set of analytic eigenfunctions in each half plane.
Specifically, Theorem~\ref{t:mwedgedef} yields an explicit correspondence between the zeros of the $\Delta_j^\pm(z)$
and the poles of the columns of $\@m^\pm(x,t,z)$, as we show next.
\end{remark}
\begin{lemma}
\label{L:merosimple}
Suppose $\Delta_{j-1}^+(z)$ has a simple zero at $z_o \in \Complex^+$.
Then $m_j^+(x,t,z)$ has at most a simple pole at $z=z_o$.
Similarly, if $\Delta_j^-(z)$ has a simple zero at $z_o^*$, $m_j^-(x,t,z)$ has at most a simple pole at $z=z_o^*$.
\end{lemma}
We therefore have:
\begin{theorem}
\label{t:chidef}
A complete set of analytic eigenfunctions is given in each half plane by
\begin{equation}
\label{e:chidef}
\chi^\pm(x,t,z) = \@m^\pm(x,t,z) \@D^\pm(z) \e^{i \@\Theta(x,t,z)}, \qquad \Im z \gtrless 0.
\end{equation}
\end{theorem}

Let $\chi_j^\pm(x,t,z)$ denote the columns of $\chi^\pm(x,t,z)$ for $j=1,\dots,4$.
Note that $\chi_1^\pm(x,t,z) = \phi_{\mp,1}(x,t,z)$ and $\chi_4^\pm(x,t,z) = \phi_{\pm,4}(x,t,z)$.
In addition, similarly to the Manakov system, we refer to $\chi^\pm_2(x,t,z)$ and $\chi^\pm_3(x,t,z)$ as
the \textit{auxiliary eigenfunctions}.

Next, we find the asymptotic behavior of the auxiliary eigenfunctions as $x \to \pm \infty$.
This will be useful in characterizing the discrete spectrum in Section~\ref{s:discretespectrum}.
We find the following for the appropriate regions of the complex plane:
\begin{lemma}
\label{L:chixasymp}
The analytic eigenfunctions $\chi^\pm(x,t,z)$ have the following asymptotic behavior for all $z$ in their respective
domains of analyticity:
\bse
\label{e:auxnewasymp}
\begin{gather}
\lim_{x \to -\infty} \chi^+(x,t,z) \e^{-i \@\Theta(x,t,z)} = \@E_-(z) \alpha_o^+(z) \@D^+(z),
\\
\lim_{x \to \infty} \chi^+(x,t,z) \e^{-i \@\Theta(x,t,z)} = \@E_+(z) \beta_o^+(z) \@D^+(z),
\\
\lim_{x \to -\infty} \chi^-(x,t,z) \e^{-i \@\Theta(x,t,z)} = \@E_-(z) \beta_o^-(z) \@D^-(z),
\\
\lim_{x \to \infty} \chi^-(x,t,z) \e^{-i \@\Theta(x,t,z)} = \@E_+(z) \alpha_o^-(z) \@D^-(z),
\end{gather}
\ese
where $\alpha_o^\pm(z)$ and $\beta_o^\pm(z)$ are given in Theorem~\ref{t:meromorphic} and $\@D^\pm(z)$
are given in Lemma~\ref{L:decompexplicit}.
\end{lemma}
The corresponding relations for the columns, which will be useful in the analysis of the discrete spectrum, are:
\bse
\label{e:chipminf}
\begin{gather}
\lim_{x \to \infty} \chi_1^+(x,t,z)\,\e^{-i\theta_1(x,t,z)} = A_{[1]} \@E_{+,1},\qquad
\lim_{x \to \infty} \chi_2^+(x,t,z)\,\e^{-i\theta_2(x,t,z)} = A_{[1,2]} \@e_2 + A_{\minorset{1,3}{1,2}} \@e_3,
\\
\lim_{x \to \infty} \chi_3^+(x,t,z)\,\e^{-i\theta_3(x,t,z)} = A_{[1,2,3]} \@e_3,\qquad
\lim_{x \to \infty} \chi_4^+(x,t,z)\,\e^{-i\theta_4(x,t,z)} = \@E_{+,4},
\\
\lim_{x \to \infty} \chi_1^-(x,t,z)\,\e^{-i\theta_1(x,t,z)} =\@E_{+,1},\qquad
\lim_{x \to \infty} \chi_2^-(x,t,z)\,\e^{-i\theta_2(x,t,z)} = A_{[2,3,4]} \@e_2,
\\
\lim_{x \to \infty} \chi_3^-(x,t,z)\,\e^{-i\theta_3(x,t,z)} = A_{\minorset{2,4}{3,4}} \@e_2 + A_{[3,4]} \@e_3,\qquad
\lim_{x \to \infty} \chi_4^-(x,t,z)\,\e^{-i\theta_4(x,t,z)} = A_{[4]}\@E_{+,4},
\\
\lim_{x \to - \infty} \chi_1^+(x,t,z)\,\e^{-i\theta_1(x,t,z)} =\@E_{-,1},\qquad
\lim_{x \to - \infty} \chi_2^+(x,t,z)\,\e^{-i\theta_2(x,t,z)} = A_{[1]} \@e_2,
\\
\lim_{x \to - \infty} \chi_3^+(x,t,z)\,\e^{-i\theta_3(x,t,z)} = - A_{\minorset{1,2}{1,3}} \@e_2 + A_{[1,2]} \@e_3,\qquad
\lim_{x \to - \infty} \chi_4^+(x,t,z)\,\e^{-i\theta_4(x,t,z)} = A_{[1,2,3]}\@E_{-,4},
\\
\lim_{x \to - \infty} \chi_1^-(x,t,z)\,\e^{-i\theta_1(x,t,z)} = A_{[2,3,4]} \@E_{-,1},\qquad
\lim_{x \to - \infty} \chi_2^-(x,t,z)\,\e^{-i\theta_2(x,t,z)} = A_{[3,4]} \@e_2 - A_{\minorset{3,4}{2,4}} \@e_3,
\\
\lim_{x \to - \infty} \chi_3^-(x,t,z)\,\e^{-i\theta_3(x,t,z)} = A_{[4]} \@e_3,\qquad
\lim_{x \to - \infty} \chi_4^-(x,t,z)\,\e^{-i\theta_4(x,t,z)} = \@E_{-,4},
\end{gather}
\ese
where again the $z$-dependence in the right-hand side
was omitted for brevity, $\{\@e_j\}_{j=1}^4$ is the standard basis for $\Real^4$, and, according to
\eref{e:Epm}, the first and last columns of the asymptotic eigenvector matrices $\@E_\pm(z)$ are given by:
$$
\@E_{\pm,1}=\@e_1+\frac{iq_\pm}{z}\@e_4\,, \qquad \@E_{\pm,4}=-\frac{iq_\pm^*}{z}\@e_1+\@e_4\,.
$$
The relations~\eref{e:chipminf} are valid for all $z$ in the region of analyticity of the corresponding columns.
Also recall that
\begin{gather*}
\chi_1^+(x,t,z) = \phi_{-,1}(x,t,z),\qquad
\chi_1^-(x,t,z) = \phi_{+,1}(x,t,z),\\
\chi_4^+(x,t,z) = \phi_{+,4}(x,t,z),\qquad
\chi_4^-(x,t,z) = \phi_{-,4}(x,t,z).
\end{gather*}
In addition, using the definition~\eref{e:chidef} of the analytic eigenfunctions together with the decompositions
of the scattering matrix in Lemma~\ref{L:decompexplicit}
and expanding the second and third columns of $\chi^\pm(x,t,z)$, we obtain:
\begin{lemma}
\label{L:chidecomp}
The auxiliary eigenfunctions have the following decompositions for $z \in \Real$:
\bse
\label{e:decomp}
\begin{gather}
\label{e:decompa}
\chi_2^+(x,t,z) = A_{[1,2]}(z) \phi_{+,2}(x,t,z)
+ A_{\minorset{1,3}{1,2}}(z) \phi_{+,3}(x,t,z)
+ A_{\minorset{1,4}{1,2}}(z) \phi_{+,4}(x,t,z),
\\
\label{e:decompb}
\chi_3^+(x,t,z) =
A_{\minorset{1,2}{2,3}}(z) \phi_{-,1}(x,t,z)
- A_{\minorset{1,2}{1,3}}(z) \phi_{-,2}(x,t,z)
+ A_{[1,2]}(z) \phi_{-,3}(x,t,z),
\\
\label{e:decompc}
\chi_2^-(x,t,z) = B_{[1,2]}(z) \phi_{-,2}(x,t,z)
+ B_{\minorset{1,3}{1,2}}(z) \phi_{-,3}(x,t,z)
+ B_{\minorset{1,4}{1,2}}(z) \phi_{-,4}(x,t,z),
\\
\label{e:decompd}
\chi_3^-(x,t,z) =
B_{\minorset{1,2}{2,3}}(z) \phi_{+,1}(x,t,z)
- B_{\minorset{1,2}{1,3}}(z) \phi_{+,2}(x,t,z)
+ B_{[1,2]}(z) \phi_{+,3}(x,t,z),
\\
\label{e:decompe}
\chi_2^+(x,t,z) = - a_{12}(z) \phi_{-,1}(x,t,z) + A_{[1]}(z) \phi_{-,2}(x,t,z),
\\
\label{e:decompf}
\chi_3^+(x,t,z) = B_{[4]}(z) \phi_{+,3}(x,t,z) - b_{43}(z) \phi_{+,4}(x,t,z),
\\
\label{e:decompg}
\chi_2^-(x,t,z) = - b_{12}(z) \phi_{+,1}(x,t,z) + B_{[1]}(z) \phi_{+,2}(x,t,z),
\\
\label{e:decomph}
\chi_3^-(x,t,z) = A_{[4]}(z) \phi_{-,3}(x,t,z) - a_{43}(z) \phi_{-,4}(x,t,z).
\end{gather}
\ese
\end{lemma}
Inverting the relations in Lemma~\ref{L:chidecomp} yields:
\begin{corollary}
\label{c:invert}
The non-analytic eigenfunctions have the following decompositions in terms of analytic eigenfunctions for $z \in \Real$:
\bse
\label{e:decompsimp}
\begin{gather}
\label{e:decompsimpa}
\phi_{+,2}(x,t,z) = \frac{1}{B_{[1]}} [b_{12} \phi_{+,1} + \chi_2^-]
= \frac{1}{A_{[1,2]}} \left[\chi_2^+ - \frac{1}{B_{[4]}} A_{\minorset{1,3}{1,2}} \chi_3^+
- \left(\frac{b_{43}}{B_{[4]}} A_{\minorset{1,3}{1,2}} + A_{\minorset{1,4}{1,2}}\right) \phi_{+,4}\right],
\\
\label{e:decompsimpc}
\phi_{+,3}(x,t,z) = \frac{1}{B_{[4]}} [\chi_3^+ + b_{43} \phi_{+,4}]
= \frac{1}{B_{[1,2]}} \left[\left(\frac{b_{12}}{B_{[1]}} B_{\minorset{1,2}{1,3}}
- B_{\minorset{1,2}{2,3}}\right) \phi_{+,1}
+ \frac{1}{B_{[1]}} B_{\minorset{1,2}{1,3}} \chi_2^-
+ \chi_3^-\right],
\\
\label{e:decompsimpe}
\phi_{-,2}(x,t,z) = \frac{1}{A_{[1]}} [a_{12} \phi_{-,1} + \chi_2^+]
= \frac{1}{B_{[1,2]}} \left[\chi_2^- - \frac{1}{A_{[4]}} B_{\minorset{1,3}{1,2}} \chi_3^-
- \left(\frac{a_{43}}{A_{[4]}} B_{\minorset{1,3}{1,2}} + B_{\minorset{1,4}{1,2}}\right) \phi_{-,4}\right],
\\
\phi_{-,3}(x,t,z) = \frac{1}{A_{[4]}} [\chi_3^- + a_{43} \phi_{-,4}]
= \frac{1}{A_{[1,2]}} \left[\left(\frac{a_{12}}{A_{[1]}} A_{\minorset{1,2}{1,3}}
- A_{\minorset{1,2}{2,3}}\right) \phi_{-,1}
+ \frac{1}{A_{[1]}} A_{\minorset{1,2}{1,3}} \chi_2^+
+ \chi_3^+\right],
\end{gather}
\ese
where the $(x,t,z)$-dependence was omitted from the right-hand side for brevity.
\end{corollary}
The decompositions in Lemma~\ref{L:chidecomp} and the inverted decompositions in
Corollary~\ref{c:invert} are the generalization of the analogous relations valid for the Manakov system~\cite{gbdkk2015,PAB2006}.

\begin{remark}
The results presented in this section (i) provide a rigorous implementation of the framework of~\cite{NMPZ1984} for the
construction of the fundamental analytic eigenfunctions,
and (ii) generalize said framework to the case of scattering problems with NZBC and eigenvalue degeneracy.
Regarding the first point, we showed rigorously that triangular decompositions of the scattering matrix yield a full set of
analytic eigenfunctions for each half plane
and that the upper and lower principal minors of the scattering matrix are all analytic in one of those half planes.
For comparison purposes, recall that \cite{NMPZ1984} only considers the case of compact support and uses Fredholm integral
equations instead of Volterra integral equations.
Thus, in that framework one can at best prove meromorphicity of the eigenfunctions
(because the singularities in the Fredholm integral equations appear implicitly)
which we instead proved to be analytic.
Moreover, the scattering coefficients themselves appear explicitly as forcing terms in those Fredholm integral equations.
Thus, to even prove meromorphicity, one would need to assume that the scattering coefficients are analytic
to begin with.
\end{remark}

\begin{remark}
In addition, here we proved that certain non-principal minors of the scattering matrix are analytic.
No such extra analyticity properties were present in problems with fewer components [namely, the scalar NLS equation and
the Manakov system].
Similarly, no such extra analyticity properties exist for the scattering problem studied in \cite{NMPZ1984}.
Indeed, such properties are a direct result of the degeneracy of the eigenvalues of the scattering problem.
No such degeneracy exists for the problem studied in \cite{NMPZ1984}, and
for the VNLS equation the degeneracy is only present when the number of components is larger than~two.
These new pieces of analytic data play a key role in the characterization of the discrete spectrum.
\end{remark}

\subsection{Asymptotic behavior as $z\to\infty$ and $z\to0$}
\label{s:asymptoticsoriginal}

In order to completely specify the inverse problem in Section \ref{s:istinverse}
it is necessary to examine the asymptotic behavior of the eigenfunctions and scattering coefficients
both as $z \to \infty$ and as $z \to 0$.
Consider the following formal expansion for $\mu_+(x,t,z)$:
\bse
\begin{gather}
\label{Emuexpansion}
\mu_+(x,t,z) = \sum_{n=0}^{\infty} \mu_n(x,t,z),
\\
\noalign{\noindent where}
\mu_0(x,t,z) = \@E_+(z),
\\
\label{Eexpansion}
\mu_{n+1}(x,t,z) = - \int_x^{\infty} \@E_+(z)\e^{i (x-y)\boldsymbol\Lambda (z)} \@E_+^{-1}(z) \Delta \@Q_+(y,t) \mu_n(y,t,z) \e^{-i (x-y)\boldsymbol\Lambda (z)} \mathrm{d} y.
\end{gather}
\ese
As with the Manakov system with NZBC, it is straightforward to prove the following:

\begin{lemma}
\label{L:asympinf}
For all $m \geq 0$, \eref{Emuexpansion} provides an asymptotic expansion for the columns of $\mu_+(x,t,z)$ as
$z \to \infty$ in the appropriate region of the complex $z$-plane, with
\bse
\label{Einfinityorder}
\begin{gather}
\label{Einfinityordera}
[\mu_{2m}]_{bd} = O(1/z^{m}), \qquad [\mu_{2m}]_{bo} = O(1/z^{m+1}),
\\
\label{Einfinityorderb}
[\mu_{2m+1}]_{bd} = O(1/z^{m+1}), \qquad [\mu_{2m+1}]_{bo} = O(1/z^{m+1}).
\end{gather}
\ese
\end{lemma}

\begin{lemma}
\label{L:asympzero}
For all $m \geq 0$, \eref{Emuexpansion} provides an asymptotic expansion for the columns of $\mu_+(x,t,z)$ as
$z \to 0$ in the appropriate region of the complex $z$-plane, with
\bse
\label{Eorderzero}
\begin{gather}
\label{Eorderzeroa}
[\mu_{2m}]_{bd} = O(z^{m}), \qquad [\mu_{2m}]_{bo} = O(z^{m-1}),
\\
\label{Eorderzerob}
[\mu_{2m+1}]_{bd} = O(z^{m}), \qquad [\mu_{2m+1}]_{bo} = O(z^{m}).
\end{gather}
\ese
\end{lemma}
The proofs of these lemmas are exactly the same as for the Manakov system with NZBC \cite{gbdkk2015}, and are therefore omitted.
Explicitly calculating the first few terms of the first and fourth columns of \eref{Emuexpansion} yields
the following for the appropriate domains of analyticity:
\bse
\label{e:mu+1inf}
\begin{gather}
\mu_{\pm,1}(x,t,z) = \begin{pmatrix} 1 \\ (i/z) \@q(x,t) \end{pmatrix} + O(1/z^2),
\qquad
\mu_{\pm,4}(x,t,z) = \begin{pmatrix} - (i/z) \@q^\dag(x,t) \^{\@e}_3 \\ \^{\@e}_3 \end{pmatrix} + O(1/z^2),
\qquad
z \to \infty,
\\
\mu_{\pm,1}(x,t,z) = \begin{pmatrix} \@q^\dag(x,t) \^{\@e}_3/q_\pm^* \\ (i/z) q_\pm \^{\@e}_3 \end{pmatrix}
+ O(z),
\qquad
\mu_{\pm,4}(x,t,z) = \begin{pmatrix} - (i/z) q_\pm^* \\ \@q(x,t)/q_\pm \end{pmatrix} + O(z),
\qquad z \to 0,
\end{gather}
\ese
where $\{\^{\@e}_j\}_{j=1}^3$ denotes the standard basis for $\Real^3$,
while doing the same for the second and third columns of \eref{Emuexpansion} yields
the following for $z \in \Real$:
\bse
\label{e:zasympreal}
\begin{gather}
\mu_{\pm,2}(x,t,z) = \@e_2 + O(1/z),
\qquad
\mu_{\pm,3}(x,t,z) = \@e_3 + O(1/z),
\qquad
z \to \infty,
\\
\mu_{\pm,2}(x,t,z) = \@e_2 + O(z),
\qquad
\mu_{\pm,3}(x,t,z) = \@e_3 + O(z),
\qquad
z \to 0,
\end{gather}
\ese
where, as before, $\{\@e_j\}_{j=1}^4$ denotes the standard basis for $\Real^4$.
The asymptotics \eref{e:mu+1inf} immediately yield the reconstruction formula:
\begin{proposition}
The solution $\@q(x,t) = (q_1(x,t),q_2(x,t),q_3(x,t))^T$ of the VNLS equation~\eref{e:vnls} with the NZBC~\eref{e:NZBC}
can be recovered from the asymptotics of the Jost eigenfunctions as follows:
\begin{equation}
\label{e:reconstruction}
q_j(x,t) = - i\lim_{z \to \infty} \left[z\,\mu_{+,j+1,1}(x,t,z)\right], \qquad j = 1,2,3,
\end{equation}
\end{proposition}

Next, we find the asymptotics of the scattering matrix entries, which
follows from the asymptotic behavior~\eref{e:mu+1inf} and \eref{e:zasympreal}
and the scattering relation \eref{e:scattering}:
\begin{lemma}
\label{L:scattering0inf}
In the appropriate regions of the $z$-plane,
\bse
\label{e:Ainf0}
\begin{gather}
\label{e:Ainf0a}
\@A(z) = \@I + O(1/z), \qquad z \to \infty,
\\
\label{e:Ainf0b}
\@A(z) = \diag (\e^{-i \Delta \theta},1,1,\e^{i \Delta \theta}) + O(z), \qquad z \to 0,
\end{gather}
\ese
where, as before, $\Delta \theta = \theta_+ - \theta_-$ and $\theta_\pm$ are as defined in \eref{e:NZBC}.
\end{lemma}
\begin{corollary}
\label{c:minorasymptotics}
All the principal minors of $\@A(z)$ tend to 1 as $z\to\infty$ in their domain of analyticity, whereas 
all non-principal minors of $\@A(z)$ are $O(1/z)$  as $z\to\infty$ 
along the real axis (for the non-analytic minors) or in their domain of analyticity (for the analytic ones).
Additionally, 
all the principal minors of $\@A(z)$ are $O(1)$ as $z\to0$ in their domain of analyticity, whereas 
all non-principal minors of $\@A(z)$ are $O(z)$  as $z\to0$ 
along the real axis (for the non-analytic minors) or in their domain of analyticity (for the analytic ones).
\end{corollary}
In particular, note from \eref{e:Ainf0} that
\begin{equation}
\label{e:a11zero}
\lim_{z \to 0} A_{[1]}(z) = \e^{-i \Delta \theta},
\end{equation}
which will allow us to directly recover the asymptotic phase difference $\Delta \theta$ from the inverse problem
via the trace formulae (see Section \ref{s:tracemain}).

As a consequence of the decompositions \eref{e:decomp} of the auxiliary eigenfunctions and the
asymptotics of the Jost eigenfunctions in \eref{e:mu+1inf} and \eref{e:zasympreal},
we find the following asymptotics for the modified auxiliary eigenfunctions for all $z \in \Real$:
\bse
\begin{gather}
\chi_2^\pm(x,t,z)\,\e^{-i\theta_2(x,t,z)} = \@e_2 + O(1/z),
\qquad
\chi_3^\pm(x,t,z)\,\e^{-i\theta_3(x,t,z)} = \@e_3 + O(1/z),
\qquad
z \to \infty,
\\
\chi_2^\pm(x,t,z)\,\e^{-i\theta_2(x,t,z)} = \e^{\mp i \Delta \theta} \@e_2 + O(z),
\qquad
\chi_3^\pm(x,t,z)\,\e^{-i\theta_3(x,t,z)} = \e^{\mp i \Delta \theta} \@e_3 + O(z),
\qquad
z \to 0,
\end{gather}
\ese
One can show that these relations hold for all $z$ in the appropriate half plane
by studying the asymptotic behavior of the fundamental tensors
$f_n$ and $g_n$ using the techniques of \cite{BDT}
and then reconstructing the behavior of the auxiliary eigenfunctions as a result.
For brevity, we omit the details.

\subsection{Behavior at the branch points}

We now discuss the behavior of the Jost eigenfunctions and the scattering matrix at the branch points $k = \pm q_o$.
As with the Manakov system,
the complication there is due to the fact that $\lambda(\pm q_o) = 0$, and therefore, at $z=\pm q_o$, the two exponentials
$\e^{\pm i\lambda x}$ reduce to the identity.
Correspondingly, at $z=\pm q_o$ the matrices $\@E_\pm(z)$ are degenerate.
Nonetheless,
the term $\@E_\pm(z)\,\e^{i (x-y)\@\Lambda(z)}\@E_\pm^{-1}(z)$ appearing in the integral equations for the Jost eigenfunctions
remains finite as $z\to\pm q_o$:
\begin{equation}
\lim_{z \to \pm q_o} \@E_\pm(z) \e^{i (x-y)\@\Lambda(z)} \@E_\pm^{-1}(z) =
\diag (1,\e^{\pm iq_o(x-y)},\e^{\pm iq_o(x-y)},1).
\end{equation}
Thus, if $\@q \to \@q_\pm$ sufficiently fast as $x \to \pm \infty$,
the integrals in \eref{e:intequations} are also convergent at $z=\pm q_o$,
and the Jost solutions admit a well-defined limit at the branch points.
Nonetheless,
$\det\phi_\pm(x,t,\pm q_o) = 0$ for all $(x,t) \in \Real^2$.
Thus, the columns of $\phi_\pm(x,t,q_o)$ [as well as those of $\phi_\pm(x,t,-q_o)$] are linearly dependent.
Comparing the asymptotic behavior of the columns of $\phi_\pm(x,t,\pm q_o)$ as $x \to \pm \infty$, we obtain
\begin{equation}
\label{e:q0norm}
\phi_{\pm,1}(x,t,q_o) = i \e^{i \theta_\pm} \phi_{\pm,4}(x,t,q_o),
\qquad
\phi_{\pm,1}(x,t,-q_o) = -i \e^{i \theta_\pm} \phi_{\pm,4}(x,t,-q_o).
\end{equation}
Next, we characterize the limiting behavior of the scattering matrix near the branch points.
It is easy to express all entries of the scattering matrix $\@A(z)$ as Wronskians:
\bse
\begin{gather}
\label{e:aijexp}
a_{j \ell}(z) = \frac{z^2}{z^2-q_o^2} W_{j\ell}(x,t,z)\e^{- 2i \theta_2(x,t,z)},
\\
\noalign{\noindent where}
W_{j\ell}(x,t,z) = \det (\phi_{-,\ell}(x,t,z),\phi_{+,j+1}(x,t,z),\phi_{+,j+2}(x,t,z),\phi_{+,j+3}(x,t,z))\,,
\end{gather}
\ese
and $j+1$, $j+2$, and $j+3$ are calculated modulo 4.
We then have the following Laurent series expansions about $z=\pm q_o$:
\begin{equation}
a_{ij}(z) = \frac{a_{ij,\pm}}{z \mp q_o} + a_{ij,\pm}^{(o)} + O(z \mp q_o),
\quad z \in \Real \setminus \{\pm q_o\}\,,
\end{equation}
where, for example,
\bse
\begin{gather}
a_{11,\pm} = \pm \frac{q_o}{2} W_{11}(x,t,\pm q_o)\,\e^{\mp 2 i q_o (x \mp q_o t)},
\\
a_{11,\pm}^{(o)} = \pm \frac{q_o}{2} \frac{\mathrm{d}}{\mathrm{d} z} W_{11}(x,t,z)\vert_{z=\pm q_o}
\e^{\mp 2 i q_o (x \mp q_o t)}
+ W_{11}(x,t,\pm q_o) \,\e^{\mp 2 i q_o (x \mp q_o t)}\,.
\end{gather}
\ese
Summarizing, the asymptotic expansions of $\@A(z)$ in neighborhoods of the branch points are
\begin{equation}
\@A(z) = \frac{1}{z \mp q_o} \@A_\pm + \@A_\pm^{(o)} + O(z \mp q_o),
\end{equation}
where $\@A_\pm^{(o)} = (a_{ij,\pm}^{(o)})$,
\begin{equation*}
\@A_\pm = a_{11,\pm} \begin{pmatrix}
1 & 0 & 0 & \mp i \e^{-i \theta_-}
\\
0 & 0 & 0 & 0
\\
0 & 0 & 0 & 0
\\
\pm i \e^{i \theta_+} & 0 & 0 & \e^{i \Delta \theta}
\end{pmatrix}
+
a_{12,\pm} \begin{pmatrix}
0 & 1 & 0 & 0
\\
0 & 0 & 0 & 0
\\
0 & 0 & 0 & 0
\\
0 & \pm i \e^{i \theta_+} & 0 & 0
\end{pmatrix}
+
a_{13,\pm} \begin{pmatrix}
0 & 0 & 1 & 0
\\
0 & 0 & 0 & 0
\\
0 & 0 & 0 & 0
\\
0 & 0 & \pm i \e^{i \theta_+} & 0
\end{pmatrix},
\end{equation*}
and $a_{1j,\pm} = \pm (q_o/2) W_{1j}(x,t,\pm q_o) \e^{\mp 2 i q_o (x \mp q_o t)}$ for $j=2,3$.
Note that the second and third rows of $\@A_\pm$ are identically zero by virtue of~\eref{e:q0norm}.

It is worth noticing that all scattering coefficients are finite at each of the branch points $\pm q_o$ iff $a_{11,\pm}=0$, i.e.,
iff $\det (\phi_{-,1}(x,t,\pm q_o),\phi_{+,2}(x,t,\pm q_o),\phi_{+,3}(x,t,\pm q_o),\phi_{+,4}(x,t,\pm q_o))=0$.
In this case, the corresponding branch point is referred to as a ``virtual level'' (cf. \cite{FT1987}, where this definition
is introduced for the scalar NLS equation). The generic situation, however, corresponds to $a_{11,\pm}\ne 0$, and in this case
all scattering coefficients have a pole at the branch points.

\section{Symmetries and discrete spectrum}

As mentioned earlier, the richness of the 3-component problem compared to the 2-component problem
manifests itself most clearly in the symmetries and the discrete spectrum.
We discuss both of them next.

\subsection{Symmetries}
\label{s:symmetries}

Recall that, in the IST with ZBC, the only symmetry of the scattering problem is the mapping $k \mapsto k^*$.
With NZBC, the symmetries are more involved, because of the presence of the two-sheeted Riemann surface.
Correspondingly, the problem admits two symmetries.
The symmetries are also complicated by the fact that, after removing the asymptotic oscillations,
the Jost solutions do not tend to the identity matrix.
Finally, for the Manakov system a further complication arises from
the presence of several analytic eigenfunctions.
As we will show next, this latter complication is compounded in the 3-component VNLS equation.


\paragraph{First symmetry: up-down.}

The first symmetry corresponds to the transformation $z \mapsto z^*$ (i.e., switching UHP and LHP),
corresponding to $(k,\lambda) \mapsto (k^*,\lambda^*)$.
Similarly to the Manakov system with NZBC, the following results are a straightforward result of the symmetries of the Lax pair and
are proved in Appendix \ref{a:symm}:
\begin{proposition}
\label{L:firstsymmetry}
If $\@v(x,t,z)$ is a fundamental matrix solution of the Lax pair,
so is $\@w(x,t,z)=\@J[\@v^{\dagger}(x,t,z^*)]^{-1}$.
\end{proposition}
\begin{lemma}
\label{L:firstsymmphi}
The Jost solutions satisfy the symmetry
\begin{equation}
\label{e:symm3}
\@J[\phi_\pm^{\dagger}(x,t,z)]^{-1} \@C(z) = \phi_\pm(x,t,z), \qquad z \in \Real,
\end{equation}
where
\begin{equation}
\label{e:firstsymmmatrix}
\@C(z) = \diag(\gamma(z),-1,-1,-\gamma(z))\,.
\end{equation}
\end{lemma}

We now use the Hodge duality to define a ``generalized cross product'' for vectors in $\Complex^4$.
Recall that
the Hodge star operator is the bijective mapping between $\Complex^4$ and the space of 3-vectors in $\Complex^4$
that is uniquely defined up to a change in sign \cite{Frankel}.
With this in mind, we define the following quantity:
\begin{definition} (Generalized cross product)
For all $\@u,\@v,\@w \in \Complex^4$, let
\begin{equation}
\label{e:lprod}
L[\@u,\@v,\@w] = \det\begin{pmatrix} u_1&u_2&u_3&u_4 \\ v_1&v_2&v_3&v_4 \\ w_1&w_2&w_3&w_4 \\
\@e_1&\@e_2&\@e_3&\@e_4 \end{pmatrix},
\end{equation}
where, as before, $\{\@e_1,\dots,\@e_4\}$ is the standard basis for $\Real^4$.
\end{definition}

Like the usual cross product in three dimensions, $L[\cdot]$ is multilinear and totally antisymmetric.
Morover, the product $L[\@v_1,\@v_2,\@v_3]$ is the Hodge dual of the wedge product $\@v_1 \wedge \@v_2 \wedge \@v_3$,
in the sense that it is simply the image of $\@v_1 \wedge \@v_2 \wedge \@v_3$ under the Hodge
star operator, where ``$\wedge$'' is the wedge product as defined in \cite{Frankel}.
The choice of orientation in~\eref{e:lprod} determines this quantity uniquely.

The key to using $L[\cdot]$ to express the symmetries is the following identity, which can be verified by direct calculation:
\begin{proposition}
Let $\@a_1,\dots,\@a_4\in\Complex^4$ and $\@A = (\@a_1,\dots,\@a_4)$.
If $\det\@A\ne0$,
\be
\@A^{-1} = (\det\@A)^{-1}\,(-L[\@a_2,\@a_3,\@a_4],L[\@a_1,\@a_3,\@a_4],-L[\@a_1,\@a_2,\@a_4],L[\@a_1,\@a_2,\@a_3])^T\,.
\label{e:Linverse}
\ee
\end{proposition}
Equation~\eref{e:Linverse} generalizes the formula
$\@A^{-1} = (\det\@A)^{-1}(\@a_2\times\@a_3,\@a_3\times\@a_1,\@a_1\times\@a_2)$
for the inverse of a $3\times3$ matrix $\@A = (\@a_1,\@a_2,\@a_3)$.
Using~\eref{e:Linverse},
one can insert~\eref{e:decomp} into~\eref{e:symm3} and extend the resulting relations
via the Schwarz reflection principle to obtain:
\begin{theorem}
\label{t:phifirstsymm}
The analytic columns of $\phi_\pm(x,t,z)$ obey the following symmetry relations:
\bse
\label{e:symm1eigen}
\begin{gather}
\label{e:symm1eigena}
\phi_{+,1}^*(x,t,z^*) = - \frac{\e^{-2 i \theta_2(x,t,z)}}{A_{[1,2]}(z) A_{[1,2,3]}(z)} \@J \, L[\chi_2^+(x,t,z),\chi_3^+(x,t,z),\phi_{+,4}(x,t,z)],
\qquad \Im z \geq 0,
\\
\label{e:symm1eigenb}
\phi_{-,1}^*(x,t,z^*) = - \frac{\e^{- 2 i \theta_2(x,t,z)}}{A_{[4]}(z) A_{[3,4]}(z)} \@J \, L[\chi_2^-(x,t,z),\chi_3^-(x,t,z),\phi_{-,4}(x,t,z)],
\qquad \Im z \leq 0,
\\
\label{e:symm1eigenc}
\phi_{+,4}^*(x,t,z^*) = - \frac{\e^{-2 i \theta_2(x,t,z)}}{A_{[3,4]}(z) A_{[2,3,4]}(z)} \@J \, L[\phi_{+,1}(x,t,z),\chi_2^-(x,t,z),\chi_3^-(x,t,z)],
\qquad \Im z \leq 0,
\\
\label{e:symm1eigend}
\phi_{-,4}^*(x,t,z^*) = - \frac{\e^{-2 i \theta_2(x,t,z)}}{A_{[1]}(z) A_{[1,2]}(z)} \@J \, L[\phi_{-,1}(x,t,z),\chi_2^+(x,t,z),\chi_3^+(x,t,z)],
\qquad \Im z \geq 0.
\end{gather}
\ese
\end{theorem}
Using the symmetry \eref{e:symm3} as in the proof of Theorem~\ref{t:phifirstsymm},
we also obtain the following:
\begin{corollary}
\label{c:realcrossproduct}
For cyclic indices $j$, $\ell$, $m$, and $n$,
\begin{equation}
\phi_{\pm,j}(x,t,z) = - \e^{2 i \theta_2(x,t,z)} \@J L [\phi_{\pm,\ell}^*(x,t,z),\phi_{\pm,m}^*(x,t,z),\phi_{\pm,n}^*(x,t,z)]/
\gamma_j(z), \qquad z \in \Real,
\end{equation}
where $\gamma_1(z)=\gamma_4(z)=1$, $\gamma_2(z) = \gamma(z)$, and $\gamma_3(z) = - \gamma(z)$.
\end{corollary}
In fact, we can write a more general result:
\begin{lemma}
\label{L:adjoint}
For any vector solutions $\@v_1$, $\@v_2$, and $\@v_3$ of the Lax pair, the following is also a solution of the Lax pair:
\begin{equation}
\label{e:adjoint}
\@v(x,t,z) = \e^{2 i \theta_2(x,t,z)} \@J L [\@v_1^*(x,t,z),\@v_2^*(x,t,z),\@v_3^*(x,t,z)], \qquad z \in \Real.
\end{equation}
\end{lemma}
Lemma~\ref{L:adjoint} is easily verified using the following identities, which hold for any vectors $\@u,\@v,\@w \in \Complex^4$:
\bse
\begin{gather}
L[\@J \@u, \@J \@v,\@J \@w] = - \@J L[\@u,\@v,\@w],
\\
2 \@J L[\@u,\@v,\@w] + L[\@u,\@v,\@w] = L[\@J \@u,\@J \@v,\@w] + L[\@u,\@J \@v,\@J\@w]
+ L[\@J \@u, \@v,\@J \@w],
\\
L[\@J\@u,\@v,\@w] + L[\@u,\@J\@v,\@w] + L[\@u,\@v,\@J\@w] =
-\@J L[\@u,\@v,\@w] - 2 L[\@u,\@v,\@w],
\\
- \@Q L[\@u,\@v,\@w] = L[\@Q^T\@u,\@v,\@w] + L[\@u,\@Q^T\@v,\@w] + L[\@u,\@v,\@Q^T\@w].
\end{gather}
\ese
\vspace*{-1ex}
\unskip
\begin{remark}
Lemma~\ref{L:adjoint} is the generalization of an analogous result for the Manakov system \cite{gbdkk2015,PAB2006}.
In~\cite{PAB2006}, the analogous result was used to construct the auxiliary eigenfunctions,
each defined in terms of 2 analytic Jost eigenfunctions of a suitably defined ``adjoint'' problem.
That approach, however, does not work in the 3-component case
(because 3 analytic adjoint Jost eigenfunctions would be necessary, whereas only 2 of them are analytic),
which made it necessary to use the approach of~\cite{BDT} and~\cite{PBT2011}.
\end{remark}

We now show how Lemma~\ref{L:firstsymmetry} affects the scattering matrix.
Recalling \eref{e:scattering}, we conclude
\begin{equation}
\label{e:symm5}
[\@A(z)]^{\dagger} = \@C(z) \@B(z) \@C^{-1}(z)\,,\qquad z\in\Real\,,
\end{equation}
with $\@C(z)$ given by~\eref{e:firstsymmmatrix}.
Componentwise, for $z \in \Real$,
\begin{gather*}
b_{11}(z) = a_{11}^*(z), \quad b_{12}(z) = - \frac{1}{\gamma(z)} a_{21}^*(z), \quad b_{13}(z) = - \frac{1}{\gamma(z)}
a_{31}^*(z), \quad b_{14}(z) = - a_{41}^*(z),
\\
b_{21}(z) = - \gamma(z) a_{12}^*(z), \quad b_{22}(z) = a_{22}^*(z), \quad b_{23}(z) = a_{32}^*(z),
\quad b_{24}(z) = \gamma(z) a_{42}^*(z),
\\
b_{31}(z) = - \gamma(z) a_{13}^*(z), \quad b_{32}(z) = a_{23}^*(z), \quad b_{33}(z) = a_{33}^*(z),
\quad b_{34}(z) = \gamma(z) a_{43}^*(z),
\\
b_{41}(z) = - a_{14}^*(z), \quad b_{42}(z) = \frac{1}{\gamma(z)} a_{24}^*(z), \quad b_{43}(z) = \frac{1}{\gamma(z)}
a_{34}^*(z), \quad b_{44}(z) = a_{44}^*(z).
\end{gather*}
Recalling \eref{e:AB}, the Schwarz reflection principle then allows us to conclude
\begin{equation}
\label{e:symmetryextend}
A_{[1]}(z) = [A_{[2,3,4]}(z^*)]^*,
\quad A_{[1,2]}(z) = [A_{[3,4]}(z^*)]^*,
\quad A_{[1,2,3]}(z) = [A_{[4]}(z^*)]^*,
\qquad \Im z \geq 0.
\end{equation}
In addition, we also have
\begin{equation}
\label{e:symm1newminors}
A_{\minorset{1,2}{1,3}}(z) = - [A_{\minorset{3,4}{2,4}}(z^*)]^*,
\quad
A_{\minorset{1,3}{1,2}}(z) = - [A_{\minorset{2,4}{3,4}}(z^*)]^*,
\qquad
\Im z \geq 0.
\end{equation}
Finally, using the symmetry~\eref{e:symm5}, we can
combine Lemma~\ref{L:adjoint} with the decompositions~\eref{e:decompsimp} to obtain:
\begin{lemma}
\label{L:chisymm1}
The auxiliary eigenfunctions obey the following symmetry relations:
\bse
\label{e:chisymm1}
\begin{gather}
\label{e:chisymm1a}
[\chi_2^+(x,t,z^*)]^* = - \frac{\e^{- 2 i \theta_2(x,t,z)}}{A_{[4]}(z) \gamma(z)} \@J L [\phi_{+,1}(x,t,z),\chi_3^-(x,t,z),\phi_{-,4}(x,t,z)], \qquad \Im z \leq 0,
\\
\label{e:chisymm1b}
[\chi_3^+(x,t,z^*)]^* = \frac{\e^{- 2 i \theta_2(x,t,z)}}{A_{[2,3,4]}(z) \gamma(z)} \@J L [\phi_{+,1}(x,t,z),\chi_2^-(x,t,z),\phi_{-,4}(x,t,z)], \qquad \Im z \leq 0,
\\
\label{e:chisymm1c}
[\chi_2^-(x,t,z^*)]^* = - \frac{\e^{- 2 i \theta_2(x,t,z)}}{A_{[1,2,3]}(z) \gamma(z)} \@J L [\phi_{-,1}(x,t,z),\chi_3^+(x,t,z),\phi_{+,4}(x,t,z)], \qquad \Im z \geq 0,
\\
\label{e:chisymm1d}
[\chi_3^-(x,t,z^*)]^* = \frac{\e^{- 2 i \theta_2(x,t,z)}}{A_{[1]}(z) \gamma(z)} \@J L [\phi_{-,1}(x,t,z),\chi_2^+(x,t,z),\phi_{+,4}(x,t,z)], \qquad \Im z \geq 0.
\end{gather}
\ese
\end{lemma}
As before, the symmetries in Lemma~\ref{L:chisymm1} are first written for $z \in \Real$
[since that is where the decompositions \eref{e:decompsimp} are valid]
and then extended to the appropriate region of the complex $z$-plane using the Schwarz reflection principle.

\paragraph{Second symmetry: in-out.}

The second symmetry of the scattering problem corresponds to the transformation $z \mapsto \z^* := q_o^2/z$
(exterior/interior of the circle $C_o$ of radius $q_o$ centered at 0),
corresponding to $(k,\lambda) \mapsto (k,-\lambda)$.
We use this symmetry to relate the values of the eigenfunctions on the two sheets (particularly, across the cuts),
where $k$ is arbitrary but fixed (on either sheet).
It is easy to show the following:
\begin{proposition}
\label{p:symm2}
If $\@v(x,t,z)$ is a solution of the Lax pair, so is
\begin{equation}
\@w(x,t,z) = \@v(x,t,\z^*).
\end{equation}
\end{proposition}
Using this result, in the appendix we prove
\begin{lemma}
\label{L:phisymm2}
The Jost eigenfunctions satisfy the following symmetry:
\begin{equation}
\label{e:phiwitha}
\phi_\pm(x,t,z) = \phi_\pm(x,t,\z) \@{\Pi}_\pm(z),\qquad z\in\Real\,,
\end{equation}
where
\begin{equation}
\label{e:symm4}
\@{\Pi}_\pm(z) = \begin{pmatrix} 0 & 0 & -iq_\pm/z \\ 0 & \@I_2 & 0 \\ iq_\pm^*/z & 0 & 0 \end{pmatrix}.
\end{equation}
\end{lemma}
Then, from \eref{e:phiwitha} and the Schwarz reflection principle,
we obtain the following symmetry relations among the analytic Jost eigenfunctions:
\bse
\label{e:second}
\begin{gather}
\phi_{\pm,1}(x,t,z) = \frac{i q_\pm^*}{z} \phi_{\pm,4}(x,t,\z^*), \quad \Im z \lessgtr 0, \\
\phi_{\pm,4}(x,t,z) = - \frac{iq_\pm}{z} \phi_{\pm,1}(x,t,\z^*), \quad \Im z \gtrless 0, \\
\phi_{\pm,2}(x,t,z) = \phi_{\pm,2}(x,t,\z^*), \quad z \in \Real, \\
\phi_{\pm,3}(x,t,z) = \phi_{\pm,3}(x,t,\z^*), \quad z \in \Real.
\end{gather}
\ese

We now discuss how this second symmetry affects the scattering matrix.
Combining the scattering relation \eref{e:scattering} with the symmetries~\eref{e:second},
we conclude
\begin{equation}
\label{e:symm6}
\@A(z) = \@{\Pi}_+^{-1}(z) \@A(\z^*) \@{\Pi}_-^{-1}(z)\,,\qquad z\in\Real\,.
\end{equation}
Componentwise, we have the following for $z \in \Real$:
\begin{gather*}
a_{11}(z) = \frac{q_-^*}{q_+^*} a_{44}(\z^*), \quad a_{12}(z) = - \frac{iz}{q_+^*} a_{42}(\z^*), \quad
a_{13}(z) = - \frac{iz}{q_+^*} a_{43}(\z^*), \quad a_{14}(z) = - \frac{q_-}{q_+^*} a_{41}(\z^*),
\\
a_{21}(z) = \frac{i q_-^*}{z} a_{24}(\z^*), \quad a_{22}(z) = a_{22}(\z^*), \quad a_{23}(z) = a_{23}(\z^*),
\quad a_{24}(z) = - \frac{i q_-}{z} a_{21}(\z^*),
\\
a_{31}(z) = \frac{i q_-^*}{z} a_{34}(\z^*), \quad a_{32}(z) = a_{32}(\z^*), \quad a_{33}(z) = a_{33}(\z^*),
\quad a_{34}(z) = - \frac{i q_-}{z} a_{31}(\z^*),
\\
a_{41}(z) = \frac{q_-^*}{q_+} a_{14}(\z^*), \quad a_{42}(z) = \frac{iz}{q_+} a_{12}(\z^*), \quad
a_{43}(z) = \frac{iz}{q_+} a_{13}(\z^*), \quad a_{44}(z) = \frac{q_-}{q_+} a_{11}(\z^*).
\end{gather*}
Similar expressions hold for the entries of $\@B(z)$ if we switch the plus and minus signs in~\eref{e:symm6}.
The analyticity properties of the scattering matrix entries in Theorem~\ref{t:gantmacher} then allow us to conclude
\begin{equation}
\label{e:symm_more}
A_{[1]}(z) = \e^{i \Delta \theta} A_{[4]}(\z^*),
\quad
A_{[1,2,3]}(z) = \e^{i \Delta \theta} A_{[2,3,4]}(\z^*), \qquad \Im z > 0,
\end{equation}
where $\Delta \theta = \theta_+ - \theta_-$ and $\theta_\pm$ are as defined in \eref{e:NZBC}.
In addition, we also obtain the following from the symmetry \eref{e:symm6}:
\begin{equation}
\label{e:symm2newminors}
A_{\minorset{1,2}{1,3}}(z) = \e^{i \Delta \theta} A_{\minorset{2,4}{3,4}}(\z^*),
\quad
A_{\minorset{1,3}{1,2}}(z) = \e^{i \Delta \theta} A_{\minorset{3,4}{2,4}}(\z^*),
\qquad
\Im z \geq 0.
\end{equation}
In addition, we can combine the symmetries \eref{e:symm1newminors} and \eref{e:symm2newminors}
to obtain another symmetry among the analytic non-principal minors:
\begin{equation}
A_{\minorset{1,2}{1,3}}(z) = - \e^{i \Delta \theta} [A_{\minorset{1,3}{1,2}}(\z)]^*,
\qquad \Im z \geq 0.
\end{equation}
We therefore see that there is only one independent analytic non-principal minor.
Finally, we combine \eref{e:symm2newminors} with the identities found using Lemma~\ref{L:relation} to obtain
\bse
\label{e:a12new}
\begin{gather}
A_{[1,2]}(z) = \e^{i \Delta \theta} A_{\minorset{2,4}{2,4}}(\z) =
  \e^{i \Delta \theta} B_{\minorset{1,3}{1,3}}(\z), \qquad z \in\Real,
\\
A_{[3,4]}(z) = \e^{-i \Delta \theta} A_{\minorset{1,3}{1,3}}(\z)
  = \e^{-i \Delta \theta} B_{\minorset{2,4}{2,4}}(\z), \qquad z\in\Real.
\end{gather}
\ese

Now that we have determined the second symmetry of the scattering matrix, we can obtain the symmetry
of the auxiliary eigenfunctions.
Specifically, we combine the second symmetry \eref{e:second} with the decompositions
\eref{e:decompsimp} to obtain:
\begin{lemma}
\label{L:chisymm2}
The auxiliary eigenfunctions obey the following symmetry relations:
\bse
\label{e:auxsymm2}
\begin{gather}
\label{e:auxsymm2a}
\chi_2^+(x,t,\z^*) = \frac{\e^{i \Delta \theta}}{A_{[3,4]}(z)} \left[A_{[4]}(z) \chi_2^-(x,t,z)
+ A_{\minorset{3,4}{2,4}}(z) \chi_3^-(x,t,z)\right],
\qquad \Im z \leq 0,
\\
\label{e:auxsymm2b}
\chi_3^+(x,t,\z^*) = \frac{\e^{i \Delta \theta}}{A_{[3,4]}(z)} \left[A_{[2,3,4]}(z) \chi_3^-(x,t,z)
- A_{\minorset{2,4}{3,4}}(z) \chi_2^-(x,t,z)\right],
\qquad \Im z \leq 0,
\\
\label{e:auxsymm2c}
\chi_2^-(x,t,\z^*) = \frac{\e^{- i \Delta \theta}}{A_{[1,2]}(z)} \left[A_{[1,2,3]}(z) \chi_2^+(x,t,z)
- A_{\minorset{1,3}{1,2}}(z) \chi_3^+(x,t,z)\right],
\qquad \Im z \geq 0,
\\
\label{e:auxsymm2d}
\chi_3^-(x,t,\z^*) = \frac{\e^{- i \Delta \theta}}{A_{[1,2]}(z)} \left[A_{[1]}(z) \chi_3^+(x,t,z)
+ A_{\minorset{1,2}{1,3}}(z) \chi_2^+(x,t,z)\right],
\qquad \Im z \geq 0.
\end{gather}
\ese
\end{lemma}
Equations~\eref{e:auxsymm2} are the generalization of the (much simpler) symmetries for the Manakov system.
Similarly to that case, these symmetries will be instrumental to
characterize the discrete spectrum.

The above relations do not include a symmetry for $A_{[1,2]}(z)$ or $A_{[3,4]}(z)$.
As we show in the appendix, these minors satisfy a different kind of in-out symmetry:
\begin{theorem}
\label{t:discrete}
The following relation holds:
\bse
\label{e:A12symm}
\begin{gather}
\e^{i\Delta\theta} A_{[1,2]}(z) A_{[1,2]}^*(q_o^2/z^*) = A_{[1]}(z) A_{[1,2,3]}(z) +  A_{\minorset{1,2}{1,3}}(z) A_{\minorset{1,3}{1,2}}(z)\,,
\qquad \Im z>0\,,
\\
\e^{-i\Delta\theta} A_{[3,4]}(z) A_{[3,4]}^*(q_o^2/z^*) = A_{[4]}(z) A_{[2,3,4]}(z) +  A_{\minorset{2,4}{3,4}}(z) A_{\minorset{3,4}{2,4}}(z)\,,
\qquad \Im z<0\,.
\end{gather}
\ese
\end{theorem}
Importantly,
unlike the symmetries for the other minors, equations~\eref{e:A12symm} are bilinear, and  they couple the values of the principal minors and those of the analytic non-principal minors.
This is another new feature of the 3-component case compared to the Manakov system,
and will have important consequences in the analysis of the discrete spectrum.

\paragraph{Reflection coefficients, their symmetries and real spectral singularities.}

A cofactor expansion along the first column of $\@A(z)$,
combined with the symmetries~\eref{e:symm5}
and written in terms of reflection coefficients, yields:
\begin{equation}
|a_{11}(z)|^2 = 1 + |a_{41}(z)|^2 + \frac{1}{\gamma(z)}(|a_{21}(z)|^2+|a_{31}(z)|^2)\,,\qquad z\in\Real\,.
\label{e:unitarity}
\end{equation}
Equation~\eref{e:unitarity} implies $a_{11}(z)\ne0$ for all $z\in\Real$ such that $\gamma(z)>0$, i.e., $z\in(-\infty,-q_o)\cup(q_o,\infty)$.
On the other hand, hence one cannot exclude spectral singularities in the scattering problem
[i.e., real zeros of $a_{11}(z)$ or of the other analytic principal minors of the scattering matrix]
for $z\in (-q_o,q_o)$.
This is the same situation as that in the Manakov system \cite{PAB2006}.

\begin{remark}
For the remainder of this work we will assume that no real spectral singularies are present.
\end{remark}
For future reference, we next introduce the reflection coefficients that will appear in the inverse problem:
\bse
\label{e:reflcoeff}
\begin{gather}
\rho_1(z) = \frac{a_{21}(z)}{a_{11}(z)},
\quad
\rho_2(z) = \frac{a_{31}(z)}{a_{11}(z)},
\quad
\rho_3(z) = \frac{a_{41}(z)}{a_{11}(z)},
\qquad
z\in \Real,
\\
\bar{\rho}_1(z)=\frac{b_{12}(z)}{b_{11}(z)},
\quad
\bar{\rho}_2(z)=\frac{b_{13}(z)}{b_{11}(z)},
\quad
\bar{\rho}_3(z)=\frac{b_{14}(z)}{b_{11}(z)},
\qquad
z\in \Real.
\end{gather}
\ese
Note that~\eqref{e:unitarity} implies $|a_{11}(z)|=1$ for all $z\in \Real$ for which all three reflection coefficients vanish
[i.e., for all $z\in\Real$ such that $\rho_1(z)=\rho_2(z)=\rho_3(z)=0$].
Thus, in particular, $|a_{11}(z)| = 1$ $\forall z\in \Real$ in the reflectionless case.

Not all of these coefficients are independent, of course.
In fact, the first symmetry for the scattering coefficients yields
\label{e:reflsymm}
\begin{equation}
\rho_1^*(z) = - \gamma(z) \bar{\rho}_1(z),
\qquad
\rho_2^*(z) = - \gamma(z) \bar{\rho}_2(z),
\qquad
\rho_3^*(z) = - \bar{\rho}_3(z),
\qquad
z\in \Real.
\end{equation}
Moreover, the second symmetry for the scattering coefficients yields the
following alternative representation for the reflection coefficients:
\begin{equation*}
\rho_1(\z) = \frac{i z}{q_+} \frac{a_{24}(z)}{a_{44}(z)},
\qquad
\rho_2(\z) = \frac{i z}{q_+} \frac{a_{34}(z)}{a_{44}(z)},
\qquad
\rho_3(\z) = - \frac{q_+^*}{q_+} \frac{a_{14}(z)}{a_{44}(z)},
\qquad
z\in \Real.
\end{equation*}
Expressing the scattering coefficients in the above relations as minors of $\@B(z)$, one obtains the following linear
system which relates the value of the reflection coefficients at $\^z$ to their values at $z$:
\begin{equation}
\label{e:rhohatrho}
q_+\@S(z)\begin{pmatrix}
\rho_1(\^z) \\
\rho_2(\^z) \\
\rho_3(\^z)
\end{pmatrix}
=|a_{11}(z)|^2 \bar{\rho}_3(z)
\begin{pmatrix}
iz\,\rho_1(z) \\
-iz\,\rho_2(z) \\
q_+^*
\end{pmatrix}\,,
\end{equation}
where
\begin{equation}
\@S(z)=\begin{pmatrix}
2(1-|a_{11}|^2)+ |a_{11}|^2\rho_1\bar{\rho}_1 & |a_{11}|^2\rho_1\bar{\rho}_2 & -iz\,|a_{11}|^2\rho_1/q_+^* \\
|a_{11}|^2\rho_2\bar{\rho}_1 &  2(1-|a_{11}|^2)+ |a_{11}|^2\rho_2\bar{\rho}_2 & -iz\,|a_{11}|^2\rho_2/q_+^* \\
iq_+^*|a_{11}|^2\bar{\rho}_1/z & iq_+^*|a_{11}|^2\bar{\rho}_2/z  & 2-|a_{11}|^2
\end{pmatrix}\,,
\end{equation}
and where the above expression for $\@S(z)$ was simplified using~\eref{e:unitarity}, rewritten in terms of the reflection coefficients as:
\begin{equation}
\label{e:unitarity2}
|\rho_3(z)|^2+\frac{1}{\gamma(z)}(|\rho_1(z)|^2+|\rho_2(z)|^2)=1-|a_{11}(z)|^{-2}\,,
\qquad
z\in\Real\,.
\end{equation}
Similarly, one obtains:
\begin{gather}
\det \@S(z) 
  = 4(|a_{11}(z)|^2-1)^2\left[ 1-|a_{11}(z)|^2|\rho_3(z)|^2\right]\,.
\label{e:detS}
\end{gather}
Note that
one cannot conclude from~\eref{e:detS} that $\det \@S(z)\ne 0$ for all $z\in \Real$.
On the other hand, $\rho_3(z)\equiv 0$ for all $z\in \Real$ implies $\rho_1(z)=\rho_2(z) =  0$ for all $z\in \Real$ for which $\det S(z)\ne 0$.
\begin{remark}
We will show later that the entire scattering matrix can be reconstructed in terms of appropriate scattering data
(namely, discrete spectrum, reflection coefficients and the value of the non-principal analytic minors along the real $z$-axis, if the latter are not identically zero).
The key ingredient are the trace formula for the analytic scattering coefficients,
which will be derived in Section~\ref{s:tracemain}.

\end{remark}

\subsection{Discrete spectrum}
\label{s:discretespectrum}

The discrete spectrum corresponds to those values $z\in\Complex$ where the columns of the fundamental analytic eigenfunctions are linearly dependent.
Using the decompositions~\eref{e:decomp}, it is easy to see that
\bse
\label{e:wrdet}
\begin{gather}
\label{e:wrdeta}
\det \chi^+(x,t,z) = A_{[1]}(z) A_{[1,2]}(z) A_{[1,2,3]}(z) \gamma(z) \e^{2 i \theta_2(x,t,z)}, \qquad \Im z \geq 0,
\\
\label{e:wrdetb}
\det \chi^-(x,t,z) = A_{[4]}(z) A_{[3,4]}(z) A_{[2,3,4]}(z) \gamma(z) \e^{2 i \theta_2(x,t,z)}, \qquad \Im z \leq 0.
\end{gather}
\ese
We therefore have:
\begin{proposition}
A point $z_o \in\Complex^+$ is a discrete eigenvalue of the scattering problem if
\be
A_{[1]}(z_o) A_{[1,2]}(z_o) A_{[1,2,3]}(z_o) = 0\,.
\label{d:discreteeigenvalues}
\ee
\end{proposition}
We thus have seven possibilities for a discrete eigenvalue, $z_o$, summarized in Table~\ref{table1}.
\begin{table}[t!]
\begin{center}
\begin{tabular}{ c | c  c  c | l }
     {} & $A_{[1]}(z_o)$ & $A_{[1,2]}(z_o)$ & $A_{[1,2,3]}(z_o)$ & ~Summary \\ \hline
    I~~ & 0 & $*$ & $*$ & ~Admissible~ [symmetric to II]   \\
    II~~ & $*$ & $*$ & 0 & ~Admissible~ [symmetric to I] \\
    III~~ & $*$ & 0 & $*$ & ~Inadmissible~ \\ \hline
    IV~~ & 0 & 0 & $*$ & ~Admissible~ [symmetric to V] \\
    V~~ & $*$ & 0 & 0 & ~Admissible~ [symmetric to IV] \\
    VI~~ & 0 & $*$ & 0 & ~Inadmissible~ \\ \hline
    VII~~ & 0 & 0 & 0 & ~Inadmissible~ \\ \hline
\end{tabular}
\end{center}
\caption{The seven possible combinations of zeros of the principal minors for the discrete spectrum.
An asterisk denotes an arbitrary nonzero value for the corresponding minor.
The rightmost column shows the corresponding results of the analysis (cf.\ Theorems~\ref{t:spectrum},
\ref{t:type2theorem} and~\ref{t:type1theorem}).}
\label{table1}
\end{table}

\begin{remark}
At each point $z_o\in\Complex^+$ such that~\eref{d:discreteeigenvalues} holds,
the fundamental analytic eigenfunctions fail to be a basis for $\Complex^{N+1}$.
The task of characterizing the discrete spectrum (which is necessary in order to obtain the soliton solutions)
is precisely to overcome this problem and identify all the relations expressing the linear dependency among all the columns
of $\chi^\pm(x,t,z)$ at such points
(which in turn is necessary to characterize the residue relations among the meromorphic matrices of the inverse problem).
For the $2\times2$ scattering problem (corresponding to the scalar NLS equation),
only one possibiliy exists: the two columns are proportional to each other.
Already in the $3\times3$ scattering problem (corresponding to the Manakov system), however,
the situation is more complicated \cite{PAB2006,gbdkk2015}.
We will see that for the $4\times4$ problem, the situation is significantly more complex than even for the Manakov system.
\end{remark}

As in the defocusing Manakov case,
we restrict ourselves to the case in which all discrete eigenvalues are simple zeros of the analytic scattering coefficients.
Also, as in the defocusing Manakov case,
one must distinguish between discrete eigenvalues on the circle $C_o$ and
``improper'' discrete eigenvalues off $C_o$:
\begin{lemma}
\label{L:L2eigen}
Let $\@v(x,t,z)$ be a nontrivial solution of the scattering problem in~\eref{e:Laxpair}.
If $\@v(x,t,z) \in L^2(\Real)$, then
$z \in C_o$.
\end{lemma}
In other words, as with the scalar defocusing NLS with NZBC and the defocusing Manakov system with NZBC,
bound states are exclusively associated with eigenvalues on $C_o$.
Thus, similarly to the defocusing Manakov system with NZBC,
discrete eigenvalues of the scattering problem not in $C_o$ do not lead to bound states \cite{PAB2006,gbdkk2015}.

\begin{remark}
Let us briefly elaborate on the consequences of Lemma~\ref{L:L2eigen}. 
Suppose $A_{[1]}(z_o) = 0$ for some $z_o \in\Complex^+$ with $|z_o| \ne q_o$.
We know that $\phi_{-,1}(x,t,z_o)$ vanishes as $x \to - \infty$.
We also know from~\eref{e:chipminf}
that $\mu_{-,1}(x,t,z_o) = \phi_{-,1}(x,t,z_o)\,\e^{-i\theta_1(x,t,z_o)}$ vanishes as $x \to \infty$.
On the other hand, this is not a contradiction, because it does not imply that $\phi_{-,1}(x,t,z_o)\to0$ as $x\to\infty$.
Indeed, it is enough for $\phi_{-,1}(x,t,z_o)$
to diverge more slowly than $\e^{i\lambda(z_o)x}$ as $x\to\infty$.
Such a possibility does not exist in the scalar case, but is allowed in the vector case,
because in this case $\phi_{-,1}(x,t,z_o)$ can simply be a linear combination of the two remaining analytic eigenfunctions,
$\chi_2^+(x,t,z_o)$ and $\chi_3^+(x,t,z_o)$.
Similar results hold for $\mu_{+,4}(x,t,z_o)$ when $z_o$ is a zero of $A_{[1,2,3]}(z)$ and for the
eigenfunctions analytic in the lower-half $z$-plane.
\end{remark}

\begin{remark}
For the remainder of this section, we consider the case where the analytic non-principal minors
in Theorem \ref{t:gantmacher}
are identically zero in their domain of analyticity.
%
The rationale is one of expediency:
In the general case (i.e., when the analytic non-principal minors are not identically zero),
the value of two analytic non-principal minors needs to be given
(in addition to that of the principal minors)
at a point of an eigenvalue quartet in order to completely specify each case.
It thus follows that for each of the seven cases in Table~\ref{table1} one must study four sub-cases,
for a total of 28 possible configurations.
Hence, the analysis of the discrete spectrum and the inverse problem in all possible configurations of the general case
would be considerably longer.
Of course 
by taking the analytic non-principal minors to be identically zero,
we are restricting ourselves to a subclass of solutions of the VNLS system.
On the other hand,
we will see that even in this case 
the discrete spectrum and the corresponding soliton solutions are still
considerably richer than for the 2-component case.
\end{remark}

The symmetries~\eref{e:auxsymm2} and~\eref{e:A12symm}
are greatly simplified when the analytic non-principal minors are identically zero.
Namely, in this case we have:
\bse
\label{e:A12symm0}
\begin{gather}
\e^{i\Delta\theta} A_{[1,2]}(z) A_{[1,2]}^*(\^z) = A_{[1]}(z) A_{[1,2,3]}(z)\,,
\qquad \Im z>0\,,
\\
\e^{-i\Delta\theta} A_{[3,4]}(z) A_{[3,4]}^*(\^z) = A_{[4]}(z) A_{[2,3,4]}(z)\,,
\qquad \Im z<0\,.
\end{gather}
\ese
as well as
\bse
\label{e:refsymm}
\begin{gather}
\label{e:refsymma}
\chi_2^+(x,t,\z^*) = \frac{A_{[4]}(z)}{A_{[3,4]}(z)} \e^{i \Delta \theta} \chi_2^-(x,t,z), \qquad \Im z < 0,
\\
\label{e:refsymmb}
\chi_3^+(x,t,\z^*) = \frac{A_{[2,3,4]}(z)}{A_{[3,4]}(z)} \e^{i \Delta \theta} \chi_3^-(x,t,z), \qquad \Im z < 0,
\\
\label{e:refsymmc}
\chi_2^-(x,t,\z^*) = \frac{A_{[1,2,3]}(z)}{A_{[1,2]}(z)} \e^{-i \Delta \theta} \chi_2^+(x,t,z), \qquad \Im z > 0,
\\
\label{e:refsymmd}
\chi_3^-(x,t,\z^*) = \frac{A_{[1]}(z)}{A_{[1,2]}(z)} \e^{-i \Delta \theta} \chi_3^+(x,t,z), \qquad \Im z > 0.
\end{gather}
\ese
Since eigenvalues on $C_o$ lead to pure dark soliton solutions,
for the remainder of this work, we only consider discrete eigenvalues off $C_o$.
Explicitly, we only consider discrete eigenvalues $z_o$ such that $\Im z_o > 0$ and $|z_o| \neq q_o$.
In Appendix~\ref{a:discrete} we use these symmetries to identify all of the possibilities for
the discrete eigenvalues:
\begin{theorem}
\label{t:spectrum}
For a discrete eigenvalue $z_o\in\Complex^+\setminus C_o$, one of the following possibilities holds:
\begin{itemize}
\setlength{\itemsep}{0cm}%
\setlength{\parskip}{0cm}%
\item[I.] $A_{[1]}(z_o) = 0$, $A_{[1,2]}(z_o) \neq 0$, $A_{[1,2,3]}(z_o) \neq 0$.
\item[II.] $A_{[1]}(z_o) \neq 0$, $A_{[1,2]}(z_o) \neq 0$, $A_{[1,2,3]}(z_o) = 0$.
\item[IV.] $A_{[1]}(z_o) = A_{[1,2]}(z_o) = 0$, $A_{[1,2,3]}(z_o) \neq 0$.
\item[V.] $A_{[1]}(z_o) \neq 0$, $A_{[1,2]}(z_o) = A_{[1,2,3]}(z_o) = 0$.
\end{itemize}
\end{theorem}
As indicated by the numbering above, we label these eigenvalues 
in accordance with Table~\ref{table1}.
In other words, Theorem~\ref{t:spectrum} says that
discrete eigenvalues of type~III, VI and VII are not allowed.
Analyzing discrete eigenvalues of type I and type IV, in Appendix~\ref{a:discrete} we obtain the following:
\begin{theorem}
\label{t:type2theorem}
(Type~I)~
Suppose $A_{[1]}(z_o)=0$, $A_{[1,2]}(z_o) \neq 0$, and $A_{[1,2,3]}(z_o) \neq 0$.
Then $|z_o|<q_o$,
and there exist constants $d_1$, $\=d_1$, $\^d_1$, and $\check{d}_1$ such that
\bse
\label{e:type2constants}
\begin{gather}
\phi_{-,1}(x,t,z_o) = d_1 \chi_2^+(x,t,z_o)/A_{[1,2]}(z_o),
\qquad
\phi_{-,4}(x,t,\z_o^*) = \check{d}_1 \chi_2^-(x,t,\z_o^*),
\\
\chi_2^-(x,t,z_o^*) = \=d_1 \phi_{+,1}(x,t,z_o^*),
\qquad
\chi_2^+(x,t,\z_o) = \^d_1 \phi_{+,4}(x,t,\z_o).
\end{gather}
\ese
These constants satisfy the following symmetry relations:
\begin{equation}
\check{d}_1 = - \frac{i z_o}{q_+^*} \frac{d_1}{A_{[1,2,3]}(z_o)},
\qquad
\=d_1 = \frac{d_1^*}{\gamma(z_o^*)},
\qquad
\^d_1 = \frac{i q_-^*}{z_o^*\gamma(z_o^*)} \frac{A_{[4]}(z_o^*)}{[A_{[1,2]}(z_o)]^*}
d_1^*.
\end{equation}
\end{theorem}
\begin{theorem}
\label{t:type1theorem}
(Type~IV)~
Suppose $A_{[1]}(z_o)=A_{[1,2]}(z_o)=0$ and $A_{[1,2,3]}(z_o) \neq 0$.
Then $|z_o|<q_o$,
and there exist constants $c_1$, $\=c_1$, $\^c_1$, and $\check{c}_1$ such that
\bse
\label{e:type1constants}
\begin{gather}
\label{e:type1constantsa}
\phi_{-,1}(x,t,z_o) = c_1 \chi_3^+(x,t,z_o)/A_{[1,2,3]}(z_o),
\qquad
\phi_{-,4}(x,t,\z_o^*) = \check{c}_1 \chi_3^-(x,t,\z_o^*),
\\
\label{e:type1constantsb}
\chi_3^-(x,t,z_o^*) = \=c_1 \phi_{+,1}(x,t,z_o^*),
\qquad
\chi_3^+(x,t,\z_o) = \^c_1 \phi_{+,4}(x,t,\z_o).
\end{gather}
\ese
These constants satisfy the following symmetry relations:
\begin{equation}
\label{e:type1constsymm}
\check{c}_1 = - \frac{i z_o}{q_+^*} \frac{A_{[1,2]}'(z_o)}{A_{[1]}'(z_o)}
\frac{c_1}{A_{[1,2,3]}(z_o)},
\qquad
\^c_1 = \frac{i q_-^*}{z_o^*\gamma(z_o^*)} c_1^*,
\qquad
\=c_1 = \frac{1}{\gamma(z_o^*)} \frac{A_{[3,4]}'(z_o^*)}{A_{[2,3,4]}'(z_o^*)}c_1^*.
\end{equation}
\end{theorem}
Examining discrete eigenvalues of type~II and~V in the same manner yields relations similar to those in
Theorems \ref{t:type2theorem} and \ref{t:type1theorem} but with the requirement that $|z_o|>q_o$.
We omit the details for brevity.

\section{Inverse problem}
\label{s:istinverse}

As usual, the inverse scattering problem is formulated in terms of an appropriate RHP which relates
eigenfunctions that are meromorphic in the upper-half $z$-plane
to eigenfunctions that are meromorphic in the lower-half $z$-plane.

\subsection{Derivation of the Riemann-Hilbert problem and reconstruction formula for the potential}
\label{s:rhpmain}

The starting point for the formulation of the inverse problem is the scattering
relation \eref{e:scattering}.
Just like the derivation of the jump condition for the Manakov system is considerably more involved
than for the scalar case, however,
so too is the derivation of the jump condition for the 3-component case more complicated than
for the Manakov system.
In both cases,
this is due to the presence of non-analytic eigenfunctions, which must be used via the decompositions
\eref{e:decomp} in terms of the fundamental analytic eigenfunctions $\chi^\pm(x,t,z)$.
As we will see, however, in this case there are additional twists.
One starts by introducing the meromorphic matrix
\begin{equation}
\@M(x,t,z) = \begin{cases}
  \@M^+(x,t,z)\,,&\Im z>0\,,\\
  \@M^-(x,t,z)\,,&\Im z<0\,.
           \end{cases}
\label{e:MRHPdef}
\end{equation}
By analogy with the 2-component case, one could think of defining
\begin{gather*}
\@M^+(x,t,z) = \left(\frac{\phi_{-,1}(z)}{A_{[1]}(z)},\frac{\chi_2^+(z)}{A_{[1,2]}(z)},\frac{\chi_3^+(z)}{A_{[1,2,3]}(z)},\phi_{+,4}(z)\right)\,\e^{-i\@\Theta(x,t,z)}, \qquad \Im z > 0,
\\
\@M^-(x,t,z) = \left(\phi_{+,1}(z),\frac{\chi_2^-(z)}{A_{[2,3,4]}(z)},\frac{\chi_3^-(z)}{A_{[3,4]}(z)},\frac{\phi_{-,4}(z)}{A_{[4]}(z)}\right)\,\e^{-i\@\Theta(x,t,z)}, \qquad \Im z < 0,
\end{gather*}
where the parametric space and time dependence of the eigenfunctions in the right-hand side was omitted for brevity.
Like in the Manakov system,
one can appropriately manipulate the scattering relation~\eref{e:scattering} to obtain
a jump condition between $\@M^\pm(x,t,z)$ on $z\in\Real$.
There is a fundamental difference between the 3-component case and the 2-component case, however:

\begin{remark}
Due to the presence of the analytic non-principal minors in the 3-component case,
if one defines the meromorphic matrices $\@M^\pm(x,t,z)$ as above, the resulting jump condition involves
a jump matrix which
does not reduce to the identity in the reflectionless case.
This condition is necessary to be able to write down soliton solutions in closed form.
In order to obtain a jump condition that satisfies the above condition,
one must modify the definition of the meromorphic matrices in the RHP.
\end{remark}

\noindent
Explicitly, we define the meromorphic matrices $\@M^\pm(x,t,z)$ in~\eref{e:MRHPdef} as follows:
\bse
\label{e:Mpmdef}
\begin{gather}
\@M^+(x,t,z) = \left(\frac{\phi_{-,1}(z)}{A_{[1]}(z)},\frac1{A_{[1,2]}(z)}\bigg(\chi_2^+(z)-\frac{A_{\minorset{1,3}{1,2}}(z)}{A_{[1,2,3]}(z)}\chi_3^+(z)\bigg),\frac{\chi_3^+(z)}{A_{[1,2,3]}(z)},\phi_{+,4}(z)\right)\,\e^{-i\@\Theta(z)},
\qquad \Im z > 0\,,
\\
\@M^-(x,t,z) = \left(\phi_{+,1}(z),\frac{\chi_2^-(z)}{A_{[2,3,4]}(z)},\frac1{A_{[3,4]}(z)}\bigg(\chi_3^-(z)-\frac{A_{\minorset{2,4}{3,4}}(z)}{A_{[2,3,4]}(z)}\chi_2^-(z)\bigg),\frac{\phi_{-,4}(z)}{A_{[4]}(z)}\right)\,\e^{-i\@\Theta(z)},
\qquad \Im z < 0\,.
\end{gather}
\ese
Obviously these modified definitions of $\@M^\pm(x,t,z)$ reduce to the previous ones when the analytic non-principal minors are identically zero.
It will be convenient to denote the columns of $\@M^\pm$ as $\@M^\pm(x,t,z)
= (M_1^\pm(x,t,z),M_2^\pm(x,t,z),M_3^\pm(x,t,z),M_4^\pm(x,t,z))$.
The symmetries \eref{e:second} and \eref{e:auxsymm2} for the Jost eigenfunctions
and auxiliary eigenfunctions, together with the symmetries for the principal and analytic non-principal minors,
then imply the following symmetry relations
for the columns of $M^\pm$ in \eref{e:Mpmdef}:
\bse
\label{e:Mpmsymm}
\begin{gather}
M_1^-(x,t,z)=\frac{iq_+^*}{z}M_4^+(x,t,\^z^*)\,, \qquad M_1^+(x,t,z)=\frac{iq_+^*}{z}M_4^-(x,t,\^z^*)\,, \\
M_j^+(x,t,z)=M_j^-(x,t,\^z^*)\,, \quad j=2,3\,,
\end{gather}
\ese
which hold for all $z\in \Complex$ at which the above quantities are analytic.
Using the definitions~\eref{e:Mpmdef}, in Appendix~\ref{a:rhp} we then prove the following:
\begin{lemma}
\label{L:jump}
The meromorphic matrix $\@M(x,t,z)$ in~\eref{e:MRHPdef}
satisfies the jump condition
\begin{equation}
\label{e:rhpnewjump}
\@M^+(x,t,z) = \@M^-(x,t,z) [\@I - \e^{- i \@K \@\Theta(x,t,z)} \@L(z) \e^{i \@K \@\Theta(x,t,z)}], \qquad z \in \Real,
\end{equation}
where $\@K = \diag(-1,1,1,-1)$ as before and
$\@L(z) = \big(L_{i,j}(z)\big)$, with
\begin{gather*}
L_{1,1}(z) = \frac{|\rho_1(z)|^2}{\gamma(z)}
+ b(z) \left[\rho_2(z) + \frac{i q_+}{z} \rho_2(\z^*)\right]
- \frac{q_+}{q_+^*} \rho_3(z) \rho_3(\z^*)
+ \frac{i q_+}{z \gamma(z)} \rho_1^*(z) \rho_1(\z^*),
\\
L_{2,1}(z) = - \rho_1(z) - \frac{i q_+}{z} \rho_1(\z^*) \rho_3(z),
\qquad
L_{3,1}(z) = - \rho_2(z) - \frac{i q_+}{z} \rho_2(\z^*) \rho_3(z),
\qquad
L_{4,1}(z) = - \rho_3(z),
\\
L_{1,2}(z) = \frac{\rho_1^*(z)}{\gamma(z)}
- a(z) \left[\frac{q_+}{q_+^*} \rho_3(\z^*) - \frac{i q_+}{z \gamma(z)} \rho_1^*(z) \rho_1(\z^*)
- b(z) \frac{i q_+}{z} \rho_2(\z^*)\right],
\\
L_{2,2}(z) = - \frac{i q_+}{z} a(z) \rho_1(\z^*),
\qquad
L_{3,2}(z) = - \frac{i q_+}{z} a(z) \rho_2(\z^*),
\qquad
L_{4,2}(z) = - a(z),
\\
L_{1,3}(z) = b(z) + \frac{i q_+}{z \gamma(z)} \left[\frac{q_+}{q_+^*} \rho_2^*(\z^*) \rho_3(\z^*)
- \frac{i q_+}{z \gamma(z)} \rho_1^*(z) \rho_1(\z^*) - \frac{i q_+}{z} b(z) \rho_2(\z^*)\right],
\\
L_{2,3}(z) = - \frac{q_+^2}{z^2 \gamma(z)} \rho_1(\z^*) \rho_2^*(\z^*),
\qquad
L_{3,3}(z) = - \frac{q_+^2}{z^2 \gamma(z)} |\rho_2(\z^*)|^2,
\qquad
L_{4,3}(z) = \frac{i q_+}{z \gamma(z)} \rho_2^*(\z^*),
\\
L_{1,4}(z) = - \frac{q_+}{q_+^*} \rho_3(\z^*) + \frac{i q_+}{z \gamma(z)} \rho_1^*(z) \rho_1(\z^*)
+ \frac{i q_+}{z} b(z) \rho_2(\z^*),
\\
L_{2,4}(z) = - \frac{i q_+}{z} \rho_1(\z^*),
\qquad
L_{3,4}(z) = - \frac{i q_+}{z} \rho_2(\z^*),
\qquad
L_{4,4}(z) = 0,
\end{gather*}
and where for brevity we defined
\begin{gather*}
a(z) = \frac{A_{\minorset{1,4}{1,2}}(z)}{A_{[1,2]}(z)} -\frac z{iq_+} \=\rho_2(\^z)\frac{A_{\minorset{1,3}{1,2}}(z)}{A_{[1,2]}(z)}\,,\qquad
b(z) = \frac{B_{\minorset{1,2}{2,3}}(z)}{B_{[1,2]}(z)} - \=\rho_1(z)\frac{B_{\minorset{1,2}{1,3}}(z)}{B_{[1,2]}(z)}\,.
\end{gather*}
\end{lemma}
Note that
$a(z)$ and $b(z)$ have explicit (albeit complicated) dependence on the reflection coefficients in \eref{e:reflcoeff}.
The expressions for $a(z)$ and $b(z)$ simplify considerably when the analytic non-principal minors are identically zero.
Also, these expressions imply that $a(z)$ and $b(z)$ vanish identically in the reflectionless case.

Using the asymptotics of the eigenfunctions and scattering matrix in Section \ref{s:asymptoticsoriginal},
we obtain the asymptotics of the meromorphic matrices $\@M^\pm(x,t,z)$ for $z$ in the appropriate half plane
in a straightforward way:
\begin{lemma}
\label{L:rhpasymp}
The meromorphic matrices $\@M^\pm(x,t,z)$ in~\eref{e:Mpmdef} have the following asymptotic behavior for $z$ in the appropriate
half plane:
\bse
\label{e:Masymp}
\begin{gather}
\@M^\pm(x,t,z) = \@I + O(1/z), \qquad z \to \infty,
\label{e:Masympinfty}
\\
\@M^\pm(x,t,z) = -(i/z) \@J \@Q_+ + O(1), \qquad z \to 0.
\end{gather}
\ese
\end{lemma}
In passing, note that the sum of the leading-order asymptotic behaviors of $\@M^\pm(x,t,z)$ as $z \to \infty$
and $z \to 0$ is $\@E_+(z)$.

\begin{remark}
As usual, the meromorphic matrices $\@M^\pm(x,t,z)$ have poles in correspondence with the discrete eigenvalues.
As in section~\ref{s:discretespectrum}, we now restrict ourselves to the case in which
the analytic non-principal minors are identically zero.
%
Hereafter, we denote by
$\{z_1,\dots,z_{N_1}\}$
and by
$\{w_1,\dots,w_{N_2}\}$
the discrete eigenvalues of type~I and~IV, respectively.
\end{remark}
The relations in \eref{e:type2constants}
and~\eref{e:type1constants}
easily yield the residue conditions for the RHP:
\begin{lemma}
\label{L:residues}
The meromorphic matrices defined in Lemma \ref{L:jump} satisfy the following residue conditions:
\begin{gather*}
\Res_{z=z_n} \@M^+ = \left(D_n(x,t) M_2^+(x,t,z_n),\@0,\@0,\@0\right),
\quad
\Res_{z=\z_n} \@M^+ = \left(\@0,\^D_n(x,t) M_4^+(x,t,\z_n),\@0,\@0\right),
\\
\Res_{z=z_n^*} \@M^- = \left(\@0,\=D_n(x,t) M_1^-(x,t,z_n^*),\@0,\@0\right),
\quad
\Res_{z=\z_n^*} \@M^- = \left(\@0,\@0,\@0,\check{D}_n(x,t) M_2^-(x,t,\z_n^*)\right),
\\
\Res_{z=w_n} \@M^+ = \left(C_n(x,t) M_3^+(x,t,w_n),\@0,\@0,\@0\right),
\quad
\Res_{z=\w_n} \@M^+ = \left(\@0,\@0,\^C_n(x,t) M_4^+(x,t,\w_n),\@0\right),
\\
\Res_{z=w_n^*} \@M^- = \left(\@0,\@0,\=C_n(x,t) M_1^-(x,t,w_n^*),\@0\right),
\quad
\Res_{z=\w_n^*} \@M^- = \left(\@0,\@0,\@0,\check{C}_n(x,t) M_3^-(x,t,\w_n^*)\right),
\end{gather*}
with
\begin{gather*}
D_n(x,t) = \frac{d_n}{A_{[1]}'(z_n)} \e^{- i (\theta_1 - \theta_2)(x,t,z_n)},
\qquad
\check{D}_n(x,t) = \frac{\check{d}_n A_{[2,3,4]}(\z_n^*)}{A_{[4]}'(\z_n^*)}
\e^{- i (\theta_1 - \theta_2)(x,t,z_n)},
\\
\=D_n(x,t) = \frac{\=d_n}{A_{[2,3,4]}'(z_n^*)} \e^{i (\theta_1 - \theta_2)(x,t,z_n^*)},
\qquad
\^D_n(x,t) = \frac{\^d_n}{A_{[1,2]}'(\z_n)} \e^{i (\theta_1 - \theta_2)(x,t,z_n^*)},
\\
C_n(x,t) = \frac{c_n}{A_{[1]}'(w_n)} \e^{- i (\theta_1 - \theta_2)(x,t,w_n)},
\qquad
\check{C}_n(x,t) = \frac{\check{c}_n A_{[3,4]}(\w_n^*)}{A_{[4]}'(\w_n^*)}
\e^{- i (\theta_1 - \theta_2)(x,t,w_n)},
\\
\=C_n(x,t) = \frac{\=c_n}{A_{[3,4]}'(w_n^*)} \e^{i (\theta_1 - \theta_2)(x,t,w_n^*)},
\qquad
\^C_n(x,t) = \frac{\^c_n}{A_{[1,2,3]}'(\w_n)} \e^{i (\theta_1 - \theta_2)(x,t,w_n^*)},
\end{gather*}
with $n=1,\dots,N_1$ in the equations involving $z_n$ and $n=1,\dots,N_2$ in those involving $w_n$.
\end{lemma}
In addition, a straightforward consequence of the symmetries of the scattering matrix and the symmetries of the
norming constants in Theorems~\ref{t:type2theorem} and~\ref{t:type1theorem} is the following:
\begin{lemma}
\label{L:residuesymmetries}
The residue conditions in Lemma \ref{L:residues} satisfy the following symmetry relations:
\begin{gather*}
\check{D}_n(x,t) = \frac{i q_+}{z_n} D_n(x,t),
\qquad
\=D_n(x,t) = \frac{D_n^*(x,t)}{\gamma(z_n^*)}, \qquad
\^D_n(x,t) = \frac{i q_+^*}{z_n^*\gamma(\^z_n)} D_n^*(x,t),
\\
\check{C}_n(x,t) =\frac{i q_+}{w_n} C_n(x,t),
\qquad
\=C_n(x,t) = \frac{C_n^*(x,t)}{\gamma(w_n^*)}, \qquad
\^C_n(x,t) = i \frac{q_+^*}{w_n^*\gamma(\^w_n)} C_n^*(x,t),
\end{gather*}
where, as before, $n=1,\dots,N_1$ for equations involving $z_n$ and $n=1,\dots,N_2$ for equations involving $w_n$.
\end{lemma}
Note that in order to simplify the above symmetries, it was necessary to consider the derivative with respect to $z$
of the symmetry relations~\eref{e:A12symm} and evaluate it at the discrete eigenvalues.

\begin{remark}
\label{R:RHP}
Based on the above results,
the inverse problem can be formulated in terms of the following Riemann-Hilbert problem:
Find a matrix-valued function $\@M(x,t,z)$
which is analytic in $\Complex\setminus\Real$ away from the discrete spectrum, satisfies
the jump condition in Lemma~\ref{L:jump}, 
the asymptotics in Lemma~\ref{L:rhpasymp} 
and the residue conditions in Lemma~\ref{L:residues} (with the symmetries in Lemma~\ref{L:residuesymmetries}).
\end{remark}
The above RHP is regularized and formally solved in the usual way;
namely, by subtracting the leading order asymptotic
behavior and any pole contributions
and then applying the Cauchy projectors
\begin{equation*}
\label{e:projector}
(P_\pm f)(z) = \frac{1}{2 \pi i} \lim_{\epsilon \to 0^+} \int_\Real \frac{f(\zeta)}{\zeta - (z \pm i \epsilon)} \d \zeta\,.
\end{equation*}
The Plemelj-Sokhotsky formulae then yield the following integral representation for the solution of the RHP:
\begin{multline}
\label{e:rhpsol}
\@M(x,t,z) = \@E_+(z)
+ \sum_{n=1}^{N_1} \left(\frac{\Res_{z=z_n} \@M^+}{z-z_n} + \frac{\Res_{z=z_n^*} \@M^-}{z-z_n^*}\right)
+ \sum_{n=1}^{N_1} \left(\frac{\Res_{z=\z_n} \@M^+}{z-\z_n} + \frac{\Res_{z=\z_n^*} \@M^-}{z-\z_n^*}\right)
\\
+ \sum_{n=1}^{N_2} \left(\frac{\Res_{z=w_n} \@M^+}{z-w_n} + \frac{\Res_{z=w_n^*} \@M^-}{z-w_n^*}\right)
+ \sum_{n=1}^{N_2} \left(\frac{\Res_{z=\w_n} \@M^+}{z-\w_n} + \frac{\Res_{z=\w_n^*} \@M^-}{z-\w_n^*}\right)
\\
- \frac{1}{2 \pi i} \int_\Real \frac{\@M^-(x,t,\zeta)}{\zeta - z} \~{\@L}(x,t,\zeta) \d \zeta,\qquad z\notin\Real\,,
\kern8em
\end{multline}
where $\~{\@L}(x,t,z) = \e^{- i \@K \@\Theta} \@L \e^{i \@K \@\Theta}$ and
$\@M(x,t,z) = \@M^\pm$ for $\Im z \gtrless 0$, as before.

To close the system, it remains to find appropriate equations for the columns of $\@M(x,t,z)$ appearing
in the residue conditions in Lemma \ref{L:residues}.
As usual,
this is done by evaluating the expression \eref{e:rhpsol} for the regular columns of the solution of the RHP at the appropriate discrete eigenvalues,
which yields the following system of algebraic equations:
\bse
\label{e:threesystem}
\begin{gather}
M_1^-(x,t,z) = \@E_{+,1}(z) + \sum_{j=1}^{N_1} \frac{D_j M_2^+(z_j)}{z-z_j}
+ \sum_{j=1}^{N_2} \frac{C_j M_3^+(w_j)}{z-w_j}
- \frac{1}{2 \pi i} \int_\Real (\@M^-\~{\@L})_1(\zeta) \frac{\d \zeta}{\zeta - z}, \quad z = z_n^*,w_n^*,
\\
M_4^+(x,t,z) = \@E_{+,4}(z) + \sum_{j=1}^{N_1} \frac{\check{D}_j M_2^-(\z_j^*)}{z-\z_j^*}
+ \sum_{j=1}^{N_2} \frac{\check{C}_j M_3^-(\w_j^*)}{z-\w_j^*}
- \frac{1}{2 \pi i} \int_\Real (\@M^-\~{\@L})_4(\zeta) \frac{\d \zeta}{\zeta - z}, \quad z = \z_n,\w_n,
\\
M_2^+(x,t,z_n) =
\@e_2 + \sum_{j=1}^{N_1} \left(\frac{\^D_j M_4^+(\z_j)}{z_n-\z_j}
+ \frac{\~D_j M_1^-(z_j^*)}{z_n-z_j^*}\right)
- \frac{1}{2 \pi i} \int_\Real (\@M^-\~{\@L})_2(\zeta) \frac{\d \zeta}{\zeta - z_n},
\\
M_3^+(x,t,w_n) =
\@e_3 + \sum_{j=1}^{N_2} \left(\frac{\^C_j M_4^+(\w_j)}{w_n-\w_j}
+ \frac{\~C_j M_1^-(w_j^*)}{w_n-w_j^*}\right)
- \frac{1}{2 \pi i} \int_\Real (\@M^-\~{\@L})_3(\zeta) \frac{\d \zeta}{\zeta - w_n},
\\
M_2^-(x,t,\z_n^*) = \@e_2
+ \sum_{j=1}^{N_1} \left(\frac{\^D_j M_4^+(\z_j)}{\z_n^*-\z_j}
+ \frac{\=D_j M_1^-(z_j^*)}{\z_n^*-z_j^*}\right)
- \frac{1}{2 \pi i} \int_\Real (\@M^-\~{\@L})_2(\zeta) \frac{\d \zeta}{\zeta - \z_n^*},
\\
M_3^-(x,t,\w_n^*) = \@e_3
+ \sum_{j=1}^{N_2} \left(\frac{\^C_j M_4^+(\w_j)}{\w_n^*-\w_j}
+ \frac{\=C_j M_1^-(w_j^*)}{\w_n^*-w_j^*}\right)
- \frac{1}{2 \pi i} \int_\Real (\@M^-\~{\@L})_3(\zeta) \frac{\d \zeta}{\zeta - \w_n^*},
\end{gather}
\ese
where again we denoted by $\{\@e_j\}_{j=1}^4$ the standard basis for $\Real^4$ and omitted the $(x,t)$-dependence
on the right-hand sides of all equations for brevity.

\begin{remark}
The integral representation~\eref{e:rhpsol} can be converted into a set of linear integral equations in the standard way.
The resulting equations formally provide, together with the algebraic equations~\eref{e:threesystem},
the solution of the inverse problem.
\end{remark}
Finally, evaluating the asymptotic behavior of~\eref{e:rhpsol} as $z\to\infty$,
and combining with~\eref{e:reconstruction}
one obtains the reconstruction formula for the solution of the initial value problem for the :

\begin{corollary}
The solution of the defocusing 3-component VNLS equation~\eref{e:vnls}
with the NZBC~\eref{ENZBC} and~\eref{e:NZBC} is obtained from the solution of the RHP in Remark~\ref{R:RHP} as follows:
\begin{multline}
q_j(x,t) = q_{+,j} - i \sum_{n=1}^{N_1} D_n(x,t) M_{j+1,2}^+(x,t,z_n)
- i \sum_{n=1}^{N_2} C_n(x,t) M_{j+1,3}^+(x,t,w_n)
\\
- \frac{1}{2 \pi} \int_\Real (\@M^- \~{\@L})_{j+1,1}(x,t,\zeta) \d \zeta,
\qquad
j=1,2,3.
\end{multline}
\end{corollary}

\begin{remark}
The question of the existence and uniqueness of the solution of the RHP
can be answered in a manner very similar to the case of the defocusing Manakov system with NZBC
\cite{gbdkk2015}.
We omit the details for brevity.
\end{remark}

\subsection{Trace formulae, asymptotic phase difference and reconstruction of the scattering matrix}
\label{s:tracemain}

It remains to find trace formulae for the analytic minors of the scattering matrix.
As we show in the appendix, these can be obtained 
by setting up appropriate scalar RHPs for these minors.
\begin{lemma}
\label{L:trace}
(Trace formulae)
If the analytic non-principal minors of the scattering matrix are identically zero,
the principal minors of the scattering matrix can be reconstructed in terms of the scattering data as follows:
\bse
\label{e:trace}
\begin{gather}
\label{e:a11trace}
A_{[1]}(z) = \prod_{n=1}^{N_1} \frac{z-z_n}{z-z_n^*}
    \prod_{n=1}^{N_2} \frac{z-w_n}{z-w_n^*}
      \exp \left( - \frac{1}{2 \pi i} \int_\Real \frac{\log[1-R(\zeta)]}{\zeta - z} \d \zeta\right),
\qquad
\Im z >0\,,
\\
\label{e:a12trace}
A_{[1,2]}(z) =  \prod_{n=1}^{N_1} \frac{z-\^z_n}{z-\^z_n^*}
    \prod_{n=1}^{N_2} \frac{z-w_n}{z-w_n^*}
   \exp \left(
     \frac1{2\pi i} \int_{|\zeta|=q_o} \frac{g(\zeta)}{\zeta-z}\d\zeta
   \right)\,,
\qquad
\Im z>0\,,
\end{gather}
\ese
where
\begin{gather*}
R(\zeta) =
|\rho_3(\zeta)|^2 + \frac1{\gamma(\zeta)}(|\rho_2(\zeta)|^2 + |\rho_1(\zeta)|^2)\,,
\qquad
\zeta\in\Real\,,
\end{gather*}
and where
$g(z) = |A_{[1]}(z)|^2$ for $\Im z \ge 0$ and
$g(z) = |A_{[1]}(z^*)|^2$ for $\Im z<0$.
\end{lemma}
The expressions of the remaining analytic minors of the scattering matrix follow immediately from the above
via the symmetries of the scattering matrix.
\begin{remark}
\label{r:tracegeneral}
When the analytic non-principal minors are not identically zero, one can exploit Corollary~\ref{c:minorasymptotics}.
A straightforward application of Cauchy's theorem then yields
the value of these extra minors at any point in their respective regions of analyticity in terms of their values along the real $z$-axis.
Explicitly,
\begin{equation*}
A_{\minorset{1,2}{1,3}}(z) =
  \frac1{2\pi i} \int_\Real A_{\minorset{1,2}{1,3}}(\zeta)\frac{\d\zeta}{\zeta - z}\,,
\qquad
\Im z > 0\,,
\end{equation*}
plus corresponding expressions for the remaining minors obtained via the symmetries.
Moreover, the reconstruction formula~\eref{e:a11trace} remains valid when these extra minors are not identically zero, 
while the function $g(z)$ in the reconstruction formula~\eref{e:a12trace} for $A_{[1,2]}(z)$ changes into:
\begin{gather*}
g(z) = 
\begin{cases}
~|A_{[1]}(z)|^2-\big|A_{\minorset{1,2}{1,3}}(z)\big|^2 \qquad & \Im z>0\,,\\
~|A_{[1]}(z^*)|^2-\big|A_{\minorset{1,2}{1,3}}(z^*)\big|^2 \qquad & \Im z<0\,.
\end{cases}
\end{gather*}
Importantly, the analysis of the discrete spectrum when the extra analytic minors are not identically zero presented in \cite{JMP2015}
shows that the pre-factors in~\eref{e:a12trace} also remain the same in this case,
provided that $z_1,\dots,z_{N_1}$ and $w_1,\dots,w_{N_2}$ continue to denote the zeros of $A_{[1]}(z)$
not in common and in common with $A_{[1,2]}(z)$, respectively.
\end{remark}
We can now compare the limit as $z \to 0$ of~\eref{e:a11trace} with~\eref{e:a11zero} to easily obtain
the asymptotic phase difference for the potential:
\begin{corollary}
The asymptotic phase difference $\Delta \theta = \theta_+-\theta_-$ is given by
\begin{equation}
\label{e:thetacondition}
\Delta \theta = -2 \sum_{n=1}^{N_1} \arg z_n - 2 \sum_{n=1}^{N_2} \arg w_n
+ \frac{1}{2 \pi} \int_\Real \log[1-R(\zeta)]\frac{\d \zeta}{\zeta},
\end{equation}
with $R(\zeta)$ as in Lemma~\ref{L:trace}.
\end{corollary}

Finally, we now show how to reconstruct the entire scattering matrix~$\@A(z)$ for $z\in\Real$ in terms of the reflection coefficients,
the discrete spectrum and extra analytic minors along the real axis.
Note first that $a_{11}(z) = A_{[1]}(z)$ and $a_{44}(z) = A_{[4]}(z) = \e^{i\Delta\theta} A_{[1]}(\^z)$ are obtained directly from the trace formulae above.
Moreover, the scattering coefficients $a_{21}(z)$, $a_{31}(z)$, and $a_{41}(z)$ can be obtained from the reflection coefficients
and $a_{11}(z) = A_{[1]}(z)$ via the definitions~\eref{e:reflcoeff}, 
and the scattering coefficients $a_{14}(z)$, $a_{24}(z)$, and $a_{34}(z)$ can be accounted for via the symmetries \eref{e:reflsymm} 
of the reflection coefficients.
It thus remains to recover the entries in the second and third columns of the scattering matrix~$\@A(z)$
from the knowledge of its first and fourth columns plus that of the minors obtained from the trace formulae.
This is done via the following result, which is proved by elementary algebra:

\begin{lemma}
The entries in the second and third column of the scattering matrix $\@A(z)$ satisfy the following linear algebraic system:
\begin{gather*}
\@C(z)\,\@y(z) = \@b(z)\,,
\\
\noalign{\noindent where}
\@y(z) =  ( a_{32}, a_{22}, a_{23}, a_{33}, a_{12}, a_{13}, a_{42}, a_{43} )^T,
\\
\@C(z) =
\begin{pmatrix}
 a_{11} & 0 & 0 & 0 &    -a_{31} & 0 & 0 & 0\\
 0 & a_{11} & 0 & 0 &    -a_{21} & 0 & 0 & 0 \\
 0 & 0 & a_{11} & 0 &    0 & -a_{21} & 0 & 0 \\
 0 & 0 & 0 & a_{11} &    0 & -a_{31} & 0 & 0 \\
 a_{44} & 0 & 0 & 0 &    0 & 0 & -a_{34} & 0\\
 0 & a_{44} & 0 & 0 &    0 & 0 & -a_{24} & 0 \\
 0 & 0 & a_{44} & 0 &    0 & 0 & 0 & -a_{24} \\
 0 & 0 & 0 & a_{44} &    0 & 0 & 0 & -a_{34}
\end{pmatrix}
\\
\@b(z) = \big( A_{\minorset{1,3}{1,2}}(z), A_{[1,2]}(z),  A_{\minorset{1,2}{1,3}}(z), e^{i\Delta \theta}A_{[3,4]}(\^z),  
  A_{\minorset{3,4}{2,4}}(z), e^{-i\Delta \theta}A_{[1,2]}(\^z), A_{\minorset{2,4}{3,4}}(z), A_{[3,4]}(z) 
\big)^T.
\end{gather*}
and where a trivial dependence on $z$ in the RHS of $\@C(z)$ and $\@y(z)$ was omitted for brevity.
\end{lemma}
Note that the vector $\@y(z)$ contains the scattering coefficients to be reconstructed, 
while $\@C(z)$ and $\@b(z)$ contain known data (see the discussion above).
Note that in two of the components of $\@b(z)$ we have used the symmetries \eref{e:a12new} to express certain non-principal minors
in terms of $A_{[1,2]}(z)$ and $A_{[3,4]}(z)$. 
Also note
\begin{equation*}
\det \@C(z) = \frac{q_+^2}{z^2} |A_{[1]}(z)|^8 \left[\rho_1(z)\rho_2(\^z)-\rho_1(\^z)\rho_2(z)\right]^2.
\end{equation*}
Thus, the entire scattering matrix can be reconstructed away from the zeros of the above determinant.

\section{Reflectionless potentials and soliton solutions}

Next we examine the case in which the reflection coefficients defined in \eref{e:reflcoeff} are
identically zero.
As before, we will assume that
the analytic non-principal minors are identically zero.

\subsection{Reflectionless potentials}
\label{s:reflsol}

As usual, the jump matrix $\@L(z)$ in
Lemma \ref{L:jump} is identically zero precisely when the scattering matrix is diagonal.
Indeed, upon examining the jump condition in Lemma~\ref{L:jump} columnwise
and using the properties of the scattering matrix, we obtain the following result:
\begin{lemma}
\label{L:scatteringreflectionless}
If the analytic non-principal minors are identically zero,
the scattering matrices $\@A(z)$ and $\@B(z)$ are diagonal if and only if $\@L(z) \equiv \@0$.
\end{lemma}
Note that, without the assumption that the analytic non-principal minors are identically zero,
we would only be able to show that the scattering matrix in the reflectionless is block-diagonal,
since we would not be able to conclude that the scattering coefficients $a_{23}(z)$, $a_{32}(z)$,
$b_{23}(z)$, and $b_{32}(z)$ are identically zero.
The analysis of such a situation is left for future work.

In the reflectionless case, the system of algebraic-integral equations \eref{e:threesystem}
simplifies to a closed system of linear equations:
\bse
\label{e:refclosed}
\begin{gather}
M_1^-(x,t,z) = \@E_{+,1}(z) + \sum_{j=1}^{N_1} \frac{D_j M_2^+(z_j)}{z-z_j}
+ \sum_{j=1}^{N_2} \frac{C_j M_3^+(w_j)}{z-w_j}, \qquad z = z_n^*,w_n^*,
\\
M_4^+(x,t,z) = \@E_{+,4}(z) + \sum_{j=1}^{N_1} \frac{\check{D}_j M_2^-(\z_j^*)}{z-\z_j^*}
+ \sum_{j=1}^{N_2} \frac{\check{C}_j M_3^-(\w_j^*)}{z-\w_j^*}, \qquad z = \z_n,\w_n,
\\
M_2^+(x,t,z_n) =
\@e_2 + \sum_{j=1}^{N_1} \left(\frac{\^D_j M_4^+(\z_j)}{z_n-\z_j}
+ \frac{\=D_j M_1^-(z_j^*)}{z_n-z_j^*}\right),
\\
M_3^+(x,t,w_n) =
\@e_3 + \sum_{j=1}^{N_2} \left(\frac{\^C_j M_4^+(\w_j)}{w_n-\w_j}
+ \frac{\=C_j M_1^-(w_j^*)}{w_n-w_j^*}\right),
\\
M_2^-(x,t,\z_n^*) = \@e_2
+ \sum_{j=1}^{N_1} \left(\frac{\^D_j M_4^+(\z_j)}{\z_n^*-\z_j}
+ \frac{\=D_j M_1^-(z_j^*)}{\z_n^*-z_j^*}\right),
\\
M_3^-(x,t,\w_n^*) = \@e_3
+ \sum_{j=1}^{N_2} \left(\frac{\^C_j M_4^+(\w_j)}{\w_n^*-\w_j}
+ \frac{\=C_j M_1^-(w_j^*)}{\w_n^*-w_j^*}\right).
\end{gather}
\ese
In the reflectionless case, the trace formulae in Lemma~\ref{L:trace} reduce to the following:
\begin{equation}
A_{[1]}(z) = \prod_{n=1}^{N_1} \frac{z-z_n}{z-z_n^*}
\prod_{n=1}^{N_2} \frac{z-w_n}{z-w_n^*},
\qquad
A_{[1,2]}(z) = \prod_{n=1}^{N_1} \frac{z-\z_n}{z-\z_n^*}
\prod_{n=1}^{N_2} \frac{z-w_n}{z-w_n^*}.
\end{equation}
In addition, the reconstruction formula \eref{e:reconstruction} is also simpler:
\begin{equation}
\label{e:reconrefl}
q_j(x,t) = q_{+,j} - i \sum_{n=1}^{N_1} D_n(x,t) M_{j+1,2}^+(x,t,z_n)
- i \sum_{n=1}^{N_2} C_n(x,t) M_{j+1,3}^+(x,t,w_n),
\qquad
j=1,2,3.
\end{equation}
We now have all the information needed to construct explicit soliton solutions.

\subsection{Pure soliton solutions}

In this section, we explore the different possibilities for reflectionless solutions.
We construct explicit soliton solutions for each type of eigenvalue and exhibit a 2-soliton interaction.

\subsubsection{Type~I}

Let $z_1$ be a type~I eigenvalue, and assume that there are no other discrete eigenvalues
(i.e., we assume that $N_1=1$ and $N_2=0$).
Then we have $A_{[1]}(z_1)=0$, $A_{[1,2]}(z_1) \neq 0$, and $A_{[1,2,3]}(z_1) \neq 0$.

When $z_1$ is the only discrete eigenvalue, the reconstruction formula~\eref{e:reconrefl} simplifies to
\begin{equation}
\label{e:type2recon}
q_j(x,t) = q_{+,j} - i D_1(x,t) M_{j+1,2}^+(x,t,z_1), \qquad j =1,2,3.
\end{equation}
In addition, the closed system of linear equations that comes from~\eref{e:refclosed} is
\bse
\label{e:type2closed}
\begin{gather}
M_1^-(x,t,z_1^*) = \@E_{+,1}(z_1^*) + \frac{D_1(x,t)}{z_1^* - z_1} M_2^+(x,t,z_1),
\\
M_2^+(x,t,z_1) = \@e_2 + \left[\frac{\=D_1(x,t)}{z_1 - z_1^*}
- \frac{i z_1^*}{q_+^*} \frac{\^D_1(x,t)}{z_1 - \z_1}\right] M_1^-(x,t,z_1^*).
\end{gather}
\ese
Substituting the solution of the system \eref{e:type2closed} into \eref{e:type2recon} yields a soliton solution.
Specifically, defining the following quantities:
\begin{equation*}
z_1=\xi_1+i\nu_1\equiv |z_1|e^{i\alpha}\,, \quad |z_1|<q_o, \quad \nu_1>0, \qquad
\~d_1=d_1/A^\prime_{[1]}(z_1)\equiv \delta_1 e^{i\phi_1},
\end{equation*}
yields the following regular soliton solution:
\begin{equation}
\label{e:qtype1}
\@q_I(x,t) =
-iU_o \, \e^{i (\xi_1 x -(\xi_1^2-\nu_1^2)t +\phi_1)} \sech U(x,t) \, \^{\@e}_1
+ q_o \e^{i (\theta_+ + \alpha)} [\cos \alpha - i \sin \alpha \tanh U(x,t)] \, \^{\@e}_3,
\end{equation}
where $\{\^{\@e}_j\}_{j=1}^3$ is the standard basis for $\Real^3$,
\begin{gather*}
U_o =\sin \alpha \sqrt{q_o^2-|z_1|^2},
\\
U(x,t) =\nu_1(x-2\xi_1t-x_1)\,, \qquad  \nu_1x_1=\ln \frac{|z_1|\delta_1}{2\nu_1\sqrt{q_o^2-|z_1|^2}}.
\end{gather*}
A typical solution of this type can be seen in Fig.~\ref{fig:type2}.
We note how a solution of this type is simply an analogue of a dark-bright soliton solution of the defocusing
Manakov system with NZBC, with bright part aligned with the first component.

\subsubsection{Type~IV}

Let $w_1$ be a type~IV eigenvalue, and assume that there are no other discrete eigenvalues
(i.e., we assume that $N_1=0$ and $N_2=1$).
Then $A_{[1]}(w_1)=A_{[1,2]}(w_1)=0$ and $A_{[1,2,3]}(w_1) \neq 0$.

When $w_1$ is the only discrete eigenvalue,
the solution that results from combining the reconstruction formula~\eref{e:reconrefl}
with the solution of the regularized RHP is
\begin{equation}
\label{e:type1recon}
q_j(x,t) = q_{+,j} - i C_1(x,t) M_{j+1,3}^+(x,t,w_1), \qquad j=1,2,3.
\end{equation}
The closed system of linear equations that comes from~\eref{e:refclosed} is
\bse
\label{e:type1closed}
\begin{gather}
M_1^-(x,t,w_1^*) = \@E_{+,1}(w_1^*) + \frac{C_1(x,t)}{w_1^* - w_1} M_3^+(x,t,w_1),
\\
M_3^+(x,t,w_1) = \@e_3 +
\left[\frac{\=C_1(x,t)}{w_1 - w_1^*}
- \frac{i w_1^*}{q_+^*} \frac{\^C_1(x,t)}{w_1 - \w_1}\right] M_1^-(x,t,w_1^*).
\end{gather}
\ese
Substituting the solution of the system~\eref{e:type1closed} into~\eref{e:type1recon} yields a soliton solution.
Specifically, defining the following quantities:
\begin{equation*}
w_1=\xi_1+i\nu_1\equiv |w_1|e^{i\alpha}\,, \quad |w_1|<q_o, \quad \nu_1>0, \qquad
\~c_1=c_1/A^\prime_{[1]}(w_1)\equiv \gamma_1 e^{i\psi_1}
\end{equation*}
yields the following regular soliton solution:
\begin{equation}
\@q_{II}(x,t) = -iU_o \, \e^{i (\xi_1 x -(\xi_1^2-\nu_1^2)t+\psi_1 )} \sech U(x,t) \, \^{\@e}_2
+ q_o \e^{i (\theta_+ + \alpha)} [\cos \alpha - i \sin \alpha \tanh U(x,t)] \, \^{\@e}_3,
\end{equation}
where $U_o$ and $U(x,t)$ are as defined in \eref{e:qtype1}, with $\~c_1$ replacing $\~d_1$, namely:
\begin{gather*}
U_o =\sin \alpha \sqrt{q_o^2-|w_1|^2},
\\
U(x,t) =\nu_1(x-2\xi_1t-x_1)\,, \qquad  \nu_1x_1=\ln \frac{|w_1|\gamma_1}{2\nu_1\sqrt{q_o^2-|w_1|^2}}.
\end{gather*}

\begin{figure}[t!]
\centerline{%
\includegraphics[scale=0.55]{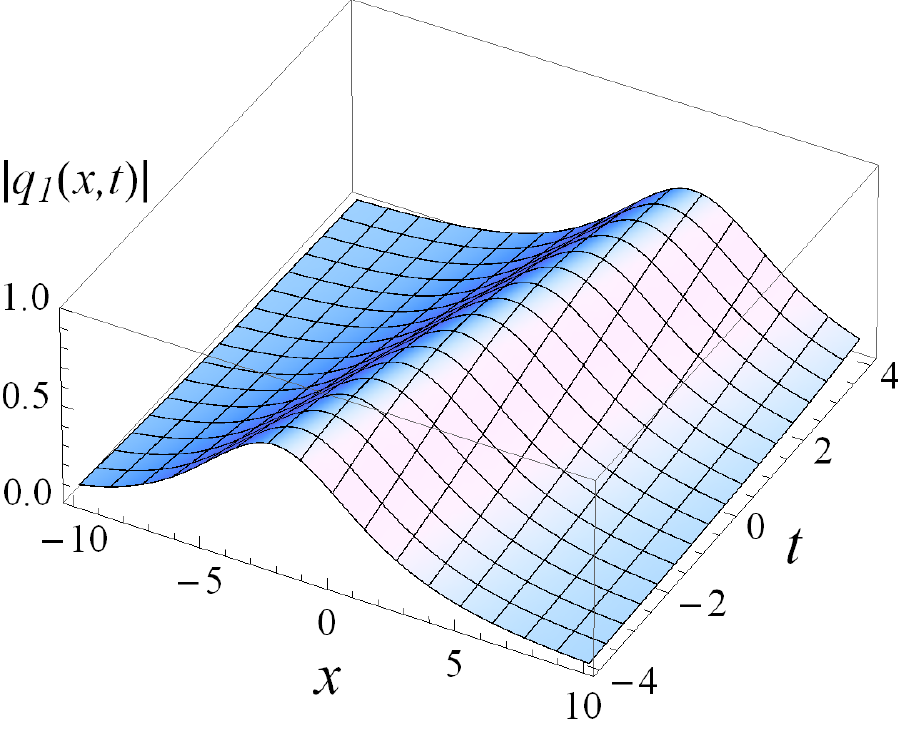}\quad
  \includegraphics[scale=0.55]{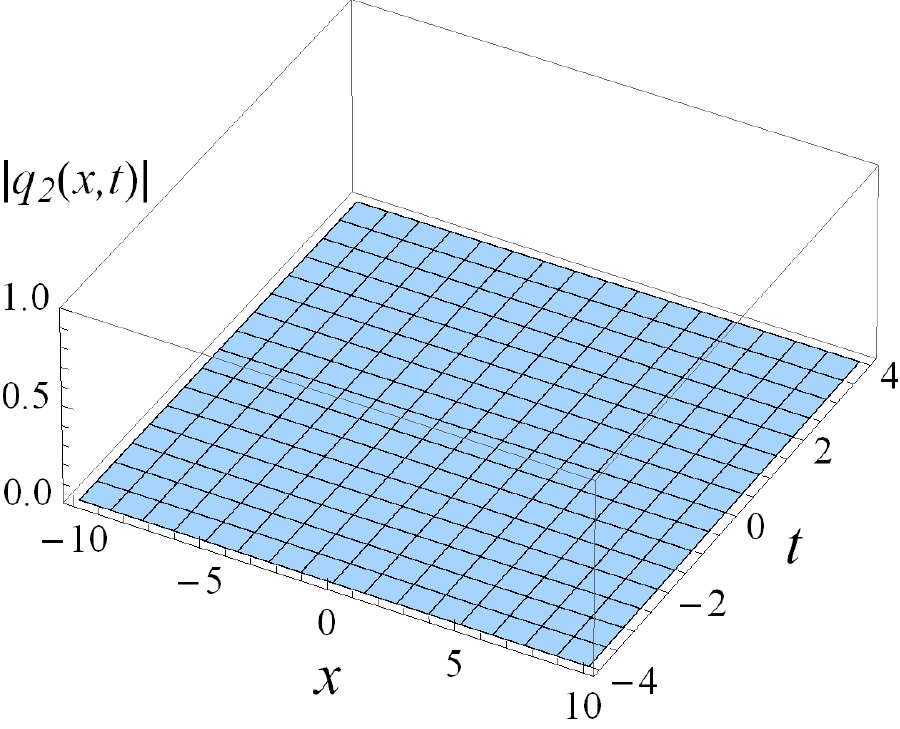}\quad
  \includegraphics[scale=0.55]{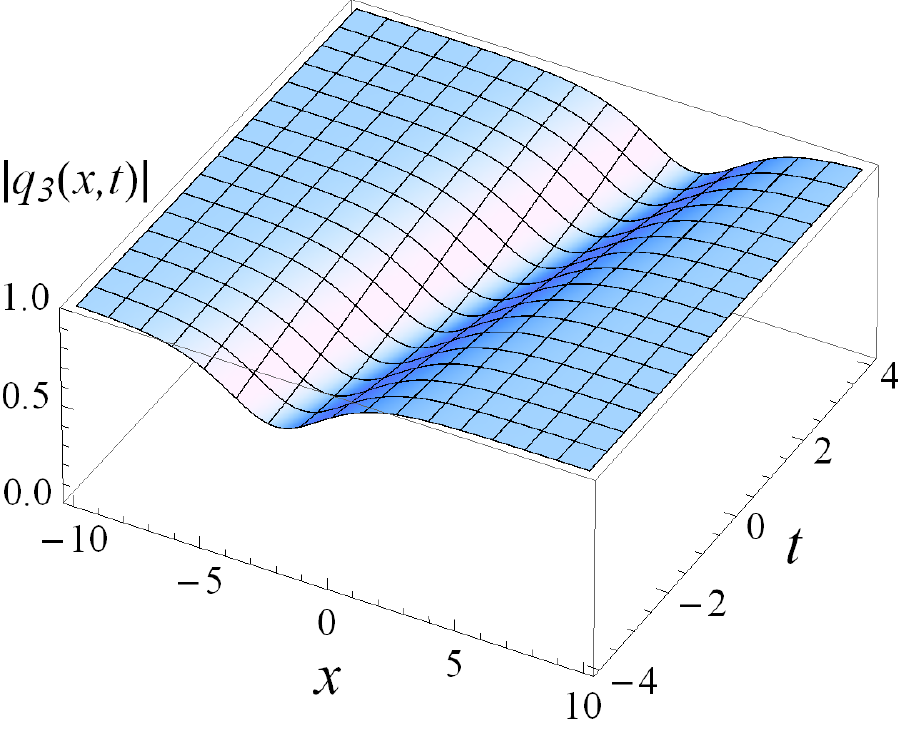}}
\caption{\footnotesize{A type~I solution of the 3-component defocusing VNLS equation with NZBC
obtained by taking $N_1=1$, $N_2=0$, $z_1 = 0.5 \e^{i \pi/4}$, $q_+=\e^{i \pi/2}$,
$\~d_1 = 1 + 2.75 i$,
resulting in a dark-bright soliton whose bright part is aligned along the first component of $\@q(x,t)$.}}
\label{fig:type2}
\bigskip
\centerline{%
  \includegraphics[scale=0.55]{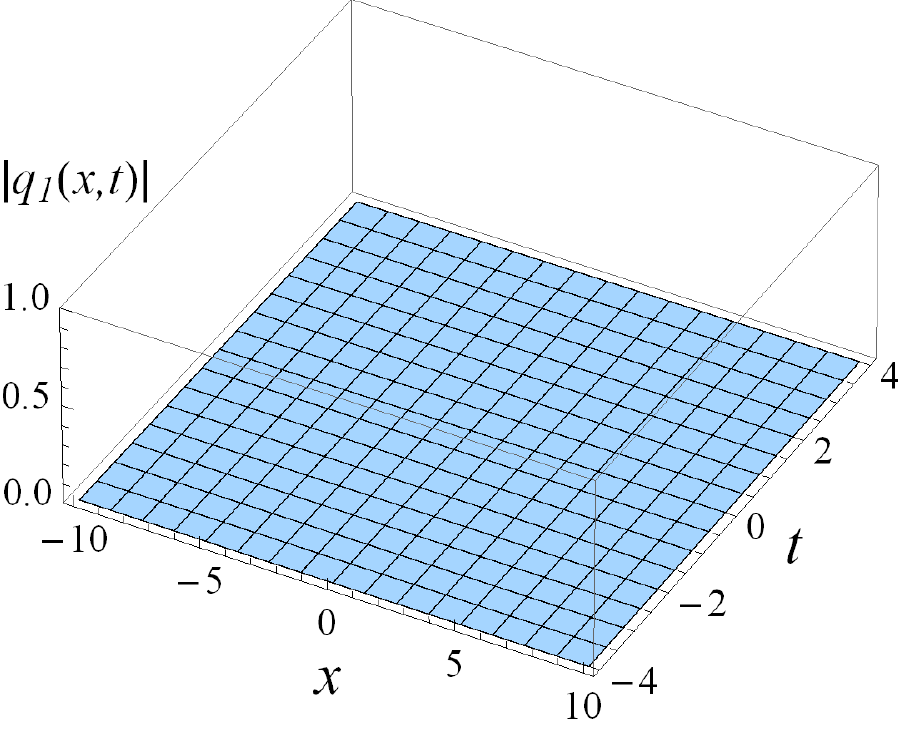}\quad
  \includegraphics[scale=0.55]{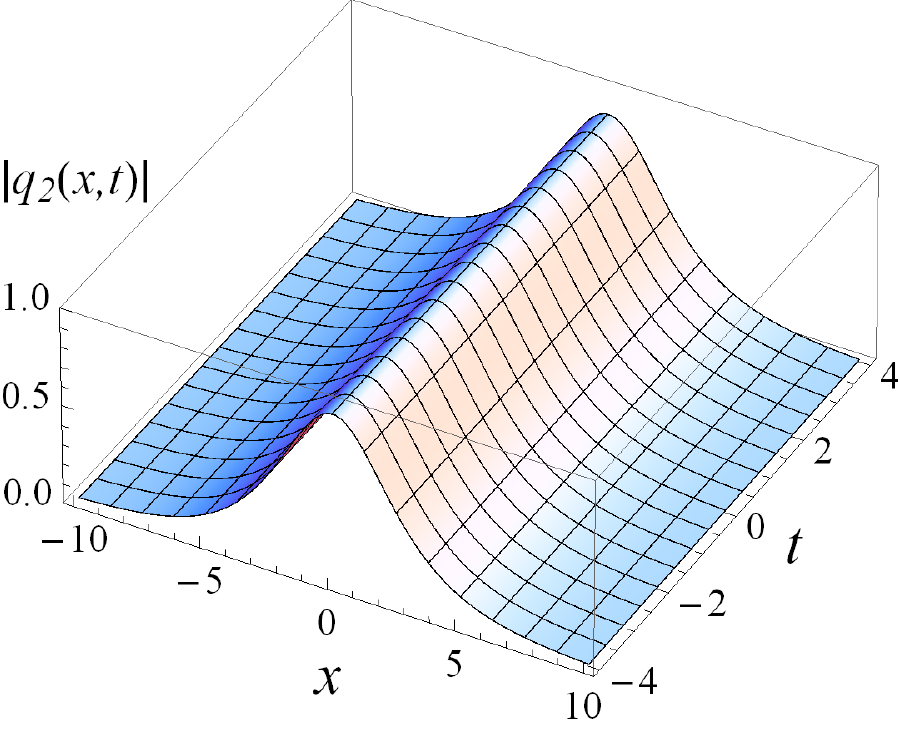}\quad
  \includegraphics[scale=0.55]{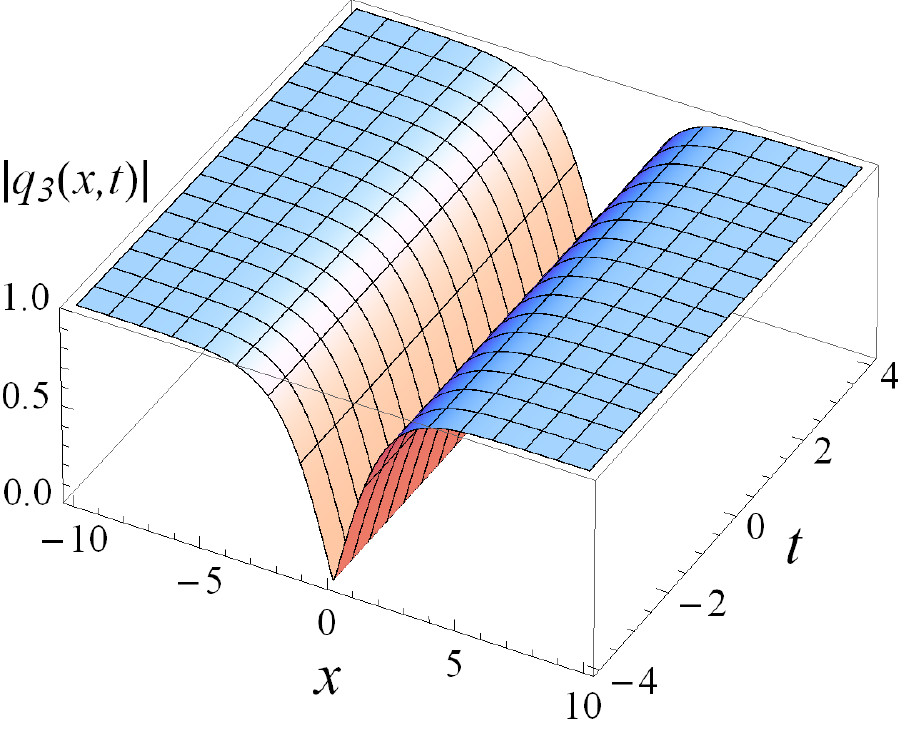}}
\caption{\footnotesize{A type~IV solution of the 3-component defocusing VNLS equation with NZBC
obtained by taking $N_1=0$, $N_2=1$, $w_1 = 0.5 \e^{i \pi/2}$, $q_+=\e^{i \pi/2}$,
$\~c_1 = 1 + 2.75 i$,
resulting in a dark-bright soliton whose bright part is aligned along the second component of $\@q(x,t)$.}}
\label{fig:type1}
\end{figure}

A typical solution of this type can be seen in Fig.~\ref{fig:type1}.
We note how a solution of this type is simply another analogue of a dark-bright soliton solution of the defocusing
Manakov system with NZBC, with bright part aligned with the second component.

\begin{figure}[t!]
\centerline{%
\includegraphics[scale=0.55]{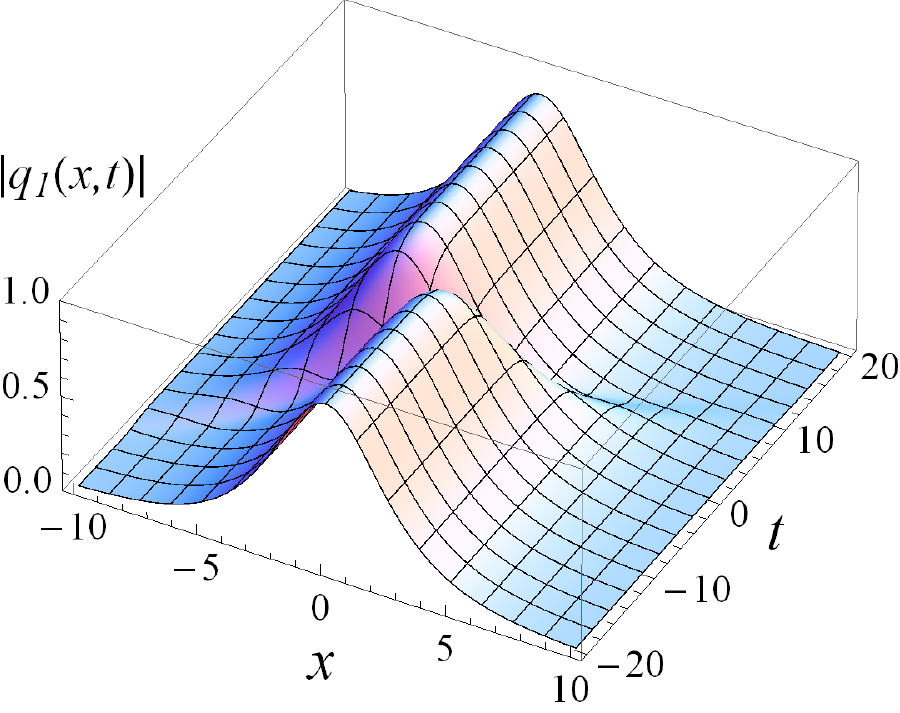}\quad
  \includegraphics[scale=0.55]{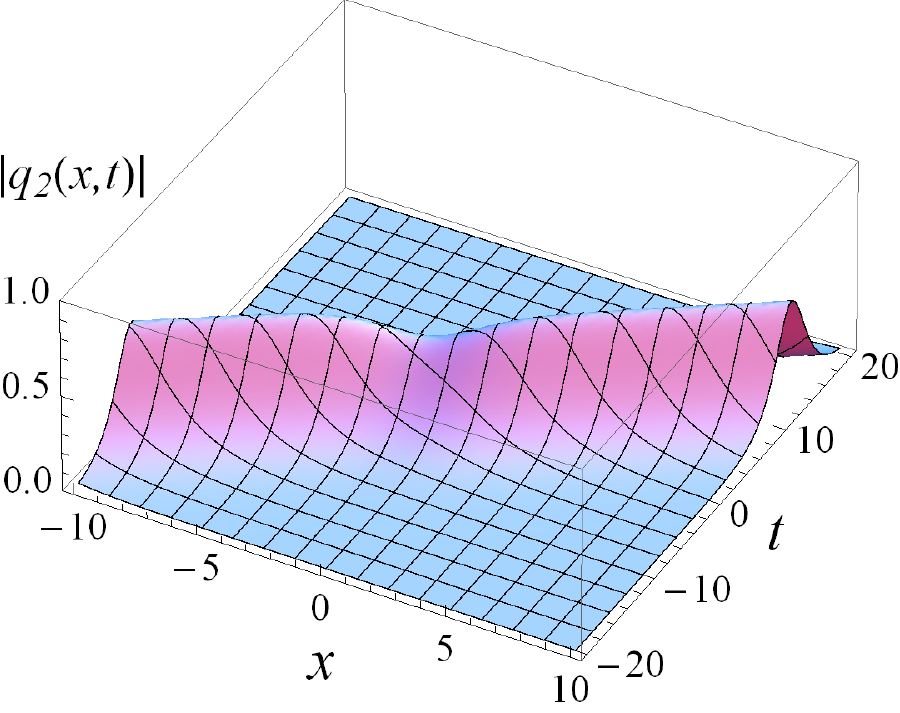}\quad
  \includegraphics[scale=0.55]{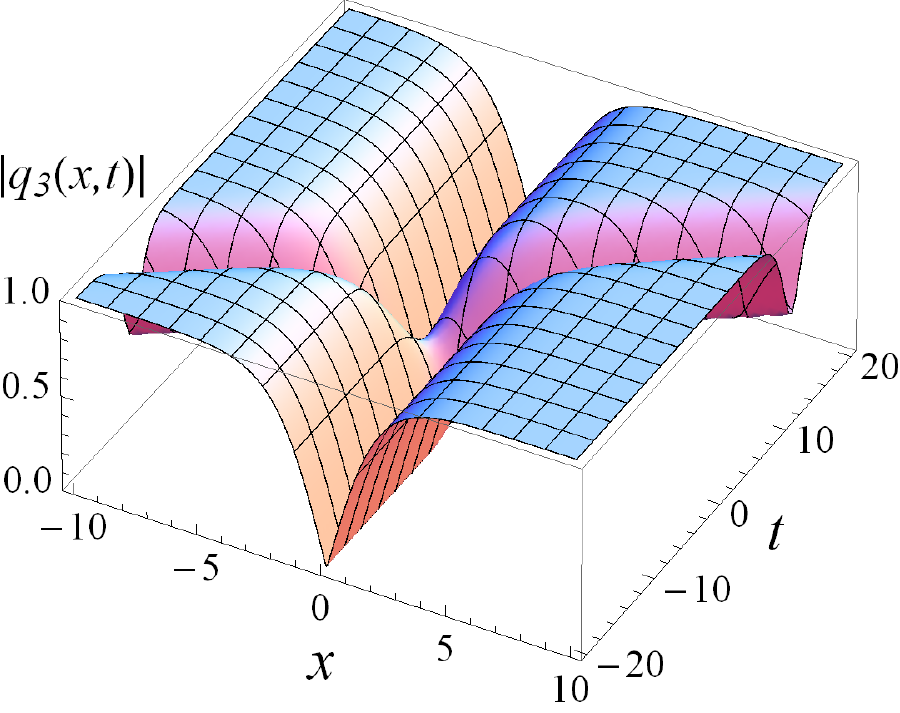}}
\caption{\footnotesize{A 2-soliton solution of the 3-component defocusing VNLS equation with NZBC
obtained by taking $N_1=N_2=1$, $z_1=0.5 \e^{i \pi/2}$, $w_1=0.75 \e^{i \pi/3}$, $q_+=\e^{i \pi/2}$,
$\~c_1=\~d_1=1+2.75 i$.
Note how the soliton interaction results in a position shift for the bright parts of the two dark-bright solitons,
even though the bright parts are aligned along components orthogonal to each other.}
}
\label{fig:multi}
\vspace*{-1ex}
\end{figure}

Of course, as in the previously-studied cases, any combination of the two types of eigenvalues is also possible.
In particular, an example of a 2-soliton solution can be seen in Fig \ref{fig:multi}.
Note how the interaction results in a position shift for the bright parts of the two dark-bright solitons,
even though the bright parts are confined along orthogonal components.


\section{Linear limit and comparison with the Manakov system}

\subsection{Small-deviation limit of the IST and comparison with direct linearization}

For the defocusing NLS equation with ZBC, it is well known that the small-amplitude limit of the IST coincides
with the direct linearization of the PDE.
Both are obtained by taking the solution of the PDE to be $O(\epsilon)$.
In turn, in the IST, this implies that the reflection coefficient is also $O(\epsilon)$.
In the case of NZBC, the corresponding situation is that in which the potential is only a small deviation
from the background (asymptotic) value.
We next show that, as one would expect, in the small deviation limit,
the solution of the VNLS equation obtained with IST coincides with the direct linearization of the PDE around
the constant background.

We first write down the linearization of the VNLS equation around a constant background and obtain its Fourier transform solution.
To this end, we need to take $\theta_+ = \theta_-$.
We can therefore use the gauge invariance of the VNLS equation to set $\theta_+ = 0$ without loss of generality.
We write $\@q_+ = \@q_o = (0,0,q_o)$, and we define
\begin{equation}
\label{e:qsmall}
\@q(x,t) = \@q_o + \@u(x,t),
\end{equation}
where $ \@u(x,t) \equiv (u_1(x,t),u_2(x,t),u_3(x,t))^T = O(\epsilon)$.
Then $\@u(x,t)$ represents a small perturbation of the background field $\@q_+$.
We insert~\eref{e:qsmall} into~\eref{e:vnls} and neglect higher powers of $\@u$ to obtain
a linearization of VNLS around the background solution:
\begin{equation}
\label{e:vnlslinear}
i \@u_t + \@u_{xx} - 2 \@q_o^T (\@u + \@u^*) \@q_o = \@0.
\end{equation}
We now find solutions of~\eref{e:vnlslinear} using the standard Fourier transforms
(except for an inessential factor of 2 in the exponent, which was chosen to simplify the comparison with the IST):
\vspace*{-1ex}
\begin{equation}
\label{e:fourier}
\@u(x,t) = \frac{1}{2 \pi} \int_\Real \^{\@u}(s,t) \e^{2 i s x} \d s,
\qquad
\^{\@u}(s,t) = 2 \int_\Real \@u(x,t) \e^{- 2 i s x} \d x.
\end{equation}
Combining~\eref{e:fourier} with~\eref{e:vnlslinear} yields
$\^{\@v}_t(s,t) = \@V(s) \^{\@v}(s,t)$, where
$\^{\@v}(s,t) = (\^{\@u}(s,t),\^u_3^*(-s,t))^T$ and
\begin{equation*}
\@V(s) =
\begin{pmatrix}
- 4 i s^2 & 0 & 0 & 0
\\
0 & - 4 i s^2 & 0 & 0
\\
0 & 0 & - i (4 s^2 + 2 q_o^2) & - 2 i q_o^2
\\
0 & 0 & 2 i q_o^2 & i (4 s^2 + 2 q_o^2)
\end{pmatrix}.
\end{equation*}
The solution of this system is
\begin{equation}
\label{e:vsol}
\^{\@v}(s,t)
=
\begin{pmatrix}
\^u_{1,0}(s) \e^{- 4 i s^2 t}
\\
\^u_{2,0}(s) \e^{-4 i s^2 t}
\\
[s - (s^2 + q_o^2)^{1/2}] A_1(s) \e^{4 i s (s^2 + q_o^2)^{1/2} \, t}
+ [s + (s^2 + q_o^2)^{1/2}] A_2(s) \e^{-4 i s (s^2 + q_o^2)^{1/2} \, t}
\\
[s + (s^2 + q_o^2)^{1/2}] A_1(s) \e^{4 i s (s^2 + q_o^2)^{1/2} \, t}
+ [s - (s^2 + q_o^2)^{1/2}] A_2(s) \e^{- 4 i s (s^2 + q_o^2)^{1/2} \, t}
\end{pmatrix},
\end{equation}
where $\^u_{j,0}(s) = \^u_j(s,0)$ and the scalar functions $A_1(s)$ and $A_2(s)$ are given by
\bse
\label{e:AA}
\begin{gather}
A_1(s) = \frac{1}{4 s (s^2 + q_o^2)^{1/2}} \left\{[-s + (s^2 + q_o^2)^{1/2}] \^u_{3,0}(s)
+ [s + (s^2 + q_o^2)^{1/2}] \^u_{3,0}^*(-s)\right\},
\\
A_2(s) = \frac{1}{4 s (s^2 + q_o^2)^{1/2}} \left\{[s + (s^2 + q_o^2)^{1/2}] \^u_{3,0}(s)
+ [-s + (s^2 + q_o^2)^{1/2}] \^u_{3,0}^*(-s)\right\}.
\end{gather}
\ese
The following symmetries are then evident by inspection:
\begin{equation}
A_1^*(-s) = - A_1(s), \qquad A_2^*(-s) = - A_2(s).
\end{equation}
We have found the solution of~\eref{e:vnlslinear} in terms of given data, and this solution
provides an approximation of the solution $\@q(x,t)$ of~\eref{e:vnls} in the small-deviation limit.

We now compute the small-deviation limit of the IST.
Note $\@Q_+ = \@Q_- = \@Q_o$, so let $\Delta\@Q(x,t) = \Delta\@Q_\pm(x,t) = \@Q(x,t) - \@Q_o$.
The integral equations~\eref{e:intequations} become
\bse
\begin{gather}
\mu_-(x,t,z) = \@E(z)
+ \int_{-\infty}^x \@E(z) \e^{i(x-y)\boldsymbol\Lambda (z)} \@E^{-1}(z) \Delta \@Q(y,t)
\mu_-(y,t,z) \e^{-i(x-y)\boldsymbol\Lambda (z)} \d y,
\\
\mu_+(x,t,z) = \@E(z)
- \int_x^\infty \@E(z) \e^{i(x-y)\boldsymbol\Lambda (z)} \@E^{-1}(z) \Delta \@Q(y,t)
\mu_+(y,t,z) \e^{-i(x-y)\boldsymbol\Lambda (z)} \d y,
\end{gather}
\ese
where $\@E_+(z) = \@E_-(z) =: \@E(z)$ since $\@Q_+ = \@Q_-$.
We therefore have
$\mu_\pm(x,t,z) = \@E(z) + O(\epsilon)$ as $\epsilon\to0$, and, recalling the scattering
relation~\eref{e:scattering},
\be
\@A(z) = \@I + \int_\Real \e^{- i y \boldsymbol\Lambda (z)} \@E^{-1}(z) \Delta \@Q(y,0) \mu_-(y,0,z)
e^{i y \boldsymbol\Lambda (z)}~ \d y\,.
\ee
Thus, $\@A(z) = \@I + O(\epsilon)$ as $\epsilon\to0$.
In particular,
by looking explicitly at the $O(\epsilon)$ portion of $\@A(z)$, one obtains
\bse
\begin{gather}
\rho_j(z) = \int_\Real q_j(y,0) \e^{-izy}~ \d y  + O(\epsilon^2), \qquad j=1,2,
\\
\rho_3(z) = \frac{1}{\gamma(z)} \int_\Real \left[q_3(y,0) - q_o - \frac{q_o^2}{z^2} (q_3^*(y,0) - q_o)\right] \d y
+ O(\epsilon^2).
\end{gather}
\ese
Consider the case in which no solitons are present.
In this case, in the solution~\eref{e:rhpsol} of the RHP we need to only keep linear terms in the reflection coefficients
$\rho_j(z)$ ($j=1,2,3$) as defined in~\eref{e:reflcoeff},
so the reconstruction formula \eref{e:reconstruction} yields the following behavior for $\@q(x,t)$
as $\epsilon \to 0$:
\begin{equation}
\label{e:reconstruct_eps}
\@q(x,t) = \@q_o
+ \frac{1}{2 \pi} \int_\Real
\left\{\left[\rho_1(\zeta)
\^{\@e}_1+\rho_2(\zeta) \^{\@e}_2\right] \e^{- i (\theta_1 - \theta_2)(x,t,\zeta)}
+ \rho_3(\zeta) \e^{- 2i \theta_1(x,t,\zeta)} \^{\@e}_3 \right\}\d \zeta
+ O(\epsilon^2),
\end{equation}
where, as before, $\{\^{\@e}_j\}_{j=1}^3$ denotes the standard basis for $\Real^3$.
Next, to compare with the linearization, we need to perform a change of variable in the above integrals.
For the first two terms in \eref{e:reconstruct_eps}, we make the substitution $\zeta = 2 s$ to obtain
the following:
\begin{equation}
\label{e:a0eq}
\int_\Real
\rho_j(\zeta) \e^{- i (\theta_1 - \theta_2)(x,t,\zeta)} \d \zeta
=
2 \int_\Real \rho_j(2 s) \e^{- 4 i s^2 t} \e^{2 i s x} \d s,
\qquad
j = 1,2.
\end{equation}
Then, comparing \eref{e:a0eq} with the Fourier transforms of the first two rows of \eref{e:vsol}
yields
\begin{equation}
\^u_{j,0}(s) = 4 \pi \rho_j(2s), \qquad j=1,2.
\end{equation}
Reverting from $\zeta$ to the original coordinates $(k,\lambda(k))$, the remaining integral
in \eref{e:reconstruct_eps} simplifies as follows:
\begin{multline}
\int_\Real\rho_3(\zeta) \e^{- 2i \theta_1(x,t,\zeta)} \d \zeta
= \int_{\Real \setminus [-q_o,q_o]} \Big\{ (k+\lambda(k)) \rho_3(k,\lambda(k)) \e^{2 i \lambda(k)(x-2kt)}
\\[-1ex]
- (k-\lambda(k)) \rho_3(-k,\lambda(k)) \e^{2 i \lambda(k)(x+2kt)} \Big\} \frac{\d k}{\lambda(k)},
\end{multline}
where the plus/minus sign results from the branching of $\lambda(k)$.
To compare with the Fourier tranform of \eref{e:vsol},
we now make the substitutions $k=(\lambda^2 + q_o^2)^{1/2}$ for $k \geq q_o$
and $k=-(\lambda^2 + q_o^2)^{1/2}$ for $k \leq q_o$ to obtain
\begin{multline}
\label{e:final_proof}
\int_\Real \rho_3(\zeta) \e^{- 2i \theta_1(x,t,\zeta)} \d \zeta
=
\int_\Real \Big\{\frac{(\lambda^2 + q_o^2)^{1/2} + \lambda}{(\lambda^2 + q_o^2)^{1/2}}
\rho_3 ((\lambda^2 + q_o^2)^{1/2},\lambda) \e^{- 4 i \lambda (\lambda^2 + q_o^2)^{1/2} \, t}
\\
- \frac{(\lambda^2 + q_o^2)^{1/2} - \lambda}{(\lambda^2 + q_o^2)^{1/2}}
\rho_3 (-(\lambda^2 + q_o^2)^{1/2},\lambda) \e^{4 i \lambda (\lambda^2 + q_o^2)^{1/2} \, t}
\Big\} \, \e^{2 i \lambda x} \d \lambda.
\end{multline}
It is evident from \eref{e:AA} and \eref{e:final_proof} that
\begin{equation}
A_1(s) = \frac{\rho_3(-(s^2 + q_o^2)^{1/2},s)}{(s^2 + q_o^2)^{1/2}},
\qquad
A_2(s) = \frac{\rho_3((s^2 + q_o^2)^{1/2},s)}{(s^2 + q_o^2)^{1/2}}.
\end{equation}
Thus, apart from the possible contribution of the discrete spectrum, the leading-order solution from the IST in the small-deviation limit
coincides exactly with the solution of the VNLS equation linearized around the background.

\subsection{Comparison with the Manakov system}
\label{s:manakovsec}

We now explore how the results of this work can be applied to the Manakov system with NZBC
[i.e., \eref{e:vnls}, where $\q(x,t)$ is now a 2-component vector].
Though a methodology to solve this problem was introduced in \cite{PAB2006} and made rigorous in \cite{gbdkk2015},
we show here that the results of this work can be used as an alternative approach to solving the
Manakov system with NZBC in the defocusing case.

We first note that all $4 \times 4$ matrices become $3 \times 3$ matrices in the case of the Manakov system.
As a result, the scattering matrix has no analytic non-principal minors.
Triangular decompositions of the now $3 \times 3$ scattering matrix $\@A(z)$ yield expressions similar to~\eref{e:triangular},
with
\bse
\begin{gather}
\alpha^+(z)\@D^+(z) =
\begin{pmatrix}
1 & - a_{12}(z) & b_{13}(z)
\\
0 & a_{11}(z) & b_{23}(z)
\\
0 & 0 & b_{33}(z)
\end{pmatrix},
\qquad
\beta^+(z)\@D^+(z) =
\begin{pmatrix}
a_{11}(z) & 0 & 0
\\
a_{21}(z) & b_{33}(z) & 0
\\
a_{31}(z) & - b_{32}(z) & 1
\end{pmatrix},
\\
\tilde{\alpha}(z)\@D^-(z) =
\begin{pmatrix}
1 & - b_{12}(z) & a_{13}(z)
\\
0 & b_{11}(z) & a_{23}(z)
\\
0 & 0 & a_{33}(z)
\end{pmatrix},
\qquad
\tilde{\beta}(z)\@D^-(z) =
\begin{pmatrix}
b_{11}(z) & 0 & 0
\\
b_{21}(z) & a_{33}(z) & 0
\\
b_{31}(z) & - a_{32}(z) & 1
\end{pmatrix},
\end{gather}
\ese
where, in this case,
\begin{equation}
\label{e:newD}
\@D^+(z) = \diag(1,a_{11}(z),b_{33}(z)), \qquad \@D^-(z) = \diag(b_{11}(z),a_{44}(z),1).
\end{equation}
We again compare with the results of \cite{PBT2011} as in Theorem \ref{t:gantmacher}
to make the following analytic extensions off the real axis:
\bse
\begin{gather}
a_{11}(z), b_{33}(z) : \quad \Im z > 0,
\qquad
b_{11}(z), a_{33}(z) : \quad \Im z < 0.
\end{gather}
\ese
These analyticity properties were found for the Manakov system in \cite{gbdkk2015}
using an alternative integral representation of the scattering matrix.
The present method of finding the analyticity properties of the scattering matrix is more easily generalizable than
the one used for the Manakov system, due to the presence of extra analytic data in higher-order cases.

We proceed as we did in Section \ref{s:FAE};
a complete set of analytic eigenfunctions is given in each half plane by
\begin{equation}
\chi^\pm(x,t,z) = \@M^\pm(x,t,z) \@D^\pm(z) \e^{i \@\Theta(x,t,z)}, \qquad \Im z \gtrless 0,
\end{equation}
with $\@D^\pm(z)$ now given in~\eref{e:newD} and the meromorphic matrices $\@M^\pm(x,t,z)$ given
by the same expressions as~\eref{e:Mdef}, where the $4 \times 4$ matrices are replaced with their corresponding
$3 \times 3$ counterparts.
Similarly to the 3-component case, the first and third columns of these analytic matrices are simply
$\phi_{\mp,1}(x,t,z)$ and $\phi_{\pm,3}(x,t,z)$,
respectively, implying that these Jost eigenfunctions are indeed analytic in the corresponding half planes.
The second column of each of the analytic matrices $\chi^\pm(x,t,z)$ is an analogue of the auxiliary eigenfunctions
defined in the 2-component case \cite{gbdkk2015,PAB2006}.
We write the second column of $\chi^+(x,t,z)$ as $\chi(x,t,z)$ and the second column of $\chi^-(x,t,z)$ as $\=\chi(x,t,z)$.
These auxiliary eigenfunctions have the following decompositions for $z \in \Real$:
\bse
\begin{gather}
\chi(x,t,z) = - a_{12}(z) \phi_{-,1}(z) + a_{11}(z) \phi_{-,2}(z) = b_{33}(z) \phi_{+,2}(z) - b_{32}(z) \phi_{+,3}(z),\kern0.45em%
\\
- \=\chi(x,t,z) = a_{33}(z) \phi_{-,2}(z) - a_{32}(z) \phi_{-,3}(z) = - b_{12}(z) \phi_{+,1}(z) + b_{11}(z) \phi_{+,2}(z),%
\end{gather}
\ese
where the $(x,t)$-dependence was omitted from the right-hand side of each equation for simplicity.
When inverted, these decompositions match exactly those found in the 2-component system \cite{gbdkk2015,PAB2006}.
From here, the results are the same as in \cite{gbdkk2015,PAB2006}.
One simply uses these decompositions to find the symmetries of the eigenfunctions and the scattering matrix.
The introduction of the so-called ``adjoint problem'' is not even necessary.

\section{Concluding remarks}

The IST presented in this work was formulated under the assumption of existence.
As usual, however, 
one can now use the reconstruction formula for the solution obtained from the inverse problem as a definition of $q(x,t)$
and prove rigorously that this function is the unique solution of the IVP.
On the other hand, it should be obvious from the discussions in the previous sections that 
the formalism is significantly more complex than that for AKNS-type systems or that for problems with zero boundary conditions.
We conclude this work with a few additional remarks.

1. While no conceptual difficulties arise in analyzing the discrete spectrum and in solving the inverse problem in the reflectionless case
when the analytic non-principal minors of the scattering matrix are not identically zero,
a comprehensive treatment of these issues and the resulting soliton solutions in this case is not entirely trivial.
When the analytic non-principal minors are non-zero, one obtains solutions
with bright component is not aligned with a single basis vector.
but has a non-zero contribution along both.
Then the interaction of these solutions results in a polarization shift of the bright components.
Some such soliton solutions were presented in \cite{PRL2015},
and a detailed treatment of a few possible cases was reported in \cite{JMP2015}.

2. At the same time, an issue that has not been resolved yet is the identification of the minimal set of scattering data 
from which the inverse problem should be defined in the case when the analytic non-principal minors are not identically zero.
On one hand, the discussion in sections~\ref{s:rhpmain} and~\ref{s:tracemain} 
might suggest that one must assign the value of the analytic non-principal minors along the real $z$-axis 
in order to uniquely specify the jump of the matrix RHP and to reconstruct all the entries of the scattering matrix.
(Things simplify somewhat in the reflectionless case, since in this case the dependence of the jump matrix 
on the extra minors disappears.
Even in this case, however, one would need the value of the extra minors along the real axis to reconstruct the whole scattering matrix.)
On the other hand, the analysis of the discrete spectrum when the extra minors are not identically zero presented in \cite{JMP2015}
shows that, at least in the reflectionless case, 
it is sufficient to specify only the value of one extra minor at one point of each quartet of discrete eigenvalues. 
This implies that one cannot independently assign the value of these minors along the real axis.
This suggests the existence of an additional trace formula, which allows one to reconstruct the extra minors over their whole domain of analyticity
in terms only of the reflection coefficients and the value of these minors at one point of each quartet of discrete eigenvalues.
Such a formula has yet to be identified, however.

3. The results of the present work establish a loose connection between the theory of integrable systems and that of algebraic combinatorics.
A connection betwen the two fields was also recently observed in two-dimensional systems in the study of the soliton solutions of
the Kadomtsev-Petviashvili equation
\cite{prl99p064103,jmp47p33514,mcs74p237,jpa36p10519,jpa41p275209,jpa37p11169,kodamawilliams2011,kodamawilliams2013}.
To the best of our knowledge, however, it is the first time that a similar connection has been reported for one-dimensional systems
and within the context of the IST.

4. The results of this work also open up a number of interesting theoretical issues in addition to the one already mentioned above.  
To mention just a few:
\begin{itemize}
\itemsep0pt
\parsep0pt
\item[(i)]
The possible existence of real spectral singularities.
It is well known that, for the scalar NLS equation, such singularities are not allowed in the defocusing case with NZBC \cite{FT1987},
but are possible in the focusing case with ZBC \cite{Zhou}.
On the other hand, the problem is open even for the defocusing Manakov system with NZBC.
\item[(ii)]
The computation of
the long-time asymptotic behavior using the Deift-Zhou method \cite{CPAM47p199,AM137p295}
(see \cite{itsustinov,vartanian1,vartanian2} for the defocusing scalar case with NZBC).
This problem is also still open even in the two-component (Manakov) case.
\item[(iii)]
The development of the IST for the multi-component focusing case with NZBC.
Although all the main ideas from the defocusing case carry over to the focusing case,
already for the Manakov system, the focusing case with NZBC was more involved than in the defocusing case,
due to the presence of four distinct fundamental domains of analyticity instead of two, and due to the fact
that the scattering problem is not self-adjoint \cite{KBK2015}.
\item[(iv)]
A detailed treatment of the $N$-component case with $N\ge4$.
Although all the formalism presented in this work (namely, tensors, triangular decomposition and symmetries)
can be extended to the general case in a straightforward manner, it should nevertheless be clear that the treatment of
the symmetries and the discrete spectrum will be significantly more involved when $N>3$.
\item[(v)]
The development of the IST for the Manakov system and multi-component VNLS equation with boundary conditions
$\@q_\pm$ which are non-parallel (though still equimodular), and more general asymmetric NZBC for which $\|\@q_+\|\ne\|\@q_-\|$.
The case of asymmetric NZBC for the scalar NLS equation
was studied in \cite{BFP2015,boitipempinelli} in the defocusing case and in \cite{demontis} for the focusing case.
A treatment of either the focusing or defocusing Manakov system with asymmetric NZBC, however, is still completely open.
\end{itemize}
It is hoped that the results of this work and the above discussion will motivate further investigations on these and related problems.

\section*{Acknowledgments}

This work was partially supported by the American Institute of Mathematics under the ``SQuaRE'' program
and by the National Science Foundation under grants DMS-1311847 and DMS-1311883.

\section*{Appendix: Proofs}
\setcounter{section}0
\renewcommand\thesubsection{A.\arabic{subsection}}
\addcontentsline{toc}{section}{Appendix: Proofs}
\gdef\rightmark{Appendix: Proofs}
\gdef\leftmark{Appendix: Proofs}
\renewcommand\theequation{A.\arabic{equation}}

This section contain the proofs of the results in the main text.

\subsection{Direct problem}
\label{a:direct}

\paragraph{Derivation of Volterra integral equations for the Jost eigenfunctions.}
We subtract off the asymptotic behavior of the potential and rewrite the first of \eref{e:Laxpair} as
\begin{equation}
\label{e:factorization}
(\phi_\pm)_x = \@X_\pm \phi_\pm + \Delta \@Q_\pm \phi_\pm,
\end{equation}
where $\Delta \@Q_\pm = \@Q - \@Q_\pm$.
Introducing the modified eigenfunctions $\mu_\pm(x,t,z)$ as in \eref{e:modified} and the
integrating factor
$\psi_\pm(x,t,z) = \e^{-i \@{\Theta}(x,t,z)} \@E_\pm^{-1}(z) \mu_\pm(x,t,z) \e^{i \@{\Theta}(x,t,z)}$,
we can then formally integrate the ODE for $\mu_\pm(x,t,z)$ to obtain the linear integral equations
\bse
\label{e:intequations}
\begin{gather}
\mu_-(x,t,z) = \@E_-(z) + \int_{-\infty}^{x} \@E_-(z) \e^{i (x-y) \@{\Lambda}(z)} \@E_-^{-1}(z) \Delta \@Q_-(y,t) \mu_-(y,t,z) \e^{- i (x-y) \@{\Lambda}(z)} \d y,
\\
\mu_+(x,t,z) = \@E_+(z) - \int_x^{\infty} \@E_+(z) \e^{i (x-y) \@{\Lambda}(z)} \@E_+^{-1}(z) \Delta \@Q_+(y,t) \mu_+(y,t,z) \e^{-i (x-y) \@{\Lambda}(z)} \d y.
\end{gather}
\ese
One can now rigorously define the Jost eigenfunctions as the unique solutions of~\eref{e:intequations} for all $z\in \Real\setminus \left\{0\right\}$
as long as $(1+|x|)(q(x,t)-q_\pm)\in L^1(\Real^\pm)$.
These results are proved much in a similar way as for the two-component case, and we refer the reader to \cite{gbdkk2015} for details.
\qed

\proof{Proof of Lemma~\ref{L:triang}.}
Combining the scattering relation \eref{e:scattering} with the definitions \eref{e:Mdef} of the fundamental
meromorphic eigenfunctions yields
$\phi_+(x,t,z) \@A(z) \alpha^\pm(z) = \phi_+(x,t,z) \beta^\pm(z)$.
Solving for $\@A(z)$ yields the desired triangular decompositions.
\qed

\proof{Proof of Lemma~\ref{L:decompexplicit}.}
Due to the constraints on $\alpha^\pm(z)$ and $\beta^\pm(z)$ given in Theorem~\ref{t:meromorphic},
the decompositions in Lemma~\ref{L:triang} are unique.
Simply finding triangular decompositions of the scattering matrix with the given constraints yields the matrices
in \eref{e:ludecomp} due to this uniqueness property.
\qed


\proof{Proof of Theorem~\ref{t:gantmacher}.}
The analyticity of the explicitly given principal minors of the scattering matrices follows directly from Corollary~\ref{c:deltas},
while the analyticity of the remaining principal minors follows from applying the identity in Lemma~\ref{L:relation}.
The analyticity of the given non-principal minors follows from an examination of the off-diagonal entries of the matrices
defined in Theorem~\ref{t:meromorphic}.
\qed

\proof{Proof of Lemma~\ref{L:merosimple}.}
The lemma is proved using the asymptotics in Theorem \ref{t:mwedgedef}.
Simply put, if any of the eigenfunctions were to have a pole of higher order, then said asymptotics could not hold.
\qed

\proof{Proof of Theorem~\ref{t:chidef}.}
The results of Lemma~\ref{L:merosimple} show that multiplication of $\@M^\pm(x,t,z)$ by the matrix
$\@D^\pm(z) \e^{i \@\Theta(x,t,z)}$, respectively, removes any pole contributions from the
$\Delta_j^\pm(z)$ ($j=1,2,3$).
The resulting columns are analytic solutions of both parts of the Lax pair, and we will see later that they
become linearly dependent only at zeros of certain principal minors of the scattering matrices.
\qed

\proof{Proof of Lemma~\ref{L:chixasymp}.}
The asymptotics~\eref{e:auxnewasymp} follow by multiplying the relations in~\eref{e:auxxasymp}
by the matrices $\@D^\pm(z)$ for the appropriate regions of analyticity and noting that the resulting
expressions still hold when $z \in Z^\pm$.
The asymptotics~\eref{e:chipminf} then follow from expanding the columns
of~\eref{e:auxnewasymp}.
\qed


\subsection{Symmetries}
\label{a:symm}

\proof{Proof of Proposition~\ref{L:firstsymmetry}.}
We obtain the results directly by noting that $k^*(z^*) = k(z)$ and $\@J\@Q=-\@Q\@J$ and
by using the facts that
$- \@J \@X^\dag(x,t,z^*) \@J = \@X(x,t,z)$
and
$- \@J \@T^\dag(x,t,z^*) \@J = \@T(x,t,z)$.
\qed

\proof{Proof of Lemma~\ref{L:firstsymmphi}.}
We use Proposition~\ref{L:firstsymmetry} with
\begin{equation}
\label{e:symm1}
\@w_\pm(x,t,z) = \@J[\phi_\pm^{\dagger}(x,t,z)]^{-1}, \qquad z \in \Real\,,
\end{equation}
and note that for all $z \in \Complex$,
$[\e^{i \@{\Theta}(x,t,z^*)}]^\dag = \e^{-i \@{\Theta}(x,t,z)}$.
It is easy to see that
\begin{equation}
\label{e:thisnew}
\@w_\pm(x,t,z) = \@J [\@E_\pm^{\dagger}(z)]^{-1} \e^{i \@{\Theta}(x,t,z)} + o(1), \qquad x \to \pm \infty.
\end{equation}
Since both $\@w_\pm$ and $\phi_\pm$ are fundamental matrix solutions of the Lax pair~\eref{e:Laxpair},
there must exist an invertible $4 \times 4$ matrix $\@C(z)$ such that~\eref{e:symm3} holds.
Comparing the asymptotics from \eref{e:thisnew} with those from \eref{e:phixasymp}, we then obtain~\eref{e:firstsymmmatrix}.
\qed

\proof{Proof of Theorem~\ref{t:phifirstsymm}.}
First, we use the symmetry \eref{e:symm3} (which contains the inverse of a transpose of a matrix)
to obtain representations of the columns of $\phi_\pm^*$
in terms of generalized cross products of other columns of $\phi_\pm$ (for $z \in \Real$).
Then, we combine these representations with the decompositions \eref{e:decompsimp}, simplify, and apply the
Schwarz reflection principle (as in the Manakov system).
\qed


\proof{Proof of Lemma~\ref{L:chisymm1}.}
We verify~\eref{e:chisymm1a}.
The rest of~\eref{e:chisymm1} are proved similarly.
First, replace $v_1$, $v_2$, and $v_3$ in~\eref{e:adjoint} with $\phi_{-,1}$, $\chi_2^+$, and $\phi_{+,4}$, respectively.
Since $z \in \Real$, we may apply the decompositions~\eref{e:decomp} to obtain
\begin{equation*}
L [\phi_{-,1}^*,(\chi_2^+)^*,\phi_{+,4}^*] =
a_{11}^* b_{34}^* L [\phi_{-,1}^*,\phi_{-,2}^*,\phi_{-,3}^*]
+ a_{11}^* b_{44}^* L [\phi_{-,4}^*,\phi_{-,1}^*,\phi_{-,2}^*],
\end{equation*}
where the $(x,t,z)$-dependence was omitted for brevity.
Combining the results of Corollary~\ref{c:realcrossproduct} with the symmetry~\eref{e:symm5}
and~\eref{e:decomp} yields the following:
\begin{equation*}
\e^{2 i \theta_2} \@J L [\phi_{-,1}^*,(\chi_2^+)^*,\phi_{+,4}^*] =
b_{11} \gamma [- a_{43} \phi_{-,4} + a_{44} \phi_{-,3}] =
b_{11} \gamma \chi_3^-,
\end{equation*}
where, again, the $(x,t,z)$-dependence was omitted for brevity.
An application of the Schwarz reflection principle completes the proof.
\qed

\proof{Proof of Lemma~\ref{L:phisymm2}.}
We use Proposition~\ref{p:symm2} and we note that
\begin{equation}
\@{\Theta}(x,t,\z^*) = \@K \@{\Theta}(x,t,z),
\end{equation}
where $\@K = \diag(-1,1,1,-1)$. For $z \in \Real$, define $\@w_\pm(x,t,z) = \phi_\pm(x,t,\z^*)$.
Since $\@w_\pm$ and $\phi_\pm$ both solve the Lax pair \eref{e:Laxpair},
there must exist invertible $4 \times 4$ matrices $\@{\Pi}_\pm(z)$ such that~\eref{e:phiwitha} holds.
Comparing the asymptotics of \eref{e:phiwitha} with the asymptotics from \eref{e:phixasymp} yields
\begin{equation}
\@E_\pm(\z^*)\e^{i \@K \@{\Theta}(x,t,z)}\@{\Pi}_\pm(z) = \@E_\pm(z)\e^{i \@{\Theta}(x,t,z)},
\end{equation}
which yields~\eref{e:symm4}.
\qed

\proof{Proof of Lemma~\ref{L:chisymm2}.}
We verify~\eref{e:auxsymm2a}.
The rest of~\eref{e:auxsymm2} are proved similarly.
First, we evaluate $\phi_{-,2}(x,t,z)$ via~\eref{e:decompsimp} at $q_o^2/z$ and then apply the symmetry~\eref{e:symm6} to obtain
\begin{equation*}
\frac{q_-}{q_+} \frac{1}{a_{44}} \left[\frac{q_+}{q_-} a_{42} \phi_{-,4} + \chi_2^+(q_o^2/z)\right] =
\frac{1}{B_{[1,2]}} \left[\chi_2^- - \frac{1}{a_{44}} B_{\minorset{1,3}{1,2}} \chi_3^- -
\left(\frac{a_{43}}{a_{44}} B_{\minorset{1,3}{1,2}} + B_{\minorset{1,4}{1,2}}\right)\phi_{-,4}\right],
\end{equation*}
where the $(x,t)$-dependence was omitted for brevity and any function other than $\chi_2^+$ with $z$-dependence is
evaluated at $z$.
We solve for $\chi_2^+(q_o^2/z)$ and use analytic continuation to obtain the desired result.
\qed

\proof{Proof of Theorem~\ref{t:discrete}.}
Evaluating~\eref{e:auxsymm2b} at $z=\z_o^*$ and combining the result with~\eref{e:symm2newminors}
and~\eref{e:symm6} yields
\[
\label{e:discreteproof2}
\chi_3^+(x,t,z_o) = \frac{1}{A_{[3,4]}(\z_o^*)} \left[A_{[1,2,3]}(z_o) \chi_3^-(\z_o^*)
- A_{\minorset{1,2}{1,3}}(z_o) \chi_2^-(\z_o^*)\right].
\]
Next, combine~\eref{e:discreteproof2} with~\eref{e:auxsymm2c} and~\eref{e:auxsymm2d} and simplify to obtain
\[
\chi_3^+(x,t,z_o) = \frac{\e^{- i \Delta \theta}}{A_{[1,2]}(z_o) A_{[3,4]}(\z_o^*)}
\left[A_{[1]}(z_o) A_{[1,2,3]}(z_o) + A_{\minorset{1,2}{1,3}}(z_o) A_{\minorset{1,3}{1,2}}(z_o)\right] \chi_3^+(x,t,z_o).
\]
The result then follows easily from the symmetries of the scattering matrix.
Note that the same result can also be derived at all points $z_o$ at which $\chi_2^+(x,t,z_o)\ne0$.

Next, since all the minors involved in~\eref{e:A12symm} are analytic in the UHP,
and the zeros of analytic functions [in this case $\chi_2^+(x,t,z_o)$] are isolated,
one can also obtain \eref{e:A12symm} at those points where $\chi_2^+(x,t,z)$ is zero
by simply taking the limit of all quantities involved.
\qed

\subsection{Discrete spectrum}
\label{a:discrete}

\proof{Proof of Lemma~\ref{L:L2eigen}.}
It is easy to show that if $\@v(x,t,k) = (v_1,v_2,v_3,v_4)^T$ is any nontrivial solution of the scattering problem,
\begin{equation}
\label{e:L2}
- i (k - k^*) \sum_{n=1}^4 |v_n(x,t,z)|^2 = \frac{\partial}{\partial x} \left[|v_1(x,t,k)|^2 - \sum_{n=2}^4 |v_n(x,t,k)|^2\right].
\end{equation}
Now, integrate~\eref{e:L2} from $-\infty$ to $\infty$. If $\@v(x,t,k) \in L^2(\Real)$, the right-hand side is zero, but since
\begin{equation*}
\int_\Real \|\@v(x,t,k)\|^2 \d x \ne 0,
\end{equation*}
this implies $k^* = k$, i.e., $z \in \Real$ or $z \in C_o$.
However, for $z \in \Real$,
the eigenfunctions do not decay as $x \to \pm \infty$, and therefore, $\@v(x,t,k)$ cannot belong to $L^2(\Real)$.
Thus, the only possibility left is $z \in C_o$.
\qed

In the proofs that follow, it will be useful to note the following:
\bse
\label{e:theta2asymp}
\begin{gather}
\lim_{x \to \infty} \left| \e^{i \theta_2(x,t,z_o)} \right| =
\begin{cases}
0, & |z_o| > q_o,
\\
\infty, & |z_o| < q_o,
\end{cases}
\\
\lim_{x \to -\infty} \left| \e^{i \theta_2(x,t,z_o)} \right| =
\begin{cases}
\infty, & |z_o| > q_o,
\\
0, & |z_o| < q_o.
\end{cases}
\end{gather}
\ese

\proof{Proof of Theorem~\ref{t:spectrum}.}
The numbering of the cases in this proof corresponds to the numbering in Table~\ref{table1}.
To prove the theorem, it is enough to exclude cases III, VI and VII.

\textit{Case~III.}~
This case is not compatible with the symmetries \eref{e:A12symm}.
In fact, when the analytic non-principal minors are identically zero, the left-hand side
of \eref{e:A12symm} is zero, while the right-hand side is not.
Therefore, we can exclude this case.

\textit{Case~VI.}~
If $A_{[1]}(z_o)=A_{[1,2,3]}(z_o)=0$ and $A_{[1,2]}(z_o) \neq 0$,
the symmetry~\eref{e:symm6} then implies $A_{[1]}(\z_o)=A_{[1,2,3]}(\z_o)=0$.
We apply the symmetries \eref{e:refsymm} to find that $\chi_2^+(x,t,\z_o)$, $\chi_3^+(x,t,\z_o)$,
$\chi_2^-(x,t,\z_o^*)$, and $\chi_3^-(x,t,\z_o^*)$ are all zero.
Suppose $A_{[1,2]}(\z_o) \neq 0$.
Then the symmetries \eref{e:refsymm} imply $\chi_2^+(x,t,z_o)$, $\chi_3^+(x,t,z_o)$, $\chi_2^-(x,t,z_o^*)$, and
$\chi_3^-(x,t,z_o^*)$ are all zero.
Using the same arguments as above, the left-hand side of \eref{e:symm1eigend} must then be zero at $z=z_o$,
which is a contradiction.
Thus, $A_{[1,2]}(\z_o) = 0$.
However, since $\chi_3^+(x,t,\z_o)$ is zero, symmetry \eref{e:refsymmd} implies $\chi_3^-(x,t,z_o^*)$ is zero.
Similarly, we find that $\chi_2^-(x,t,z_o^*)$ is zero.
We arrive at the same problem as before [namely, that the left-hand side of \eref{e:symm1eigend} must have a zero
at $z=z_o$].
Therefore, we can exclude this case also.

\textit{Case~VII.}~
Suppose $A_{[1]}(z_o)=A_{[1,2]}(z_o)=A_{[1,2,3]}(z_o)=0$.
We note that at least one auxiliary eigenfunction must be nonzero at each point in the eigenvalue quartet
consisting of $z_o$, $z_o^*$, $\z_o$, and $\z_o^*$.
Indeed,
suppose $\chi_2^+(x,t,z_o)=\chi_3^+(x,t,z_o)=\@0$. We can then introduce $\hat{\chi}_2^+(x,t,z)=\chi_2^+(x,t,z)/A_{[1,2]}(z)$
and $\hat{\chi}_3^+(x,t,z)=\chi_3^+(x,t,z)/A_{[1,2]}(z)$, which are then finite at $z=z_o$.

Since the left-hand side of \eref{e:wrdeta} has a triple zero at $z=z_o$,
\begin{equation*}
\det (\phi_{-,1}(x,t,z_o),\hat{\chi}_{2}^+(x,t,z_o),\hat{\chi}_{3}^+(x,t,z_o),\phi_{+,4}(x,t,z_o)) = \@0.
\end{equation*}
Suppose now $\hat{\chi}_{2}^+(x,t,z_o)=\@0$.
Then the cross product in the right-hand side of~\eref{e:symm1eigend} will have a triple zero at $z=z_o$.
Since the zeros of the minors are simple, we conclude $\phi_{-,4}(x,t,z_o^*)=\@0$, which is a contradiction.
Thus, $\hat{\chi}_{2}^+(x,t,z_o) \neq \@0$.
Similarly, $\hat{\chi}_{3}^+(x,t,z_o) \neq \@0$.

Then there exist constants (not all zero) such that
\begin{equation*}
b_o \phi_{-,1}(x,t,z_o) - b_1 \hat{\chi}_{2}^+(x,t,z_o) - b_2 \hat{\chi}_{3}^+(x,t,z_o) - b_3 \phi_{+,4}(x,t,z_o) = \@0.
\end{equation*}
If $b_o=0$, then the left-hand side of~\eref{e:symm1eigena} will have a zero at $z=z_o$, which is a contradiction.
Thus, $b_o \neq 0$.
Using \eref{e:symm1eigen}, we find that
\begin{equation*}
[\phi_{-,4}(z_o^*)]^* = - \frac{b_3 \e^{-2i \theta_2(z_o)}}{2 b_o A_{[1]}'(z_o) A_{[1,2]}'(z_o)}
\@J L[\phi_{+,4}(z_o),\hat{\chi}_{2}^+(z_o),\hat{\chi}_{3}^+(z_o)]
=
\frac{b_3 A_{[1,2,3]}'(z_o)}{b_o A_{[1]}'(z_o)} [\phi_{+,1}(z_o^*)]^*,
\end{equation*}
where the $(x,t)$-dependence was omitted for brevity.
However, this is a bound state, which contradicts Lemma \ref{L:L2eigen} since $|z_o| \neq q_o$.
Thus, at least one of $\chi_2^+(x,t,z_o)$ and $\chi_3^+(x,t,z_o)$ must be nonzero,
and similar results follow for the remaining auxiliary eigenfunctions.

Now, we assume that the auxiliary eigenfunctions are not zero at any of the points in the eigenvalue quartet.
The case in which any one of the auxiliary eigenfunctions is zero at a point in the eigenvalue quartet
will be considered next.

Again, recall that the left-hand sides of the symmetries~\eref{e:chisymm1} must be analytic for all $z$ in their
corresponding domains of analyticity.
Specifically, since $\chi_2^-(x,t,z)$ and $\chi_3^-(x,t,z)$ are analytic for all $z$ in the lower-half plane,
we have
\begin{equation*}
L[\phi_{-,1},\chi_3^+,\phi_{+,4}](x,t,z_o)=
L[\phi_{-,1},\chi_2^+,\phi_{+,4}](x,t,z_o) = \@0\,.
\end{equation*}
Then a nontrivial linear combination of the eigenfunctions in each of these cross products must be zero.
This statement is strengthened by Lemma~\ref{L:L2eigen}, which states that for eigenvalues off the circle $C_o$,
bound states cannot exist.
A bound state would result if $\phi_{-,1}(x,t,z_o)$ were proportional to $\phi_{+,4}(x,t,z_o)$,
so we may conclude that there exist constants such that
\begin{equation}
\label{e:appm}
\chi_2^+(x,t,z_o) = c_o \phi_{-,1}(x,t,z_o) + c_1 \phi_{+,4}(x,t,z_o),
\quad
\chi_3^+(x,t,z_o) = d_o \phi_{-,1}(x,t,z_o) + d_1 \phi_{+,4}(x,t,z_o).
\end{equation}
Here is where the asymptotics \eref{e:chipminf} of the eigenfunctions as $x \to \pm \infty$ and the
exponential behavior \eref{e:theta2asymp} of the auxiliary eigenfunctions are important.
Suppose $|z_o| > q_o$.
Then the left-hand sides of \eref{e:appm} vanish as $x \to \infty$,
while the right-hand sides do not unless $c_o=d_o=0$.
Explicitly, we have
\begin{equation}
\label{e:type7consta}
\chi_2^+(x,t,z_o) = c_1 \phi_{+,4}(x,t,z_o),
\qquad
\chi_3^+(x,t,z_o) = d_1 \phi_{+,4}(x,t,z_o).
\end{equation}
No additional information is obtained by taking the limit as $x \to - \infty$. Similarly, we obtain
\begin{equation}
\label{e:contradictionproof}
\chi_2^-(x,t,z_o^*) = \=c_1 \phi_{+,1}(x,t,z_o^*),
\qquad
\chi_3^-(x,t,z_o^*) = \=d_1 \phi_{+,1}(x,t,z_o^*),
\end{equation}
where $\=c_1$ and $\=d_1$ are constants.

Note that since
we are assuming the auxiliary eigenfunctions are nonzero at $z_o$, then $c_1d_1\ne 0$ and $\=c_1\=d_1\ne0$. Moreover, \eref{e:type7consta} imply
that
$\chi_2^+(x,t,z_o)$ and $\chi_3^+(x,t,z_o)$ are proportional to each other:
\begin{equation}
\label{e:type7prop}
\chi_2^+(x,t,z_o)=\frac{c_1}{d_1}\chi_3^+(x,t,z_o)\,.
\end{equation}


We now need to again examine the symmetries \eref{e:symm1eigen}.
As before, the left-hand side of each symmetry is an analytic function in its corresponding half plane, but
note that the right-hand sides of \eref{e:symm1eigena} and \eref{e:symm1eigend} have denominators
with double zeros at $z=z_o$.
In order for the left-hand sides to actually be analytic at $z=z_o$,
the numerators of the right-hand sides must also have double
zeros at $z=z_o$.
Namely, we have
\begin{equation*}
L'[\phi_{-,1},\chi_2^+,\chi_3^+](x,t,z_o) =
L'[\chi_2^+,\chi_3^+,\phi_{+,4}](x,t,z_o) = \@0,
\end{equation*}
where the prime denotes differentiation with respect to $z$.
Combining this information with \eref{e:type7consta} then yields
\begin{equation*}
L[\phi_{-,1},d_1 \chi_{2,z}^+ - c_1 \chi_{3,z}^+, \phi_{+,4}](x,t,z_o) = \@0.
\end{equation*}
Suppose $d_1 \chi_{2,z}^+(x,t,z_o) = c_1 \chi_{3,z}^+(x,t,z_o)$.
It is easy to see using simple algebra that combining this assumption with the expression for
$\phi_{+,1}(x,t,z)$ obtained using the symmetry \eref{e:symm1eigena} yields $\phi_{+,1}(x,t,z_o^*) = \@0$,
which is a contradiction since $\phi_{+,1}(x,t,z)$ is one of the Jost eigenfunctions.
Then as a result of this, there exist constants (not all zero) such that
\begin{equation*}
b_o \phi_{-,1}(x,t,z_o) + b_1 [d_1 \chi_{2,z}^+(x,t,z_o) - c_1 \chi_{3,z}^+(x,t,z_o)]
+ b_2 \phi_{+,4}(x,t,z_o) = \@0.
\end{equation*}
It is easy to conclude that $b_1 \neq 0$ by using Lemma \ref{L:L2eigen}.
Suppose $b_o=0$.
Then as above, we combine this assumption with the expression for $\phi_{+,1}(x,t,z)$ obtained
from \eref{e:symm1eigena} to obtain
$\phi_{+,1}(x,t,z_o^*) = \@0$.
Again, this is a contradiction, so we may rescale these new constants and write
\begin{equation}
\label{e:phim1}
\phi_{-,1}(x,t,z_o) = b_1 [d_1 \chi_{2,z}^+(x,t,z_o) - c_1 \chi_{3,z}^+(x,t,z_o)] + b_2 \phi_{+,4}(x,t,z_o).
\end{equation}

The expressions in \eref{e:type7consta} and \eref{e:phim1} are of great importance.
Specifically, combining them with the symmetries \eref{e:symm1eigen} and \eref{e:chisymm1}
yields expressions for $\chi_2^-(x,t,z_o^*)$ and $\chi_3^-(x,t,z_o^*)$ in terms of $\phi_{+,1}(x,t,z_o^*)$.
We note, however, that combining these results with \eref{e:contradictionproof} yields
$\phi_{+,1}(x,t,z_o^*)$ in terms of $\phi_{-,4}(x,t,z_o^*)$, which is a bound state.

Finally, let us consider the case when only one of the two auxiliary eigenfunctions is zero at $z_o$. Let us assume
$\chi_2^+(x,t,z_o)=0$ while $\chi_3^+(x,t,z_o)\ne 0$ (the other case can obviously be dealt with analogously), and suppose
$|z_o|>q_o$. We then
introduce $\hat{\chi}_2^+(z)=\chi_2^+(z)/A_{[1,2]}(z)$ which is finite at $z=z_o$.
Following the same logic as above, one can show that
\begin{equation}
\label{e:case7third}
\chi_3^+(x,t,z_o)=d_1\phi_{+,4}(x,t,z_o)\,, \qquad d_1\ne 0\,.
\end{equation}
From \eref{e:symm1eigena} and \eref{e:symm1eigenc} we then find that
\begin{align}
L[\hat{\chi}_2^+,\chi_3^+,\phi_{+,4}](x,t,z_o)=0\,, \label{e:case7fourth} \\
L[\phi_{-,1},\hat{\chi}_2^+,\chi_3^+](x,t,z_o)=0 \label{e:case7fifth}\,,
\end{align}
and both zeros need to be simple (otherwise the left-hand side of \eref{e:symm1eigena} and \eref{e:symm1eigenc}
would be zero).
Now, \eref{e:case7fourth} is satisfied in virtue of \eref{e:case7third}, while \eref{e:case7fifth} implies that
there exist constants (not all zero) such that
$$
b_o\phi_{-,1}(x,t,z_o)-b_1\hat{\chi}_2^+(x,t,z_o)-b_2\chi_3^+(x,t,z_o)=0\,.
$$
$b_o$ cannot be zero, otherwise $b_1\hat{\chi}_2^+(x,t,z_o)+b_2\chi_3^+(x,t,z_o)=0$, together with \eref{e:case7third}, would imply
that the zero in \eref{e:case7fourth} has multiplicity two.
Consequently, we can write
$$
\phi_{-,1}(x,t,z_o)=\frac{b_1}{b_o}\hat{\chi}_2^+(x,t,z_o)+\frac{b_2}{b_o}\chi_3^+(x,t,z_o)\,.
$$
Since $|z_o|>q_o$, the right-hand side of the above vanish as $x \to \infty$ (cf the
exponential behavior \eref{e:theta2asymp} of the auxiliary eigenfunctions),
while the right-hand side does not, hence we arrive at a contradiction.

All the arguments above can be repeated for $|z_o| < q_o$, and one arrives at a contradiction as well.
Therefore, we can exclude this case.
The remaining cases do not lead to contradictions.
\qed

\proof{Proof of Theorem~\ref{t:type2theorem}.}
Since $A_{[1]}(z_o)=A_{[2,3,4]}(z_o^*)=0$ and the denominators on the right-hand sides of~\eref{e:refsymmb}
and~\eref{e:refsymmd} are not zero at $z=z_o^*$ and $z=z_o$, respectively, we conclude
\begin{equation*}
\chi_3^+(x,t,\z_o) = \chi_3^-(x,t,\z_o^*) = \@0.
\end{equation*}
The symmetry~\eref{e:A12symm0} implies immediately that $A_{[1,2]}(\z_o)=0$.
%
The left-hand side of symmetry \eref{e:chisymm1b} is zero at $z=\z_o^*$,
so $\{\phi_{+,1},\chi_2^-,\phi_{-,4}\}$ is a linearly
dependent set at $z=\z_o^*$.
If $\chi_2^-(x,t,\z_o^*) = \@0$, then the symmetry \eref{e:refsymmc} implies $\chi_2^+(x,t,z_o) = \@0$.
We already know that this cannot occur, so $\chi_2^-(x,t,\z_o^*)$ is nonzero.
Similarly, Eqs. \eref{e:chisymm1d}, \eref{e:refsymma}, and \eref{e:symm1eigenb} imply that
$\chi_2^+(x,t,\z_o)$ is nonzero as well.

Equations \eref{e:wrdet} then imply that there exist appropriate constants (not all zero) such that
\begin{gather*}
\beta_o \phi_{-,1}(x,t,z_o) + \beta_1 \chi_2^+(x,t,z_o) + \beta_2 \chi_3^+(x,t,z_o)
+ \beta_3 \phi_{+,4}(x,t,z_o) = \@0,
\\
\=\beta_o \phi_{+,1}(x,t,z_o^*) + \=\beta_1 \chi_2^-(x,t,z_o^*) + \=\beta_2 \chi_3^-(x,t,z_o^*)
+ \=\beta_3 \phi_{-,4}(x,t,z_o^*) = \@0.
\end{gather*}
If $\beta_o = 0$, then the set $\{\chi_2^+,\chi_3^+,\phi_{+,4}\}$ will be linearly dependent at $z=z_o$.
Then the left-hand side of the symmetry \eref{e:symm1eigena} will be zero at $z=z_o$, which is a contradiction.
Similarly, if $\beta_1 = 0$, the set $\{\phi_{+,1},\chi_3^-,\phi_{-,4}\}$ will be linearly dependent at
$z=z_o^*$, and the left-hand side of \eref{e:chisymm1a} will be zero at $z=z_o^*$, which is another contradiction.

In addition, since $\chi_3^+(x,t,\z_o)=\chi_3^-(x,t,\z_o^*)=\@0$,
Eqs. \eref{e:chisymm1b} and \eref{e:chisymm1d} imply the existence of appropriate constants
(not all zero) such that
\begin{gather*}
\~\beta_o \phi_{+,1}(x,t,\z_o^*) + \~\beta_1 \chi_2^-(x,t,\z_o^*) + \~\beta_2 \phi_{-,4}(x,t,\z_o^*) = \@0,
\\
\^\beta_o \phi_{-,1}(x,t,\z_o) + \^\beta_1 \chi_2^+(x,t,\z_o) + \^\beta_2 \phi_{+,4}(x,t,\z_o) = \@0.
\end{gather*}
We obtain a bound state if either $\~\beta_1 = 0$ or $\^\beta_1 = 0$, and we can use an argument
similar to the one above to show that $\~\beta_2 \neq 0$.

Knowing that certain norming constants cannot be zero allows us to rescale them and write
\bse
\label{e:2proof}
\begin{gather}
\label{e:2proofa}
\phi_{-,1}(x,t,z_o) = d_1 \chi_2^+(x,t,z_o) + d_2 \chi_3^+(x,t,z_o) + d_3 \phi_{+,4}(x,t,z_o),
\\
\label{e:2proofb}
\chi_2^-(x,t,z_o^*) = \=d_1 \phi_{+,1}(x,t,z_o^*) + \=d_2 \chi_3^-(x,t,z_o^*) + \=d_3 \phi_{-,4}(x,t,z_o^*),
\\
\label{e:2proofc}
\phi_{-,4}(x,t,\z_o^*) =
\check{d}_1 \chi_2^-(x,t,\z_o^*) + \check{d}_2 \phi_{+,1}(x,t,\z_o^*),
\\
\label{e:2proofd}
\chi_2^+(x,t,\z_o) =
\^d_1 \phi_{+,4}(x,t,\z_o) + \^d_2 \phi_{-,1}(x,t,\z_o),
\end{gather}
\ese
where each $d_j$, $\=d_j$, etc. is a constant.
Applying the second symmetry to \eref{e:2proofc} and \eref{e:2proofd} and then comparing
with the rest of \eref{e:2proof} yields
\bse
\begin{gather}
\check{d}_1 = - \frac{i z_o}{q_-^*} \frac{A_{[1,2]}(z_o)}{A_{[1,2,3]}(z_o)} \e^{i \Delta \theta} d_1,
\qquad \check{d}_2 = - \e^{- i (\theta_+ + \theta_-)} d_2,
\qquad \check{d}_3 = 0,
\\
\^d_1 = \frac{i q_+^*}{z_o^*} \frac{A_{[4]}(z_o^*)}{A_{[3,4]}(z_o^*)} \e^{i \Delta \theta} \=d_1,
\qquad \^d_2 = - \frac{i q_-}{z_o^*} \frac{A_{[4]}(z_o^*)}{A_{[3,4]}(z_o)} \e^{i \Delta \theta} \=d_2,
\qquad \^d_3 = 0.
\end{gather}
\ese
Next, we evaluate \eref{e:symm1eigenb} at $z=z_o^*$, take the complex conjugate, and
apply \eref{e:2proofb} and the symmetries of the scattering matrix to obtain
\begin{equation}
\label{e:phiconj2}
\phi_{-,1}(x,t,z_o) = - \frac{\=d_1^* \e^{2 i \theta_2(x,t,z_o)}}{A_{[1,2]}(z_o)A_{[1,2,3]}(z_o)}
\@J L [\phi_{+,1}(x,t,z_o^*),\chi_3^-(x,t,z_o^*),\phi_{-,4}(x,t,z_o^*)]^*.
\end{equation}
Comparing \eref{e:phiconj2} with \eref{e:chisymm1a} yields
\begin{equation}
\label{e:phiconj2a}
\phi_{-,1}(x,t,z_o) = \frac{\gamma(z_o)}{A_{[1,2]}(z_o)} \=d_1^* \chi_2^+(x,t,z_o).
\end{equation}
Then, comparing \eref{e:phiconj2a} with \eref{e:2proofa} yields
\begin{equation}
\=d_1 = \frac{[A_{[1,2]}(z_o)]^*}{\gamma(z_o^*)} d_1^*, \qquad d_2 = 0.
\end{equation}
We conclude $\=d_2 = 0$.
Similarly, $\^d_2 = \check{d}_2 = 0$.
We can then rescale the arbitrary constant $d_1$ to obtain the desired symmetries.

Finally, we show that $|z_o|<q_o$.
We do so by combining the asymptotics \eref{e:chipminf} with \eref{e:theta2asymp}.
We rewrite the first of \eref{e:type2constants} as
\begin{equation*}
\phi_{-,1}(x,t,z_o) = c_1 m_3^+(x,t,z_o) \e^{i \theta_2(x,t,z_o)}.
\end{equation*}
Suppose $|z_o|>q_o$.
Then the left-hand side of this equation vanishes as $x \to - \infty$, while the right-hand side vanishes as $x \to \infty$.
This is a bound state, which cannot occur.
We arrive at no such contradiction when $|z_o|<q_o$.
\qed

\proof{Proof of Theorem~\ref{t:type1theorem}.}
Note that the left-hand sides of~\eref{e:refsymma} and~\eref{e:refsymmc} will have poles at $z=z_o^*$ and $z=z_o$,
respectively, unless the following is true:
\begin{equation*}
\chi_2^-(x,t,z_o^*) = \chi_2^+(x,t,z_o) = \@0.
\end{equation*}
This time, the symmetry~\eref{e:A12symm0} implies that $A_{[1,2]}(\z_o) \neq 0$.

Note that~\eref{e:chisymm1a} implies that the set $\{\phi_{+,1},\chi_3^-,\phi_{-,4}\}$ is linearly dependent
at $z=z_o^*$, while~\eref{e:chisymm1c} implies that the set $\{\phi_{-,1},\chi_3^+,\phi_{+,4}\}$ is linearly
dependent at $z=z_o$.
As above, $\chi_3^-(x,t,z_o^*)$ and $\chi_3^+(x,t,z_o)$ are both nonzero.
Then there exist appropriate constants (not all zero) such that
\begin{gather*}
\alpha_o \phi_{+,1}(x,t,z_o^*) + \alpha_1 \chi_3^-(x,t,z_o^*) + \alpha_2 \phi_{-,4}(x,t,z_o^*) = \@0,
\\
\=\alpha_o \phi_{-,1}(x,t,z_o) + \=\alpha_1 \chi_3^+(x,t,z_o) + \=\alpha_2 \phi_{+,4}(x,t,z_o) = \@0.
\end{gather*}
It is obvious that $\alpha_1 \neq 0$, as a bound state would result otherwise.
Suppose $\=\alpha_o = 0$.
Then as above, the left-hand side of~\eref{e:symm1eigena} will have a zero at $z=z_o$, which is a contradiction.

It is easy to see from~\eref{e:symm1eigen} that neither $\chi_2^+(x,t,\z_o)$ nor $\chi_3^+(x,t,\z_o)$ can be zero.
(the left-hand side of~\eref{e:symm1eigend} would be zero at $z=\z_o$ otherwise)
and that neither $\chi_2^-(x,t,z)$ nor $\chi_3^-(x,t,z)$ can be zero at $z=\z_o^*$ (the left-hand side of~\eref{e:symm1eigenc}
would be zero at $z=\z_o^*$ otherwise).
Then~\eref{e:wrdet} implies that there exist appropriate constants (not all zero) such that
\begin{gather*}
\~\alpha_o \phi_{+,1}(x,t,\z_o^*) + \~\alpha_1 \chi_2^-(x,t,\z_o^*) + \~\alpha_2 \chi_3^-(x,t,\z_o^*)
+ \~\alpha_3 \phi_{-,4}(x,t,\z_o^*) = \@0,
\\
\^\alpha_o \phi_{-,1}(x,t,\z_o) + \^\alpha_1 \chi_2^+(x,t,\z_o) + \^\alpha_2 \chi_3^+(x,t,\z_o)
+ \^\alpha_3 \phi_{+,4}(x,t,\z_o) = \@0.
\end{gather*}
Suppose $\~\alpha_3 = 0$.
Then the left-hand side of~\eref{e:symm1eigend} has a zero at $z=\z_o$, which is a contradiction.
Similarly, $\^\alpha_2 \neq 0$.

Knowing that certain norming constants cannot be zero allows us to rescale them and conclude that
\bse
\begin{gather}
\label{e:1proof}
\phi_{-,1}(x,t,z_o) = c_1 \chi_3^+(x,t,z_o) + c_2 \phi_{+,4}(x,t,z_o),
\\
\label{e:1proofb}
\chi_3^-(x,t,z_o^*) = \=c_1 \phi_{+,1}(x,t,z_o^*) + \=c_2 \phi_{-,4}(x,t,z_o^*),
\\
\label{e:phiminus4proof}
\phi_{-,4}(x,t,\z_o) =
\check{c}_1 \chi_3^-(x,t,\z_o) + \check{c}_2 \phi_{+,1}(x,t,\z_o) + \check{c}_3 \chi_2^-(x,t,\z_o),
\\
\label{e:1proofd}
\chi_3^+(x,t,\z_o^*) =
\^c_1 \phi_{+,4}(x,t,\z_o^*) + \^c_2 \phi_{-,1}(x,t,\z_o^*) + \^c_3 \chi_2^+(x,t,\z_o^*),
\end{gather}
\ese
where each $c_j$, $\=c_j$, etc. is a constant.
Applying~\eref{e:second} and~\eref{e:refsymm} to~\eref{e:1proof} yields
\begin{equation}
\label{e:type1proof2}
\frac{i q_-^*}{z_o} \phi_{-,4}(x,t,\z_o^*) =
\frac{A_{[1,2]}'(z_o)}{A_{[1]}'(z_o)} c_1 \e^{i \Delta \theta} \chi_3^-(x,t,\z_o^*)
- \frac{i q_+}{z_o} c_2 \phi_{+,1}(x,t,\z_o^*).
\end{equation}
Comparing~\eref{e:type1proof2} with~\eref{e:phiminus4proof} yields the first of~\eref{e:type1constsymm}
as well as
\begin{equation}
\check{c}_2 = - \e^{i (\theta_+ + \theta_-)} c_2,
\qquad
\check{c}_3 = 0.
\end{equation}
Similarly, applying~\eref{e:second} and~\eref{e:refsymm}~\eref{e:1proofb} yields
\begin{equation}
\label{e:nexttolast}
\^c_1 = \frac{i q_-^*}{z_o^*} \frac{A_{[2,3,4]}'(z_o^*)}{A_{[3,4]}'(z_o^*)} \=c_1,
\qquad
\^c_2 = - \frac{i q_+}{z_o^*} \frac{A_{[2,3,4]}'(z_o^*)}{A_{[3,4]}'(z_o^*)} \=c_2,
\qquad
\^c_3 = 0.
\end{equation}
Next, we evaluate~\eref{e:symm1eigend} at $z=\z_o^*$, take the complex conjugate, and
apply~\eref{e:1proofd} and the symmetries of the scattering matrix to obtain
\begin{equation}
\label{e:phiconj}
\phi_{-,4}(x,t,\z_o^*) = - \frac{\^c_1^* \e^{2 i \theta_2(x,t,z_o)}}{A_{[3,4]}(\z_o^*)A_{[2,3,4]}(\z_o^*)}
\@J L [\phi_{-,1}(x,t,\z_o),\chi_2^+(x,t,\z_o),\phi_{+,4}(x,t,\z_o)]^*.
\end{equation}
Comparing~\eref{e:phiconj} with~\eref{e:chisymm1d} yields
\begin{equation}
\label{e:phiconja}
\phi_{-,4}(x,t,\z_o^*) = \frac{z_o^2}{q_o^2} \frac{\gamma(z_o)}{A_{[3,4]}(\z_o^*)} \^c_1^* \chi_3^-(x,t,\z_o^*).
\end{equation}
Then, comparing~\eref{e:phiconja} with~\eref{e:phiminus4proof} yields
\begin{equation}
\check{c}_1 = \frac{z_o^2}{q_o^2} \frac{\gamma(z_o)}{A_{[3,4]}(\z_o^*)} \^c_1^*,
\qquad
\check{c}_2 = 0.
\end{equation}
Note that $\check{c}_2 = 0$ implies $c_2 = 0$.
Following a similar method with~\eref{e:symm1eigenc}, we obtain
\begin{equation}
\label{e:last}
\^c_1 = \frac{q_o^2}{(z_o^*)^2} \frac{A_{[1,2]}(\z_o)}{\gamma(z_o^*)} \check{c}_1^*,
\qquad
\^c_2 = 0.
\end{equation}
In addition, note that $\^c_2 = 0$ implies $\=c_2 = 0$.
Finally, the rest of~\eref{e:type1constsymm} is obtained by combining~\eref{e:last} with the first of~\eref{e:type1constsymm},
the first of~\eref{e:nexttolast}, and the symmetries of the scattering matrix.
We can then rescale the arbitrary constant $c_1$ to obtain the desired symmetries.

The proof that $|z_o|<q_o$ is similar to that in Theorem~\ref{t:type2theorem}.
\qed

\subsection{Riemann-Hilbert problem, reflectionless potentials and soliton solutions}
\label{a:rhp}

\proof{Proof of Lemma~\ref{L:jump}.}
We suppress all $(x,t,z)$-dependence when doing so introduces no confusion.
In addition, Lemma~\ref{L:relation} will be used throughout to write some minors of $\@B(z)$
in terms of the corresponding minors of $\@A(z)$.
Using the scattering relation \eref{e:scattering}, we obtain
\begin{equation*}
\phi_{+,4} = - \frac{a_{14}}{a_{44}} \phi_{+,1} - \frac{a_{24}}{a_{44}} \phi_{+,2} -\frac{a_{34}}{a_{44}} \phi_{+,3}
+ \frac{\phi_{-,4}}{a_{44}}.
\end{equation*}
Combining this with~\eref{e:decompsimpa} and~\eref{e:decompsimpc} yields
\bse
\label{a:jumpequations}
\begin{multline}
\label{e:phi4}
\phi_{+,4} = - \left[
\frac{a_{14}}{a_{44}} + \frac{a_{24}}{a_{44}} \frac{b_{12}}{b_{11}}
+ \frac{a_{34}}{a_{44}}
\left(\frac{b_{12}}{b_{11}} \frac{B_{\minorset{1,2}{1,3}}}{B_{[1,2]}} - \frac{B_{\minorset{1,2}{2,3}}}{B_{[1,2]}}\right)
\right] \phi_{+,1}
- \frac{a_{24}}{a_{44}} \frac{\chi_2^-}{A_{[2,3,4]}}
\\
- \frac{a_{34}}{a_{44}} \frac{1}{A_{[3,4]}}
\left[\chi_3^- - \frac{A_{\minorset{2,4}{3,4}}}{A_{[2,3,4]}} \chi_2^-\right]
+ \frac{\phi_{-,4}}{A_{[4]}}.
\end{multline}
We combine this with~\eref{e:scattering},~\eref{e:decompsimpa}, and~\eref{e:decompsimpc} to obtain
\begin{multline}
\frac{\phi_{-,1}}{A_{[1]}} =
\left[
1 + \frac{a_{21}}{a_{11}} \frac{b_{12}}{b_{11}} + \frac{a_{31}}{a_{11}}
\left(\frac{b_{12}}{b_{11}} \frac{B_{\minorset{1,2}{1,3}}}{B_{[1,2]}} - \frac{B_{\minorset{1,2}{2,3}}}{B_{[1,2]}}\right)
- \frac{a_{41}}{a_{11}} \left(\frac{a_{14}}{a_{44}} + \frac{a_{24}}{a_{44}} \frac{b_{12}}{b_{11}}
+ \frac{a_{34}}{a_{44}}
\left(\frac{b_{12}}{b_{11}} \frac{B_{\minorset{1,2}{1,3}}}{B_{[1,2]}} - \frac{B_{\minorset{1,2}{2,3}}}{B_{[1,2]}}\right)\right)
\right] \phi_{+,1}
\\
+ \left[\frac{a_{21}}{a_{11}} - \frac{a_{41}}{a_{11}} \frac{a_{24}}{a_{44}}\right] \frac{\chi_2^-}{A_{[2,3,4]}}
+ \left[\frac{a_{31}}{a_{11}} - \frac{a_{41}}{a_{11}} \frac{a_{34}}{a_{44}}\right]
\frac{1}{A_{[3,4]}} \left[\chi_3^- - \frac{A_{\minorset{2,4}{3,4}}}{A_{[2,3,4]}} \chi_2^-\right]
+ \frac{a_{41}}{a_{11}} \frac{\phi_{-,4}}{A_{[4]}}.
\end{multline}
We then combine~\eref{e:decompa} and~\eref{e:decompb} with~\eref{e:phi4} to obtain
\begin{multline}
\frac{1}{A_{[1,2]}} \left[\chi_2^+ - \frac{A_{\minorset{1,3}{1,2}}}{A_{[1,2,3]}} \chi_3^+ \right] =
\left[
\frac{b_{12}}{b_{11}} - \left(\frac{b_{43}}{b_{44}} \frac{A_{\minorset{1,3}{1,2}}}{A_{[1,2]}}
+ \frac{A_{\minorset{1,4}{1,2}}}{A_{[1,2]}}\right)\left(\frac{a_{14}}{a_{44}}
+ \frac{a_{24}}{a_{44}} \frac{b_{12}}{b_{11}}
+ \frac{a_{34}}{a_{44}}
\left(\frac{b_{12}}{b_{11}} \frac{B_{\minorset{1,2}{1,3}}}{B_{[1,2]}}
- \frac{B_{\minorset{1,2}{2,3}}}{B_{[1,2]}}\right)\right)
\right] \phi_{+,1}
\\
+ \left[1 - \frac{a_{24}}{a_{44}} \left(\frac{b_{43}}{b_{44}} \frac{A_{\minorset{1,3}{1,2}}}{A_{[1,2]}}
+ \frac{A_{\minorset{1,4}{1,2}}}{A_{[1,2]}}\right)\right] \frac{\chi_2^-}{A_{[2,3,4]}}
- \frac{a_{34}}{a_{44}} \left[\frac{b_{43}}{b_{44}} \frac{A_{\minorset{1,3}{1,2}}}{A_{[1,2]}}
+ \frac{A_{\minorset{1,4}{1,2}}}{A_{[1,2]}}\right] \frac{1}{A_{[3,4]}}
\left[\chi_3^- - \frac{A_{\minorset{2,4}{3,4}}}{A_{[2,3,4]}} \chi_2^-\right]
\\
+ \left[\frac{b_{43}}{b_{44}} \frac{A_{\minorset{1,3}{1,2}}}{A_{[1,2]}}
+ \frac{A_{\minorset{1,4}{1,2}}}{A_{[1,2]}}\right] \frac{\phi_{-,4}}{A_{[4]}}.
\end{multline}
Finally, combining~\eref{e:decompc} and~\eref{e:decompd} with~\eref{e:phi4} yields
\begin{multline}
\label{e:chi3}
\frac{\chi_3^+}{A_{[1,2,3]}} =
\left[
\frac{b_{12}}{b_{11}} \frac{B_{\minorset{1,2}{1,3}}}{B_{[1,2]}}
- \frac{B_{\minorset{1,2}{2,3}}}{B_{[1,2]}}
+ \frac{b_{43}}{b_{44}} \left(\frac{a_{14}}{a_{44}}
+ \frac{a_{24}}{a_{44}} \frac{b_{12}}{b_{11}}
+ \frac{a_{34}}{a_{44}}
\left(\frac{b_{12}}{b_{11}} \frac{B_{\minorset{1,2}{1,3}}}{B_{[1,2]}}
- \frac{B_{\minorset{1,2}{2,3}}}{B_{[1,2]}}\right)\right)
\right] \phi_{+,1}
\\
+ \frac{a_{24}}{a_{44}} \frac{b_{43}}{b_{44}} \frac{\chi_2^-}{A_{[2,3,4]}}
+ \left[1 + \frac{a_{34}}{a_{44}} \frac{b_{43}}{b_{44}}\right] \frac{1}{A_{[3,4]}}
\left[\chi_3^- - \frac{A_{\minorset{2,4}{3,4}}}{A_{[2,3,4]}} \chi_2^-\right]
- \frac{b_{43}}{b_{44}} \frac{\phi_{-,4}}{A_{[4]}}.
\end{multline}
\ese
Collecting~\eref{a:jumpequations} and writing the expressions in terms of the reflection coefficients
in \eref{e:reflcoeff}, the jump condition \eref{e:rhpnewjump} follows.
\qed

\proof{Proof of Lemma~\ref{L:trace}.}
We first derive the trace formula for $A_{[1]}(z)$.
A cofactor expansion along the first column of $\@A(z)$, combined with the symmetries~\eref{e:symm5}, yields
\begin{equation}
\label{a:arhos}
\frac{1}{a_{11}(z) b_{11}(z)} = 1 - \frac{1}{|a_{11}(z)|^2} \left[|a_{41}(z)|^2 - \frac{z^2}{z^2 - q_o^2}
\left(|a_{31}(z)|^2 + |a_{21}|^2\right)
\right].
\end{equation}
\eqref{a:arhos} evaluated for $z\in \Real$.
We now combine this expression with the definitions of the reflection coefficients in \eref{e:reflcoeff}
and take the logarithm of the result to obtain
\begin{equation}
\label{e:logs}
\log a_{11}(z) - \log [1/b_{11}(z)] =
- \log \left\{1 - |\rho_3(z)|^2 - \frac{z^2}{z^2 - q_o^2} [|\rho_2(z)|^2 + |\rho_1(z)|^2]\right\}.
\end{equation}
We define the following quantities:
\begin{equation}
\label{e:betas}
\beta_1(z) = a_{11}(z) \prod_{n=1}^{N_1} \frac{z-z_n^*}{z-z_n} \prod_{n=1}^{N_2} \frac{z-w_n^*}{z-w_n},
\qquad
\beta_2(z) = (1/b_{11}(z)) \prod_{n=1}^{N_1} \frac{z-z_n^*}{z-z_n} \prod_{n=1}^{N_2} \frac{z-w_n^*}{z-w_n}.
\end{equation}
Each quantity in \eref{e:betas} has no zeros or poles and approaches 1 as $z \to \infty$ in its appropriate domain
of analyticity.
Combining Eq. \eref{e:logs} with the definitions \eref{e:betas} yields
\begin{equation}
\label{e:rhpa11}
\log \beta_1(z) - \log \beta_2(z) =
- \log \left\{1 - |\rho_3(z)|^2 - \frac{z^2}{z^2 - q_o^2} [|\rho_2(z)|^2 + |\rho_1(z)|^2]\right\}.
\end{equation}
Equation~\eref{e:rhpa11} is the jump condition of a scalar RHP.
Since the unknown quantities $\log \beta_1(z)$ and $\log \beta_2(z)$ have no poles and are $O(1/z)$ as $z \to \infty$,
we can apply the Cauchy projector $P^+$ from~\eref{e:projector} and solve for $a_{11}(z)$ to
obtain \eref{e:a11trace}.


The trace formula for $A_{[1,2]}(z)$ is also obtained by formulating and solving a suitable scalar RHP. 
To this end, we introduce the sectionally meromorphic function
$$
f(z) =  \begin{cases}
A_{[1,2]}(z) \qquad & z\in D_1\,,\\
1/A^*_{[1,2]}(z)  \qquad & z\in D_2\,,\\
1/A_{[3,4]}(z)  \qquad & z\in D_3\,,\\
A^*_{[3,4]}(z)  \qquad & z\in D_4\,,
\end{cases}
$$
where $D_1$ and $D_2$ denote respectively the exterior and the interior of the semicircle $C_o^+$ of radius $q_o$ in the UHP,
while $D_3$ and $D_4$ denote respectively the interior and the exterior of the  semicircle $C_o^-$ of radius $q_o$ in the LHP.
Note $f(z)$ has zeros/poles at the zeros of $A_{[1,2]}(z)$ in the UHP and of $A_{[3,4]}(z)$ in the LHP, 
and $\lim_{z\to\infty}f(z)=1$.
Using the symmetry \eqref{e:symmetryextend} one can easily show that $f(z)$ has no jump across the real axis, while the jump
across the circle is provided by the symmetry relations \eqref{e:A12symm}. Specifically, combining \eqref{e:A12symm} with \eqref{e:symmetryextend}
and \eqref{e:symm_more}, one can express the jump of $f(z)$ across the circle $C_o = C_o^+\cup C_o^-$ as:
\bse
\label{e:logsbar}
\begin{gather}
f^+(z)/f^-(z)=|A_{[1]}(z)|^2-|A_{\minorset{1,2}{1,3}}(z)|^2 \qquad z\in C_o^+\,,\\
f^+(z)/f^-(z)=|A_{[1]}(z^*)|^2-|A_{\minorset{1,2}{1,3}}(z^*)|^2 \qquad z\in C_o^-\,,
\end{gather}
\ese
where $C_o^\pm$ are both oriented clockwise.
Note that the jump reduces to the function $g(z)$ appearing in~\eref{e:a12trace} 
when the analytic non-principal minors are identically zero.
We then proceed in the same way as for the derivation of \eref{e:a11trace}.
Namely, we first factor out the zeros/poles, by introducing the Blaschke factors
\begin{equation}
\label{e:betasbar}
\=\beta_1(z) = A_{[1,2]}(z) \prod_{n=1}^{N_1} \frac{z-\z_n^*}{z-\z_n} \prod_{n=1}^{N_2} \frac{z-w_n^*}{z-w_n},
\qquad
\=\beta_2(z) = (1/A_{[3,4]}(z)) \prod_{n=1}^{N_1} \frac{z-\z_n^*}{z-\z_n} \prod_{n=1}^{N_2} \frac{z-w_n^*}{z-w_n}.
\end{equation}
Each quantity in~\eref{e:betasbar} has no zeros or poles and approaches 1 as $z \to \infty$ in its appropriate
region of analyticity.
Combining the log of~\eref{e:logsbar} with~\eref{e:betasbar}, applying the Cauchy projector
$P^+$, and simplifying yields \eref{e:a12trace}.
\qed

\proof{Proof of Lemma~\ref{L:scatteringreflectionless}.}
In the reflectionless case, the jump matrix $\@L(z)$ defined in \eref{e:rhpnewjump} is identically zero.
In this case, Eq. \eref{e:phi4} implies $a_{34}(z) = a_{24}(z) = a_{14}(z) \equiv 0$.
Combining this information with the symmetries of the scattering matrices shows that the following entries must
also be identically zero:
$a_{21}(z)$, $a_{31}(z)$, $a_{41}(z)$, $b_{12}(z)$, $b_{13}(z)$, $b_{14}(z)$, $b_{41}(z)$, $b_{42}(z)$, $b_{43}(z)$.
Lemma~\ref{L:relation} implies
\begin{equation*}
a_{12}(z) = - B_{\minorset{1,3,4}{2,3,4}}(z), \qquad a_{13}(z) = B_{\minorset{1,2,4}{2,3,4}}(z).
\end{equation*}
Then $a_{12}(z)$ and $a_{13}(z)$ must also be identically zero.
Similar results follow for $a_{42}(z)$, $a_{43}(z)$, $b_{21}(z)$, $b_{31}(z)$, $b_{24}(z)$, and $b_{34}(z)$ by using
the same method.
Next, the assumption the analytic non-principal minors are identically zero implies that $a_{23}(z) a_{44}(z) \equiv 0$.
Since the scattering matrix is continuous, we conclude $a_{23}(z) \equiv 0$.
The symmetries of the scattering matrix imply the same result for $a_{32}(z)$, $b_{23}(z)$, and $b_{32}(z)$.
Thus, the scattering matrix and its inverse are diagonal.

The converse is trivial.
\qed


\bibliographystyle{amsplain}

\makeatletter
\def\journal#1&#2,#3 (#4){\begingroup
\let\journal=\d@mmyjournal{\frenchspacing\sl #1\/\unskip\,}
{\bf\ignorespaces #2}\rm, #3 (#4)\endgroup}
\def\d@mmyjournal{\errmessage{Reference foul up: nested \journal macros}}
\def\title#1{{``#1''}}
\def\@biblabel#1{#1.}
\makeatother

%
%

\end{document}